\documentclass{jpp}

\usepackage{graphicx}
\usepackage[makeroom]{cancel}

\usepackage{natbib}
\usepackage{amsmath}
\usepackage{amssymb}
\usepackage{color}
\usepackage{bm}
\usepackage{authblk}

\ifCUPmtlplainloaded \else
  \checkfont{eurm10}
  \iffontfound
    \IfFileExists{upmath.sty}
      {\typeout{^^JFound AMS Euler Roman fonts on the system,
                   using the 'upmath' package.^^J}%
       \usepackage{upmath}}
      {\typeout{^^JFound AMS Euler Roman fonts on the system, but you
                   dont seem to have the}%
       \typeout{'upmath' package installed. JPP.cls can take advantage
                 of these fonts, if you use 'upmath' package.^^J}%
      }
  \else
  \fi
\fi

\ifCUPmtlplainloaded \else
  \checkfont{msam10}
  \iffontfound
    \IfFileExists{amssymb.sty}
      {\typeout{^^JFound AMS Symbol fonts on the system, using the
                'amssymb' package.^^J}%
       \usepackage{amssymb}%
       \let\le=\leqslant  
       \let\ge=\geqslant  
      }{}
  \fi
\fi

\ifCUPmtlplainloaded \else
  \checkfont{msam10}
  \iffontfound
    \IfFileExists{amssymb.sty}
      {\typeout{^^JFound AMS Symbol fonts on the system, using the
                'amssymb' package.^^J}%
       \usepackage{amssymb}%
       \let\le=\leqslant  
       \let\ge=\geqslant  
      }{}
  \fi
\fi

\ifCUPmtlplainloaded \else
\IfFileExists{amsbsy.sty}
             {\typeout{^^JFound the 'amsbsy' package on the system, using it.^^J}%
     \usepackage{amsbsy}}
    {}
\fi

\newsavebox{\astrutbox}
\sbox{\astrutbox}{\rule[-5pt]{0pt}{20pt}}

\newcommand{\xhat}{\mbox{$\hat{\mathbf{x}}$}}
\newcommand{\yhat}{\mbox{$\hat{\mathbf{y}}$}}
\newcommand{\zhat}{\mbox{$\hat{\mathbf{z}}$}}

\newcommand{\appref}[1]{Appendix~\ref{#1}}   
                                   
\newcommand{\eqr}[1]{Eq.\thinspace(#1)}
\newcommand{\pfrac}[2]{\frac{\partial #1}{\partial #2}}

\newcommand{\mvec}[1]{\mathbf{#1}}

\newcommand{\gcs}{\nabla}
\newcommand{\gvs}{\nabla_{\mvec{v}}}

\newcommand{\gke}{\texttt{Gkeyll}}

\title[FPC Analysis of Perpendicular Collisionless Shock]{A Field-Particle Correlation Analysis of a Perpendicular Magnetized Collisionless Shock}

\author{James Juno\thanks{Email address
for correspondence: james-juno@uiowa.edu}$^{1}$, Gregory~G.~Howes$^{1}$, Jason M.~TenBarge$^{2}$, Lynn B.~Wilson III$^{3}$, Anatoly Spitkovsky$^2$, Damiano Caprioli$^4$, Kristopher G.~Klein$^5$, Ammar Hakim$^6$}
  
\affiliation{$^1$Department of Physics and Astronomy,University of Iowa, Iowa City IA 54224, USA\\
[\affilskip]
$^2$Department of Astrophysical Sciences, Princeton University, Princeton, NJ 08544, USA\\
[\affilskip]
$^3$ NASA Goddard Space Flight Center, Heliophysics Division, Greenbelt, MD 20771, USA\\
[\affilskip]
$^4$ Department of Astronomy and Astrophysics, University of Chicago, Chicago, Illinois 60637, USA\\
[\affilskip]
$^5$ Lunar and Planetary Laboratory, University of Arizona, Tucson, AZ 85719, USA\\
[\affilskip]
$^6$ Princeton Plasma Physics Laboratory, Princeton, NJ 08543, USA
}

\pubyear{2020}
\volume{?}
\pagerange{?}
\date{?; revised ?; accepted ?. - To be entered by editorial office}

\begin{document}
\maketitle

\begin{abstract}
Using the field-particle correlation technique, we examine the particle energization in a 1D-2V continuum Vlasov--Maxwell simulation of a perpendicular magnetized collisionless shock.
The combination of the field-particle correlation technique with the high fidelity representation of the particle distribution function provided by a direct discretization of the Vlasov equation allows us to ascertain the details of the exchange of energy between the electromagnetic fields and the particles in phase space.
We identify the velocity-space signatures of shock-drift acceleration of the ions and adiabatic heating of the electrons due to the perpendicular collisionless shock by constructing a simplified model with the minimum ingredients necessary to produce the observed energization signatures in the self-consistent Vlasov-Maxwell simulation.
We are thus able to completely characterize the energy transfer in the perpendicular collisionless shock considered here and provide predictions for the application of the field-particle correlation technique to spacecraft measurements of collisionless shocks.
\end{abstract}


\section{Introduction}

Shock-waves, disturbances propagating faster than the largest local wave speed, are ubiquitous in space and astrophysical plasmas. 
From supernova remnants to the Earth's bow shock, we observe a variety of plasma environments in which these shock-waves efficiently convert the bulk kinetic energy of their supersonic flows into other forms of energy, \emph{e.g.}, plasma heat, accelerated particles, and electromagnetic radiation.
Critically, a commonality amongst this variety of plasma environments is that they are weakly collisional, \emph{i.e.}, the collisional mean free path is much larger than the relevant plasma length scales, such as the gyroradius.
Thus, this energy conversion must be mediated by collisionless interactions.
Here, collisionless energy transfer refers to the myriad of mechanisms for transferring energy between the particles in the plasma and the electromagnetic fields, or vice versa, from wave-particle resonances to instabilities.

Diagnosing collisionless energy transfer is a grand challenge in plasma physics, as the plasma has many routes at its disposal for converting energy from one form to another. 
For collisionless shocks, a number of processes have been identified as potential energy transfer mechanisms, and the efficiency of each of these mechanisms is strongly dependent upon factors such as the shock geometry and the fast magnetosonic Mach number $M_{f} = U_{shock}/v_{f}$, where $U_{shock}$ is the shock velocity and $v_{f}$ is the fast magnetosonic wave velocity.
In lower Mach number shocks, dispersive radiation and wave-particle interactions provide an effective resistivity through the shock ramp and the requisite dissipation routes for the shock's energy conversion \citep[e.g.,][and references therein]{Kennel:1985, Balogh:2013}.

As the shock velocity increases though, particles can be reflected in the shock transition, further complicating the energy exchange \citep{Schwartz:1983,Burgess:1984,Scholer:1990,Guo:2013}.
Example energization mechanisms driven by particle reflection in higher Mach number shocks include \emph{shock surfing acceleration} \citep{Sagdeev:1966,Sagdeev:1973,Lever:2001,Shapiro:2003}, \emph{shock-drift acceleration} \citep{Paschmann:1982, Sckopke:1983, Anagnostopoulos:1994, Anagnostopoulos:1998, Ball:2001, Anagnostopoulos:2009, Park:2013}, \emph{diffusive shock acceleration} \citep{Fermi:1949, Fermi:1954, Blandford:1978, Ellison:1983, Blandford:1987, Decker:1988, Malkov:2001, Caprioli:2010b}, and the \emph{``fast Fermi'' mechanism} \citep{Leroy:1984, Wu:1984, Savoini:2010}.
And the picture is not made simpler by the ways in which these different energization mechanisms may interact.
The transition from particles gaining energy via shock-drift acceleration to particles gaining energy via diffusive shock acceleration \citep{Caprioli:2014d}, the onset of upstream kinetic instabilities generated by the reflected particles \citep{Schwartz:1985,Schwartz:1992,Schwartz:1995,Omidi:2010, Wilson:2010, Wilson:2012, Turner:2013, Wilson:2013b, Wilson:2014a, Wilson:2014b}, the electromagnetic fluctuations resulting from these instabilities themselves contributing to the energetics of the shock via processes such as magnetic pumping \citep{Lichko:2017, Lichko:2020}, and the prospect of shock reformation due to the reflected particles, especially reflected ions \citep{Leroy:1983,Kucharek:1991,Giacalone:1992}, all complicate the energy exchange between the plasma and electromagnetic fields through the collisionless shock.
Disentangling the competition between these processes remains challenging.

To ascertain the details of the energy transfer in collisionless shocks, we perform a first principles, continuum kinetic simulation of a perpendicular collisionless shock and utilize the field-particle correlation technique \citep{Klein:2016,Klein:2017a,Klein:2017b,Howes:2017,Howes:2018,Li:2019,Chen:2019,Klein:2020,Horvath:2020} to characterize this energy exchange directly in phase space.
We consider a reduced dimensionality and simplified geometry to isolate the available energization mechanisms available to the plasma, focusing on the energization mechanisms of shock-drift acceleration, for the ions, and adiabatic heating, for the electrons.
We emphasize that, while these processes have been studied previously using kinetic simulations and the particle-in-cell numerical method in higher dimensionality and greater generality \citep[e.g.,][]{Park:2013, XGuo:2014a, XGuo:2014b, Park:2015, Xu:2020}, this study is the first direct diagnosis of the energy transfer in a collisionless shock in phase space and identification of the velocity-space signatures of shock-drift acceleration and adiabatic heating.

This Eulerian perspective---focusing on individual regions of phase space for determining the details of the energy exchange, in contrast to the more commonly used Lagrangian perspective of integrating particle trajectories to identify how individual particles are energized---is of high utility for interpreting spacecraft data.
For example, using Magnetospheric Multiscale mission measurements of the electron distribution function in the Earth's turbulent magnetosheath and the field-particle correlation technique, \citet{Chen:2019} found both the velocity-space signature of electron Landau damping and determined that the observed turbulence was principally dissipating via electron Landau damping.
In this regard, the work presented here is the beginning of a broader program of study to identify the velocity-space signatures of energization mechanisms in collisionless shocks and deploy the field-particle correlation technique for the analysis of energy exchange using \emph{in situ} measurements of collisionless shocks.

To understand the observed velocity-space signatures and connect the resulting signatures to known mechanisms for plasma energization, we construct simplified analytical models for ions and electrons being energized by similar processes absent the complications of a fully self-consistent shock, computing the field-particle correlation on the particle distribution functions predicted by these idealized models.
These simplified models allow us to proceed pedagogically and connect the two distinct pictures, the Eulerian point-of-view for identifying where in phase space the particles are being energized and the Lagrangian point-of-view for analyzing how individual particles gain and lose energy.
And while significant intuition is gained from the Lagrangian perspective, this novel Eulerian perspective provided by the field-particle correlation technique has some advantages, chief among them the ability to easily distinguish how different regions of phase space are being energized.
We will show how the field-particle correlation technique allows us to easily separate the energy exchange occurring between the electromagnetic fields and the multi-component distribution functions (\emph{e.g.}, incoming beam vs.~reflected ions in the shock foot and ramp) which frequently characterize collisionless shock dynamics.
We may thus distinguish the different effects the same electromagnetic fields are having on different parts of the distribution function, from how the cross-shock electric field decelerates the incoming bulk flow and accelerates reflected ions, to how the motional electric field supporting the upstream $\mvec{E} \times \mvec{B}$ motion energizes both the reflected ions and bulk electrons.

The rest of the paper is organized as follows. In Section~\ref{sec:compModel} we provide details of the simulations performed, followed by a broad overview of the results of the simulations examining the shock structure in the electromagnetic fields and electron and ion distribution functions. 
We then present an overview of the central analysis tool of this paper, the field-particle correlation technique, in Section~\ref{sec:fpc}.
We apply the field-particle correlation technique to obtain the key results of the paper in Sections~\ref{sec:ions} and \ref{sec:electrons}: the velocity-space signatures of (i) shock-drift acceleration of the ions and (ii) adiabatic heating of the electrons.
We conclude in Section~\ref{sec:conclusions} with a discussion of the implications of the results presented for spacecraft observations and future avenues of research applying field-particle correlations to a larger range of shock parameters.
    
\section{Computational Model and Overview of Results}\label{sec:compModel}

To perform a self-consistent simulation of a perpendicular collisionless shock, we employ the continuum Vlasov-Maxwell solver in the \gke~framework \citep{Juno:2018,HakimJuno:2020}.
We emphasize that, unlike traditional particle based approaches such as the particle-in-cell method, \gke~directly discretizes the Vlasov-Maxwell system of equations on a phase-space grid to obtain a high fidelity representation of the distribution function, free of the shot noise introduced by finite sized particles.
In other words, we solve the following system of equations with a grid-based method for every species $s$ in the plasma,
\begin{align}
    & \pfrac{f_s}{t} + \gcs \cdot (\mvec{v} f_s) + \gvs \cdot \left ( \frac{q_s}{m_s}[\mvec{E} + \mvec{v} \times \mvec{B}] f_s \right ) = 0 \notag \\
    & \frac{\partial \mvec{B}}{\partial t} + \nabla\times\mvec{E} = 0, \qquad \epsilon_0\mu_0\frac{\partial \mvec{E}}{\partial t} - \nabla\times\mvec{B} = -\mu_0\mvec{J}, \notag  \\
    & \nabla\cdot\mvec{E} = \frac{\varrho_c}{\epsilon_0}, \qquad \nabla\cdot\mvec{B} = 0, \notag
\end{align}
where $f_s = f_s(\mvec{x}, \mvec{v})$ is the particle distribution function for species $s$, $q_s$ and $m_s$ are the charge and mass of species $s$ respectively, $\mvec{E} = \mvec{E}(\mvec{x})$ and $\mvec{B} = \mvec{B}(\mvec{x})$ are the electric and magnetic fields respectively, and the coupling between the electromagnetic fields and particles is given by velocity moments of the particle distribution function,
\begin{align}
    \varrho_c = \sum_s q_s \int f_s \thinspace d\mvec{v}, \qquad \mvec{J} & = \sum_s q_s \int \mvec{v} f_s \thinspace d\mvec{v}, \notag
\end{align}
i.e., the charge and current density.
This approach has been previously leveraged in the study of electrostatic collisionless shocks \citep{Pusztai:2018,Sundstrom:2019}, allowing for a detailed study of the phase-space dynamics which result from the evolution of the shock.
Comparisons to the particle-in-cell method for the study of kinetic instabilities have clearly demonstrated the advantages of using a continuum representation to eliminate  discrete particle noise in the particle velocity distributions \citep{Skoutnev:2019, Juno:2020}. 

Here, a perpendicular shock refers to the orientation of the magnetic field with respect to the shock normal. 
Since the magnetic field is perpendicular to the shock normal, in one spatial dimension, we require only the two velocity components perpendicular to the magnetic field to fully describe the dynamics of the system, i.e., 1D-2V.
The particular 1D geometry we choose is the one spatial coordinate along the shock normal in the $x$ direction, with the initial magnetic field in the $z$ direction, $\mvec{B} (t = 0) = B_0 \mvec{\hat{z}}$.
For completeness, in this dimensionality and field geometry, the Vlasov-Maxwell system of equations is
\begin{align}
    & \pfrac{f_s}{t} + v_x \pfrac{f_s}{x} + \frac{q_s}{m_s} \left ( [E_x + v_y B_z] \pfrac{f_s}{v_x} + [E_y - v_x B_z] \pfrac{f_s}{v_y} \right ) = C[f_s] \notag \\
    & \pfrac{B_z}{t} = -\pfrac{E_y}{x}, \notag \\
    & \pfrac{E_x}{t} = -\frac{J_x}{\epsilon_0}, \qquad \pfrac{E_y}{t} = -c^2 \pfrac{B_z}{x} - \frac{J_y}{\epsilon_0}, \notag
\end{align}
where we have added a collision operator $C[f_s]$ on the right-hand side of of the Vlasov equation\footnote{The addition of a collision operator on the right-hand side of the Vlasov equation introduces some semantic ambiguity of the name of this equation, which, with the inclusion of a collision operator, is now formally the Boltzmann equation and the system of equations the Boltzmann-Maxwell system of equations---see \citet{Henon:1982} for a discussion of this linguistic history. Because the collision operator is principally employed for numerical reasons and to provide velocity-space regularization, we will continue to refer to the equation system of interest as the Vlasov-Maxwell system of equations to emphasize our focus on collisionless physics.}.

The electrons and ions are initialized with the same supersonic flow directed in the negative $x$ direction towards a reflecting wall, which leads to a shock wave that propagates in the positive $x$ direction in our simulation.
Note that the particles reflect from the wall, but the ``reflecting wall'' boundary condition for the electromagnetic fields is a conducting wall boundary condition in the traditional sense, with zero normal magnetic field and zero tangential electric field.
This method of initialization is often called the ``reflecting-wall'' setup\footnote{In contrast to the ``injection'' set-up where an injection boundary condition would be employed on each side of the domain and the plasma blocks would collide and form a shock in the middle of the domain.} and has been previously employed in numerous particle-in-cell studies of collisionless shocks \citep[e.g.,][]{Papadopoulos:1971, Spitkovsky:2005, Spitkovsky:2007}.

Detailed parameters are as follows: the reflecting wall for the particles and conducting wall for the electromagnetic fields are at $x = 0$, and plasma is injected with a copy boundary condition\footnote{For \gke, this means that the value in the layer of cells beyond the $x = 25 d_i$ edge (the ``ghost'' or ``halo'' layer) is exactly equal to the value in the layer of cells at the $x = 25 d_i$ edge, for all the quantities being evolved: the distribution functions for the electrons and ions, and the electromagnetic fields. Because the plasma is initialized with a flow propagating in the negative $x$ direction, this boundary condition leads to a continuous injection of plasma from the right wall with the corresponding electric field and magnetic field to support the $\mvec{E} \times \mvec{B}$ flow.} at $x = 25 d_i$, where $d_i =  c/\omega_{pi}$ is the ion collisionless skin depth.
Here, $c$ is the speed of light, and $\omega_{pi} = \sqrt{e^2 n_0/\epsilon_0 m_i}$ is the ion plasma frequency.
Note that the subscript $0$ denotes the upstream value, e.g., $n_0$ is the upstream density and $B_0$ is the upstream magnetic field magnitude.
We use a reduced mass ratio between the ions and electrons, $m_i/m_e = 100$.
The total plasma beta, $\beta = 2 \mu_0 n_0 (T_{e_0} + T_{i_0})/B_0^2 = 2$, with the ion beta, $\beta_i = 1.3$, and electron beta, $\beta_e = 0.7$.
Both the ions and electrons are non-relativistic, with $v_{te}/c = 1/16$, where $v_{ts} = \sqrt{2 T_{s_0}/m_s}$.
With this choice of electron beta and $v_{te}/c$, the ratio of the electron plasma frequency, $\omega_{pe} = \sqrt{e^2 n_0/\epsilon_0 m_e}$, to the electron cyclotron frequency, $\Omega_{ce} = -e B_0/m_e$, is $\omega_{pe}/\Omega_{ce} \sim 13.4$.
The in-flow velocity in the simulation frame to initialize the perpendicular, electromagnetic shock is $U_x = -3 v_A$, where $v_A = B_0/\sqrt{\mu_0 n_0 m_i}$ is the ion Alfv\'en speed.
Note that the in-flow velocity is negative because the plasma initially flows in the negative $x$ direction.
Since the plasma is initialized with a flow in a background magnetic field, we initialize the corresponding electric field to support this flow, $\mvec{E} = -U_x \hat{\mvec{x}} \times \mvec{B} = U_x B_0 \mvec{\hat{y}}$.

For the grid resolution in configuration space, we use $N_x = 1536$ grid cells, corresponding to $\Delta x \sim d_e/6 \sim 3.7 \lambda_D$, where $d_e = c/\omega_{pe}$ and $\lambda_D = v_{te}/(\sqrt{2} \omega_{pe})$ are the electron inertial length and electron Debye length respectively, and we employ piecewise quadratic Serendipity elements for the discontinuous Galerkin basis expansion \citep{Arnold:2011}.
In velocity space, the electron extents are $\pm 8 v_{te}$, and the ion extents are $\pm 16 v_{ti}$, with zero-flux boundary conditions at the velocity-space limits and $N_{v_x} = N_{v_y} = 64$ grid cells for both species, corresponding to $\Delta v = v_{te}/4$ for the electrons and $\Delta v = v_{ti}/2$ for the ions.
The basis expansion in velocity space is also piecewise quadratic Serendipity elements.
For further details about the algorithm and the choice of basis expansion, we refer the reader to \citet{Juno:2018} and \citet{HakimJuno:2020}.

We have run the simulation with a small amount of collisions to regularize velocity space. 
In this case, we choose an electron-electron collision frequency, $\nu_{ee} = 0.01 \Omega_{ci}$, much less than the ion cyclotron frequency, $\Omega_{ci} = e B_0/m_i$, with the ion-ion collision frequency correspondingly smaller based on the square root of the mass ratio, $\nu_{ii} = 0.001 \Omega_{ci}$.
Note that because the ions are hotter than the electrons, they should formally be even more collisionless than the electrons; however, these collisionalities are larger than typical solar wind collisionalities \citep{Wilson:2018} and are not chosen to be realistically small, but instead chosen to be just large enough to provide regularization of velocity-space structure given finite velocity resolution. 
Details on the implementation of the collision operator and its conservation properties can be found in \citet{Hakim:2019}.

\begin{figure}
    \centering
    \includegraphics[width=.95\textwidth]{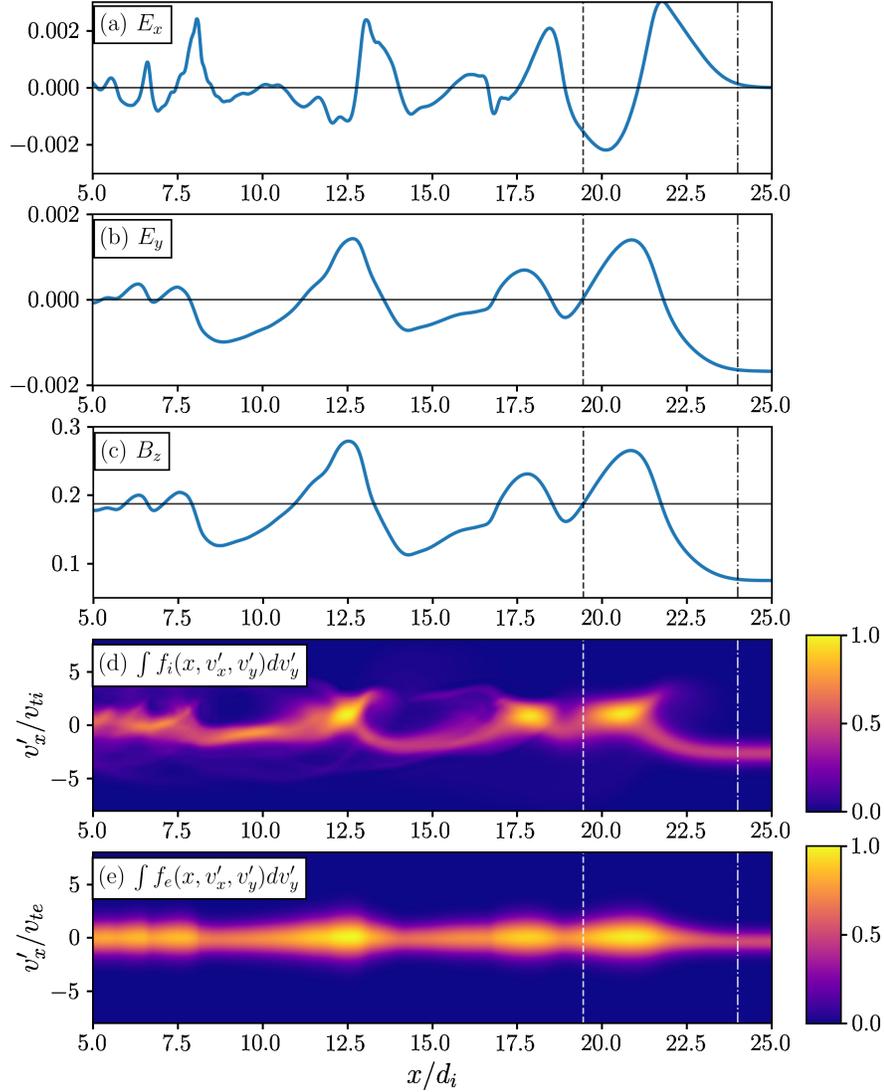}
    \vspace{-1.0cm}
    \caption{The x-electric field (a), y-electric field (b), z-magnetic field (c), ion distribution function integrated in $v'_y$ (d), and electron distribution function integrated in $v'_y$ (e) after the perpendicular shock has formed and propagated through the simulation domain, $t = 11 \Omega_{cp}^{-1}$. Note that the distribution functions are plotted in the simulation frame $f_s(x, v_x', v_y')$ for each species $s$. We have marked an approximate transition from upstream of the shock to the shocked plasma (dashed-dotted line), and likewise an approximate transition from the shock to the dowstream region (dashed line). To mark the oscillation of the electromagnetic fields and the sloshing of energy between the fields and particles in the downstream region, we have used a solid black line to mark the approximate compression of the magnetic field, along with $\mvec{E} = 0$. We expect the y-electric field to roughly oscillate about zero in the frame of the simulation, as the ``reflecting-wall'' set-up is performed in the frame of the downstream plasma, where the $\mvec{E} \times \mvec{B}$ velocity is zero.}
    \label{fig:perpShockFields}
\end{figure}

In Figure~\ref{fig:perpShockFields}, we show the electromagnetic fields and particle distribution functions for the electrons and ions in $x-v_x$ phase space after the perpendicular shock has formed and propagated through the simulation domain, $t_{end} = 11 \Omega_{ci}^{-1}$.
Although the downstream of the shock is fairly oscillatory as the energy injected into the plasma by the shock sloshes back and forth between the electromagnetic fields and particles, we can estimate the compression ratio of this low Mach number shock from the average downstream magnetic field to be roughly $r \sim 2.5$ (solid black line in $B_z$ plot in Figure~\ref{fig:perpShockFields}).
With this estimate for the compression ratio, we calculate the shock velocity in the simulation frame to be $U_{shock} = U_x/(r-1) = 2 v_A$.
Note that in this reflecting wall set-up, the simulation frame is equivalent to the frame in which the downstream plasma is at rest.

Thus, combining the velocity of the incoming flow with the velocity of the shock in the downstream frame, this self-consistently produced perpendicular shock is a $M_A = 5$ shock, where $M_A$ is the Alfv\'en Mach number.
Equivalently, using the definitions for the sound speed and magnetosonic speed given by
\begin{align}
    c_s & = \sqrt{\frac{\gamma_i T_i + \gamma_e T_e}{m_i}}, \\
    v_{f} & = \sqrt{c_s^2 + v_A^2},
\end{align}
where $\gamma_i = \gamma_e = 1 + 2/\textrm{VDIM} = 2$ since the simulation domain has two velocity dimensions, we find this shock has fast magnetosonic Mach number, $M_{f} \approx 2.89$.
With these plasma parameters and this magnetosonic Mach number, we note that this shock is super-critical $M_f> M_{f_{crit}} \simeq 2$ \citep{Wilson:2016a}, similar to the Earth's bow shock, and thus bodes well for the ultimate goal of predicting velocity-space signatures of energization mechanisms in spacecraft observations of heliospheric shocks.

\begin{figure}
    \centering
    \includegraphics[width=\textwidth]{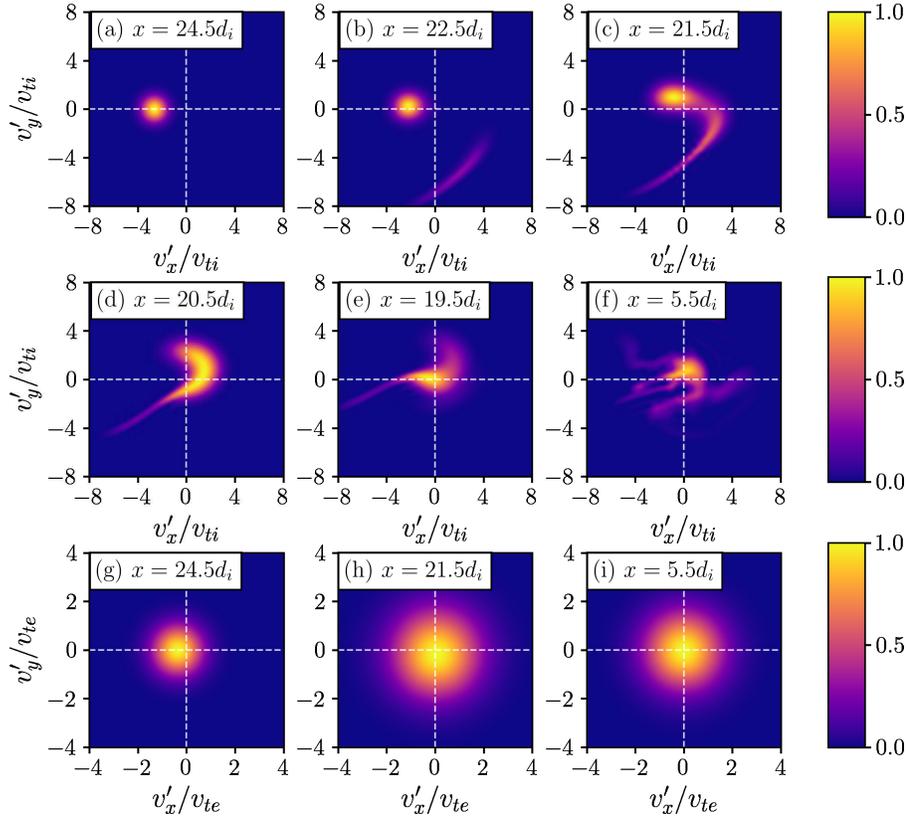}
    \caption{The ion (panels (a)--(f)) and electron (panels (g)--(i)) distribution functions in the simulation frame (downstream frame), $f_s(x, v_x', v_y')$ for each species $s$, plotted at various points through the shock at $t = 11 \Omega_{cp}^{-1}$. As we move from upstream, $x = 24.5 d_i$, through the shock ramp, $x = 21.5 d_i$, we can identify the reflected ion population as well as a broadening of the electron distribution function.}
    \label{fig:distThroughShock}
\end{figure}

In this regard, we focus our attention now on the particle distribution functions and the phase-space structure generated through the shock.
While the particle distribution functions in $x-v_x$ phase space shown in Figure~\ref{fig:perpShockFields} are illustrative of the dynamics through the shock, showing a reflected population of ions (d) and a clear compression of the electrons (e), we can gain further insight into the dynamics of this shock by looking at the distribution function in $v_x-v_y$ at fixed points in configuration space through the shock.
In Figure~\ref{fig:distThroughShock}, we plot the ion ((a)--(f)) and electron ((g)--(i)) distribution functions in velocity space at several points through the shock, from upstream through the ramp to downstream.

As an example of the wealth of data contained in the distribution function, we draw special attention to the ion distribution function in the shock ramp. 
As we move from upstream into the shock, at the beginning of the ramp at $x=22.5d_i$, we begin to see a small population of reflected ions, forming a small ``crescent'' distribution in the lower right quadrant of $v_x-v_y$ space. 
Further up the ramp at $x=21.5d_i$, we observe that the incoming ion beam begins to be deflected by the fields in the shock transition, generating a ``boomerang'' distribution that smoothly connects the decelerated incoming ion beam with the reflected ion population.
It is this reflected population, in agreement with previous studies of super-critical shocks \citep[e.g.,][]{Ball:2001, Balogh:2013}, which dominates the ion energization and provides a segue to our key result: diagnosing the velocity-space signatures of particle energization in this perpendicular electromagnetic shock.
To obtain these velocity-space signatures, we now describe our tool of choice for our analysis of the high quality distribution function data provided by the continuum kinetic simulation: the field-particle correlation technique. 


\section{The Field-Particle Correlation Technique}\label{sec:fpc}
From combining the Vlasov equation and Maxwell's equations, we can obtain a conservation equation for the total energy of the kinetic plasma \citep[\emph{e.g.},][]{Klein:2017b},
\begin{align}
  W =  \int d\mvec{x} \left(\frac{\epsilon_0}{2} |\mvec{E}|^2 + \frac{1}{2\mu_0}|\mvec{B}|^2\right)
  + \sum_s  \int d\mvec{x} \int d\mvec{v} \ \frac{1}{2}m_s v^2 f_s.
  \label{eq:vmcons}
\end{align}
The first integral represents the energy of electromagnetic fields in the plasma and the second accounts for the combined microscopic kinetic energy\footnote{Note that the microscopic kinetic energy of a plasma species $s$ includes contributions from the kinetic energy of bulk plasma flows (associated with the first-moment of the distribution) as well as thermal and non-thermal energy contained in the second moment of the particle velocity distribution $f_s(\mvec{v})$. One cannot extract energy from the thermal component, of course, and the irreversible, entropy increasing conversion of free energy in the non-thermal component to thermal energy is dictated by the physics of nonequilibrium thermodynamics in this kinetic system.} of all plasma species $s$. 
In the absence of particle collisions, the net microscopic kinetic energy of a given plasma species may only be changed through collisionless interactions between the particles of that species and the electromagnetic fields. 

To explore the energy transfer between fields and particles, we define the \emph{phase-space energy density} for a particle species $s$ by $w_s(\mvec{x},\mvec{v},t) \equiv m_s v^2 f_s(\mvec{x},\mvec{v},t)/2$ in the non-relativistic limit.  
Multiplying the Vlasov equation by $m_s v^2/2$, we obtain an expression for the rate of change of this phase-space energy density,
\begin{align}
  \frac{\partial w_s(\mvec{x},\mvec{v},t)}{\partial t} = - \mvec{v}\cdot \nabla  w_s  -
  q_s\frac{v^2}{2}  \mathbf{E} \cdot \frac{\partial f_s}{\partial \mathbf{v}}
   - q_s\frac{v^2}{2} \left(\mathbf{v} \times \mathbf{B}\right)
      \cdot \frac{\partial f_s}{\partial \mathbf{v}}.
  \label{eq:dws} 
\end{align}
This equation describes the mechanisms that govern how  the energy density in the 3D-3V phase space $(\mvec{x},\mvec{v})$ evolves, where each term has a clear physical interpretation. 

The first term on the right-hand side of \eqref{eq:dws} describes how $w_s(\mvec{x},\mvec{v},t)$  changes due to particle advection from other spatial regions, giving rise in fluid theory to the energy change through pressure forces and heat fluxes\footnote{In the context of fluid theory, it has been shown that these pressure forces can mediate the conversion of bulk flow kinetic energy to random kinetic energy in the velocity distribution \citep{Yang:2017b}. This distinction between energy converstion and energization is further discussed in \appref{app:energizationVsEnergyConversion}.}. 
Because this term describes the advection of particle kinetic energy as particles move from one spatial position to another, when integrated over the full plasma volume, this term yields zero net change of the total kinetic energy of particle species $s$, $\mathcal{W}_s= \int d\mvec{x} \int d\mvec{v} \ \frac{1}{2}m_s v^2 f_s$. 
The third term on the right-hand side of \eqref{eq:dws} describes the magnetic forces on the particles. 
Although this term can move kinetic energy from one location in velocity space to another, when integrated over all velocity space, this term does zero net work on the particles, as expected for the magnetic force. 

The second term on the right-hand side of \eqref{eq:dws} describes the work done on the plasma species $s$ by the electric field. 
When \eqref{eq:dws}  is integrated over all velocity space and all physical space to obtain the rate of change of the total kinetic energy $\mathcal{W}_s$ of a particle species $s$, the first and third terms have zero net contribution
\citep{Klein:2016,Howes:2017}, yielding
\begin{align}
  \frac{\partial \mathcal{W}_s}{\partial t} = - \int d\mvec{x} \int d\mvec{v}
\  q_s \frac{v^2}{2}   \frac{\partial f_{s}}{\partial \mvec{v}} \cdot \mvec{E}
 = \int d\mvec{x}\left( \int d\mvec{v} \ q_s
  \mvec{v}  f_{s} \right) \cdot \mvec{E}= \int d\mvec{x} \ \mvec{j}_s \cdot \mvec{E},
  \label{eq:dwsdt}
\end{align}
This expression makes clear that the change in species energy $\mathcal{W}_s$ is due to work done on that species by the electric field, $\mvec{j}_s \cdot \mvec{E}$. 

In our exploration of particle energization at collisionless shocks, we choose to focus on the second term in \eqref{eq:dws} to investigate the energization of the particles by the electric field. 
The form of that term demonstrates that the rate of particle energization can be computed at a single-point in physical space $\mvec{x}_0$ by measuring the electric field at that position $\mvec{E}(\mvec{x}_0)$ and the particle velocity distribution at the same position $f_s(\mvec{x}_0,\mvec{v})$. 
This fundamental fact underlies the field-particle correlation (FPC) technique \citep{Klein:2016, Howes:2017, Klein:2017b}, where the unnormalized correlation (essentially a time-average) of the product of the electric field $\mvec{E}(\mvec{x}_0)$ and a term that depends on the particle velocity distribution $f_s(\mvec{x}_0,\mvec{v})$ over some correlation interval $\tau$ is computed by
\begin{align}
    C_\mvec{E}(\mvec{x}_0,\mvec{v},t,\tau) =  \frac{1}{\tau} \int_{t-\tau/2}^{t+\tau/2} \left[-q_s\frac{v^2}{2} 
    \frac{\partial f_s(\mvec{x}_0,\mvec{v},t')}{\partial \mvec{v}} \right] \cdot \mvec{E}(\mvec{x}_0,t')\thinspace  dt'. \label{eq:FPC-full}
\end{align}

The resulting correlation $C_\mvec{E}(\mvec{x}_0,\mvec{v},t,\tau)$ directly measures the rate of change of phase-space energy density at position $\mvec{x}_0$ as a function of 3V particle velocity space $\mvec{v}$, producing a \emph{velocity-space signature} that is characteristic of the mechanism of energization and can be used to identify a particular, locally-occurring energization process, \emph{e.g.}, Landau damping \citep{Howes:2017, Klein:2017b} and cyclotron damping \citep{Klein:2020}.
We note that as part of this identification, further analysis may be required to ascertain certain details; for example, if one obtains velocity-space signatures corresponding to the presence of Landau damping, the resonant velocity the velocity-space signature is concentrated around is necessary to determine what wave modes are Landau damping in the plasma, as one can find similar structure whether a Langmuir wave \citep{Howes:2017} or kinetic Alfvén wave \citep{Klein:2017b, Horvath:2020} is undergoing Landau damping.
However, even with this caveat, a key advantage of the FPC method to diagnose particle energization is that it requires only measurements at a single spatial point $\mvec{x}_0$ to determine the energization by the electric field. 
An appropriately instrumented \emph{single} spacecraft mission can provide the requisite full 3V particle velocity distribution $f_s(\mvec{x}_0,\mvec{v},t)$ and electric field  $\mvec{E}(\mvec{x}_0,t)$ at the spacecraft position $\mvec{x}_0$ as a function of time. 
Thus, the velocity-space signatures determined here using kinetic numerical simulations, our key results, may be directly sought using spacecraft observations.

In the case of particle energization as a consequence of the dissipation of weakly collisional plasma turbulence, the rate of particle energization represented by the second term in \eqref{eq:dws} generally includes two distinct contributions: (i) an often large-amplitude oscillatory component that leads to zero net energization which is associated with undamped wave motion, and (ii) a typically smaller amplitude secular component that corresponds to the net collisionless transfer of energy from the fields to the particles \citep{Klein:2016,Howes:2017}. 
By an appropriate choice of the correlation interval $\tau$, the oscillatory energy transfer is largely eliminated by the time-average, exposing the secular energy transfer associated with the collisionless damping of the turbulent fluctuations.  
For the perpendicular collisionless shock in this study, the shock is quasi-stationary in the shock-rest frame of reference, with smooth electromagnetic fields through the shock as seen in Figure~\ref{fig:perpShockFields}, and thus we need not time-average the correlation, but instead take the instantaneous correlation (the limit $\tau \rightarrow 0$). 
We will thus suppress the dependence of the correlation on $\tau$ henceforth.
We note that the FPC with $\tau=0$ is simply the instantaneous rate of change of the phase-space energy density, $\partial w_s/\partial t$, due to work done on the particles by the electric field. 
If kinetic instabilities were to arise upstream or within the shock transition region, or if the shock itself were to become nonstationary, then it is likely that taking a correlation interval $\tau$ longer than either the unstable wave period or the shock reformation time would be necessary to recover a meaningful velocity-space signature of the net particle energization.

In addition, we adopt two final modifications of the FPC analysis that are well suited for the study of collisionless shocks: (i) we separate the contributions to the rate of energization by the different components of the electric field, $E_x$ and $E_y$; and (ii) we replace $v^2$ in \eqr{\ref{eq:FPC-full}} by the component associated with the electric field, e.g., using $v_x^2$ for the correlation using $E_x$. 
We refer the reader to \appref{app:v2FPC} for a discussion of the validity and usefulness of this transformation.
Therefore, the form of the FPCs implemented here for a position $x=x_0$ in our 1D-2V \gke ~simulation is given by 
\begin{align}
    C_{E_x} (x_0, v_x, v_y, t) & = -q_s \frac{v_x^2}{2} E_x(x_0,t) \pfrac{f_s(x_0, v_x, v_y, t)}{v_x}, \label{eq:vxCorrelation} \\
    C_{E_y} (x_0, v_x, v_y, t) & = -q_s \frac{v_y^2}{2} E_y(x_0,t) \pfrac{f_s(x_0, v_x, v_y, t)}{v_y}.  \label{eq:vyCorrelation}
\end{align}

An issue which cannot be overemphasized in performing the FPC analysis of a collisionless shock is making a judicious choice of the frame of reference in which to calculate  \eqr{\ref{eq:vxCorrelation}} and \eqr{\ref{eq:vyCorrelation}} \citep{Goodrich:1984}. 
We choose to evaluate the correlations in the frame of reference in which the shock is at rest (the shock-rest frame, unprimed variables), as opposed to the frame of reference of the simulation, in which the plasma is at rest downstream of the shock  (the simulation frame or downstream frame, primed variables)\footnote{Note that in both cases, because the one dimensional spatial coordinate is aligned with the shock normal, both frames of reference are normal incidence frames.}. 
For clarity, the shock velocity in the simulation frame is given by $\mvec{U}_{shock} = U_{shock} \xhat = 2 v_A \xhat$.
It is critical not only that the velocity coordinates are transformed to the shock-rest frame, $\mvec{v}=\mvec{v}'-\mvec{U}_{shock}$, but also that the electromagnetic fields are appropriately Galilean transformed to the shock-rest frame,
\begin{align}
    \mvec{E} = \mvec{E}' + \mvec{U}_{shock} \times \mvec{B}'
\end{align}
and $\mvec{B}=\mvec{B}'$.

We note that our discussion and application of the FPC to the collisionless shock is principally concerned with how the plasma is energized via the electromagnetic fields and focuses on the phase-space dynamics governed by the electric field term in the Vlasov equation. 
As mentioned previously, plasmas additionally convert bulk kinetic to thermal energy, and vice versa, via other terms in \eqref{eq:dws} such as the $\mvec{v} \cdot \nabla$ term, which gives rise to pressure forces and heat fluxes. 
To explore how these other physical mechanisms impact the flow of energy through 3D-3V phase space, one can perform complementary correlations with these other terms. 
Correlating with the magnetic term in the Lorentz force allows the determination of how the magnetic field leads to changes in $w_s(\mvec{x},\mvec{v},t)$ as a function of velocity $\mvec{v}$---\emph{e.g.}, energy can be moved between different degrees of freedom by the magnetic field, even though the net energy change (integrated over velocity space) must always be zero.
Similarly, if spatial gradients of $f_s(\mvec{x},\mvec{v},t)$ are available, the velocity-space signatures of the work done on the particles by the pressure tensor can be determined. 
Of course, computing the total rate of change of the phase-space energy density  $w_s(\mvec{x},\mvec{v},t)$ at a particular point in configuration and velocity space requires all terms of \eqref{eq:dws}.

Our focus here, however, is on the term in the Vlasov equation which produces net energization of a plasma species $s$, the electric field term in \eqref{eq:dws}. 
In fact, as shown in \appref{app:exb}, these additional terms such as the $\mvec{v} \times \mvec{B}$ term, can have a cancellation effect on the evolution of the phase-space energy density, so that the net energization due to, for example, an $\mvec{E} \times \mvec{B}$ drift is identically zero, as it should be.
As such, we are well justified in formulating the FPC to focus only on the net energization and avoid obfuscating the signatures of energization with the additional motion of phase-space energy density due to these other terms in the Vlasov equation. 
For a further discussion of energization versus energy conversion, we refer the reader to \appref{app:energizationVsEnergyConversion}.


\section{Field-Particle Correlation Analysis: Ions}\label{sec:ions}

\subsection{Velocity-Space Signature of Ion Energization}
\begin{figure}
 \begin{center}
     \includegraphics[width=0.5\textwidth]{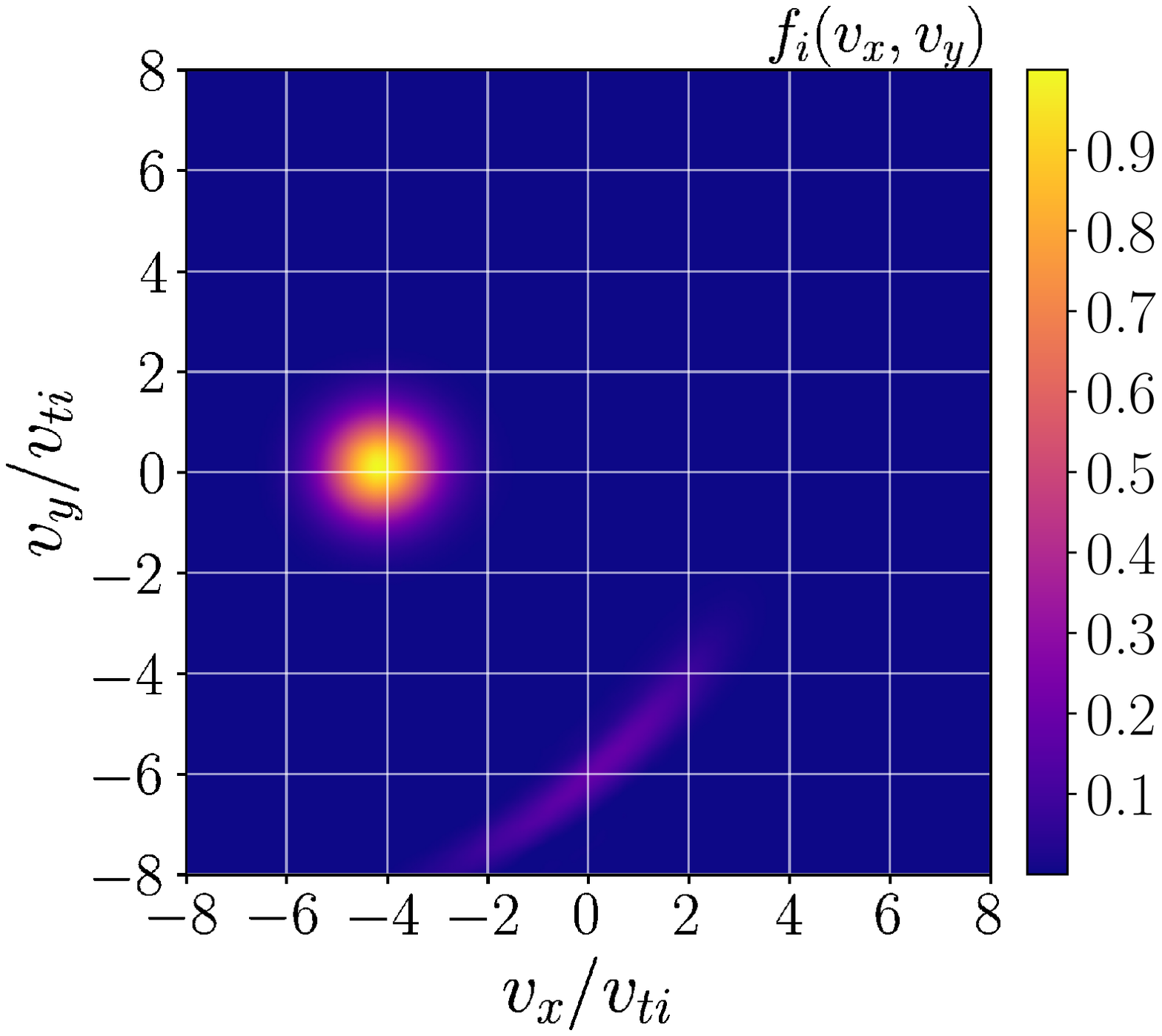}
   \end{center}
 \vskip -2.0in
\hspace*{1.35in} (a)
\vskip +1.85in
\begin{center}
     \includegraphics[width=0.49\textwidth]{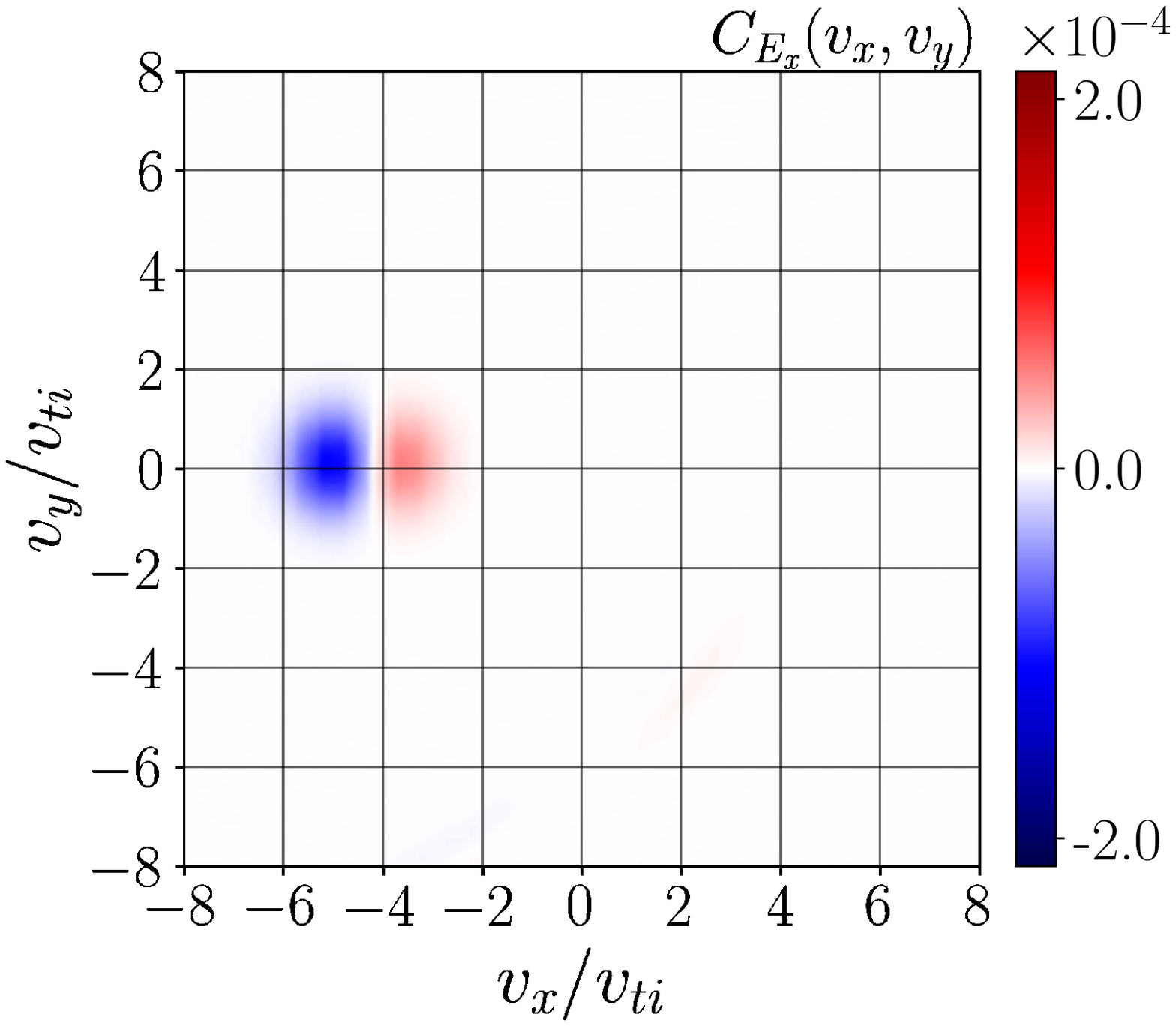}
      \includegraphics[width=0.49\textwidth]{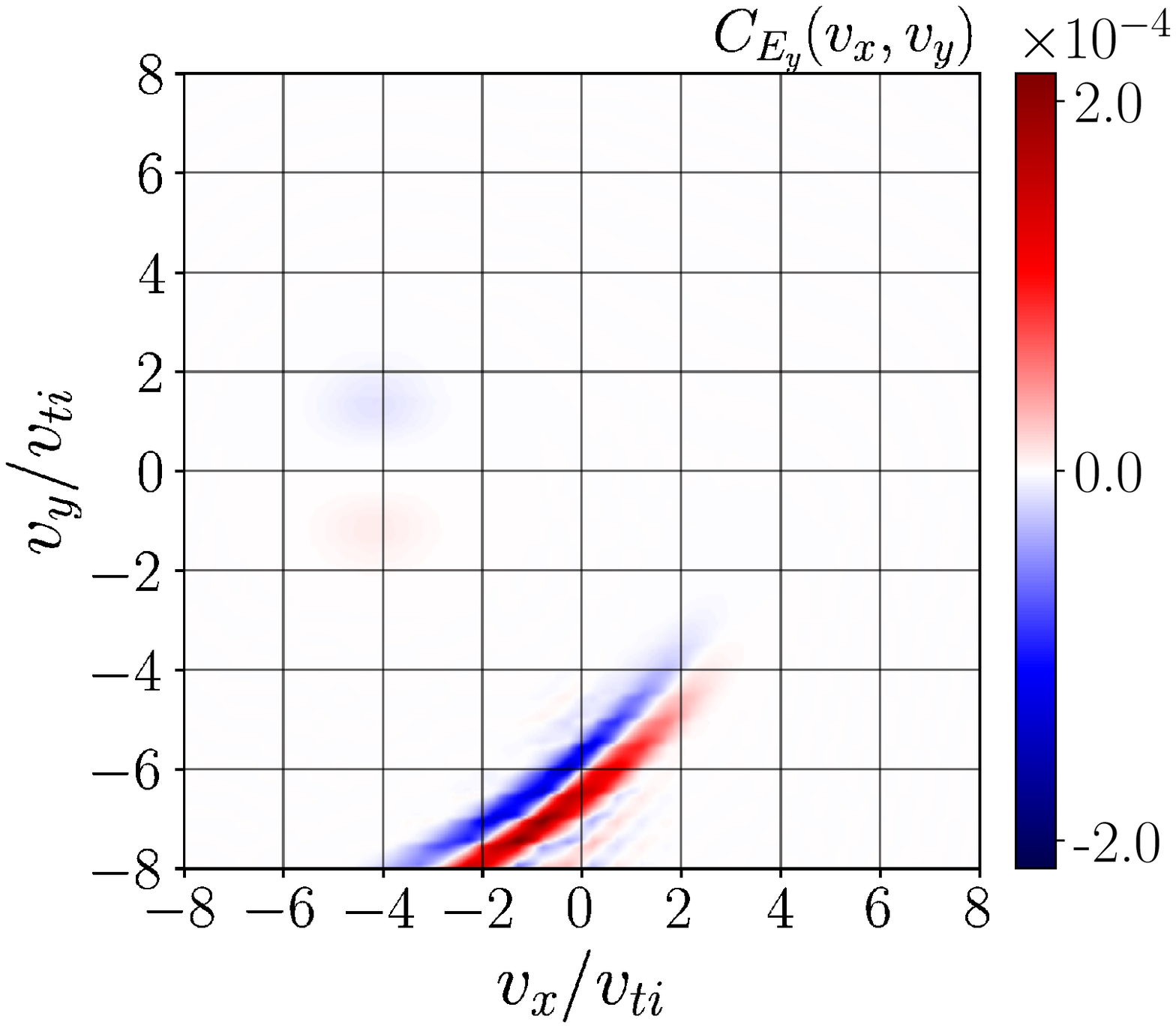}
        \end{center}
 \vskip -2.0in
\hspace*{0.05in} (b)\hspace*{2.45in} (c)
\vskip +1.85in
    \caption{The ion distribution function  $f_i(v_x,v_y)$ (a), and the $C_{E_x}$ (b) and $C_{E_y}$ (c) components of the FPC, \eqr{\ref{eq:vxCorrelation}} and \eqr{\ref{eq:vyCorrelation}}, computed  at $x = 22.9 d_i$ from the self-consistent \gke~simulation. Note that the FPC is computed in the shock-rest frame. While the bulk incoming ions are slowed down by the cross-shock electric field, $E_x$, we see the distribution of reflected ions gain energy due the motional electric field, $E_y$, which supports the incoming supersonic $\mvec{E} \times \mvec{B}$ flow.}
    \label{fig:ficexcey}
\end{figure}

In Figure~\ref{fig:ficexcey}, we plot the ion distribution function $f_i(v_x,v_y)$ (a), and the $C_{E_x}$ (b) and $C_{E_y}$ (c) components of the FPC, \eqr{\ref{eq:vxCorrelation}} and \eqr{\ref{eq:vyCorrelation}}, at $x = 22.9 d_i$.
The ion distribution function includes both a component from the incoming beam of ions upstream, as well as the aforementioned ``crescent'' population of reflected ions.
The $E_x$ contribution to the FPC, $C_{E_x}(v_x,v_y)$, at position $x = 22.9 d_i$, panel (b), shows that the incoming ion beam is being acted upon strongly by the cross-shock electric field $E_x$, but that $E_x$ has little effect on the reflected ion population at this position.  
On the other hand, the $E_y$ contribution to the FPC, $C_{E_y}(v_x,v_y)$, at position $x = 22.9 d_i$, panels (c), shows that the reflected ions principally interact with this component of the electric field, i.e., the motional electric field which supports the incoming $\mvec{E} \times \mvec{B}$ flow.

To understand this visual representation of the rate of ion energization over velocity space, recall that the FPC determines the rate of change of the phase-space energy density of a particular plasma species,  $w_s(\mvec{x},\mvec{v},t) = m_s |\mvec{v}|^2 f_s(\mvec{x},\mvec{v},t)/2$ due to the electric field. 
The phase-space energy density of the ions, $w_i$, can only change if the number of ions in that volume of phase space changes. 
Therefore, nested blue and red crescents in Figure~\ref{fig:ficexcey}(c) indicate that ions are accelerated by $E_y$ from the blue region to the red region. 
Conservation of particle number requires that the number of ions lost from the blue region is the same as the number gained in the red region, but because that red region is at higher velocity $v_y$, the net effect, obtained by integrating $C_{E_y}$ over velocity space $(v_x,v_y)$,  is an increase in the ion phase-space energy density $w_i$.
We also note that the observed $C_{E_y}$ signature is a larger amplitude than the observed $C_{E_x}$, such that $E_y$ dominates the energy exchange at this particular point in space.
Furthermore, the FPC method computes the rate of change of energy density, so the rate of energization \emph{per ion} in the low density population of reflected ions is much higher in amplitude than the loss of energy \emph{per ion} by the much more dense incoming beam.

As a first attempt to understand this signature, consider that the gradient length scale of the collisionless shock in our simulation is $L_{shock} \sim \rho_i$, where $\rho_i = v_{t_i}/\Omega_{ci}$ is the ion Larmor, or gyro-, radius.
Therefore, ions encountering this gradient in the magnetic field will not necessarily have closed orbits and smoothly transition downstream. 
Depending on an ion's gyrophase when it encounters this magnetic field gradient, the ion's new Larmor orbit may cause the ion to move back upstream, where the magnetic field magnitude is smaller.
The increased Larmor radius of this reflected ion in the upstream region then allows the ion to gain energy along the motional electric field supporting the incoming $\mvec{E} \times \mvec{B}$ motion.
This energization of the reflected ion population via $E_y$ is consistent with the well-known energization mechanism \emph{shock-drift acceleration} \citep{Paschmann:1982, Sckopke:1983, Anagnostopoulos:1994, Anagnostopoulos:1998, Ball:2001, Anagnostopoulos:2009, Park:2013}.
But to understand why shock-drift acceleration would produce the particular velocity-space signature observed in panel (c) of Figure~\ref{fig:ficexcey}, we turn to a simplified analytic model to connect the well-known Lagrangian picture for shock-drift acceleration with the new Eulerian perspective granted by the FPC.


\subsection{Shock-Drift Acceleration in an Idealized Perpendicular Shock}\label{sec:ion-SPM}
We consider now a simplified reduction of the electromagnetic fields observed in our self-consistent simulation to a step function in the magnetic field,
\begin{align}
   B_z(x) = \left\{ \begin{array}{cc}
   B_{u} & x \ge 0\\
   B_{d} & x <0\\
   \end{array} \right.
\end{align}
with amplitude jump $B_{d}/B_{u}=4$.
We will also continue to work exclusively in the shock-rest frame, where to a good approximation the motional electric field, $E_y$, is a constant through the entire shock. 
The value of the constant $E_y$, as well as the ion and electron plasma betas, are chosen so that the shock velocity is similar to the self-consistent simulation, $M_A = 4.9$ and $M_f = 3.0$.
This reduced model corresponds to the limit  $L_{shock}/\rho_i \ll 1$ and allows us to decompose the ion motion more easily between upstream and downstream gyro- and $\mvec{E} \times \mvec{B}$ motion.
To mimic the geometry of the self-consistent simulation, we take $E_y < 0$ and $B_z > 0$ so that the inflow $\mvec{E} \times \mvec{B}$ is in the negative $x$ direction.

\begin{figure}
    \begin{center}
      \includegraphics[width=0.48\textwidth]{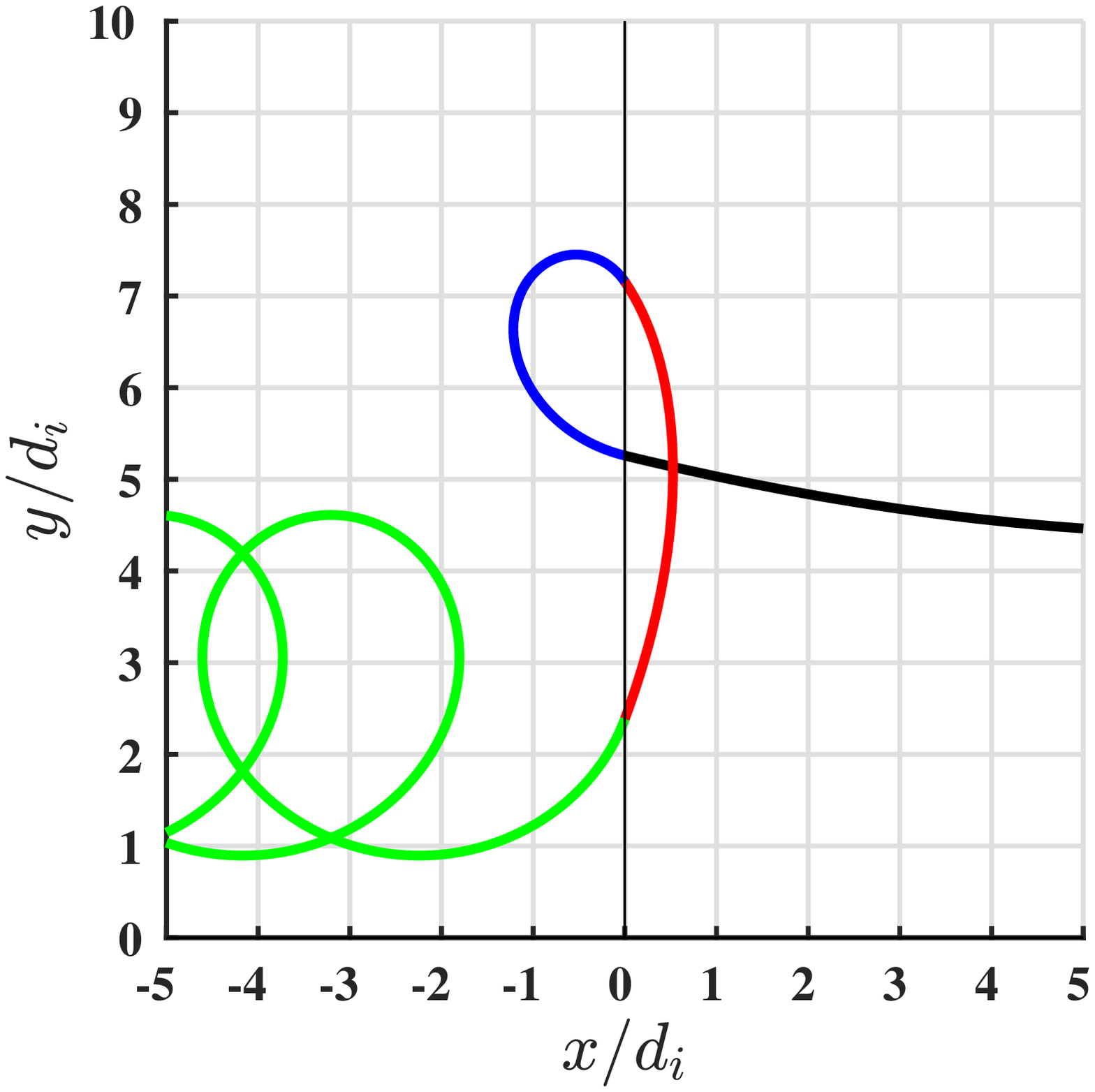}
      \includegraphics[width=0.48\textwidth]{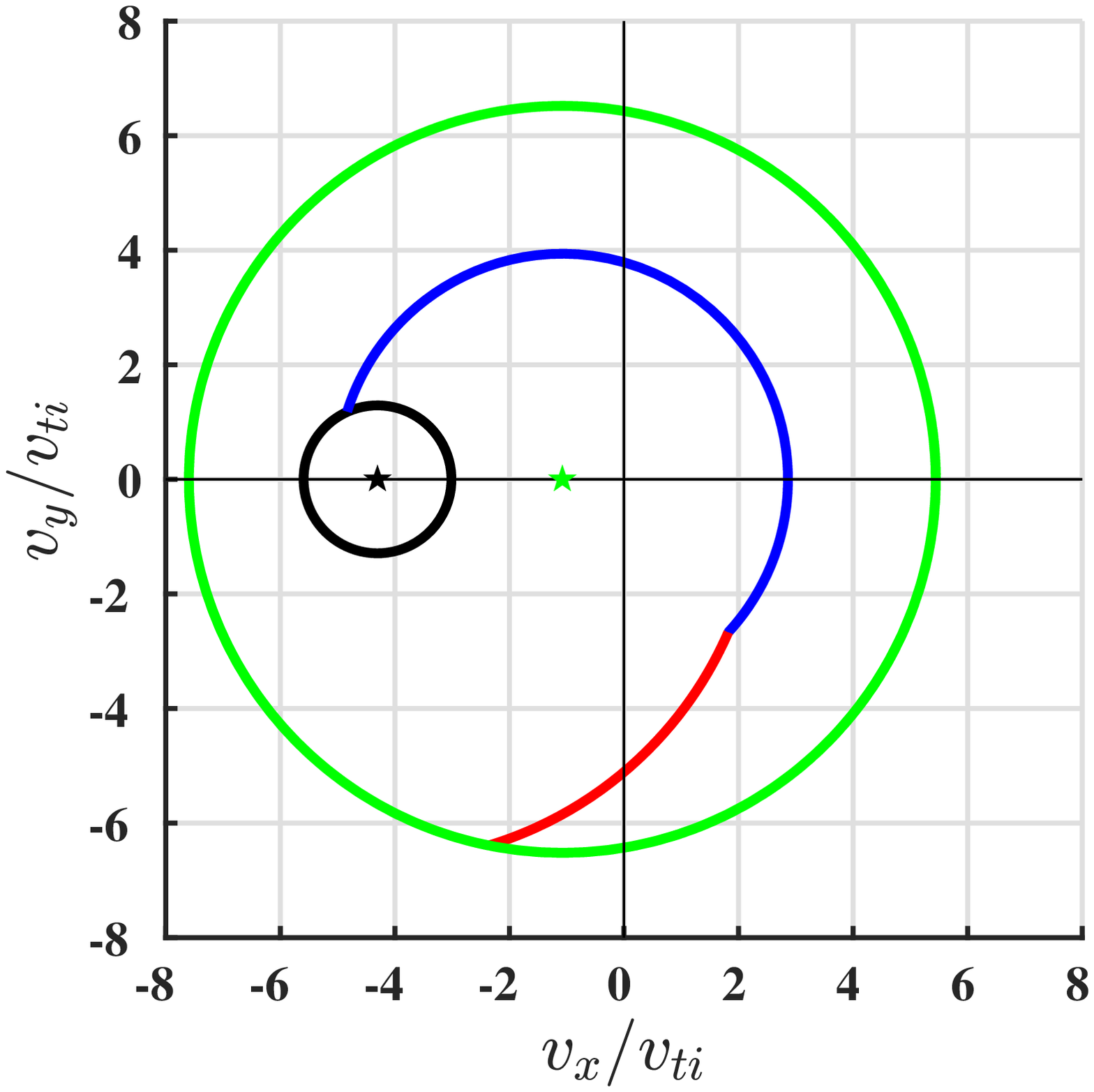}
   \end{center}
 \vskip -2.4in
\hspace*{0.05in} (a)\hspace*{2.45in} (b)
\vskip +2.25in
   \caption{(a) Real space trajectory of an ion as it traverses the shock front and (b) velocity-space trajectory.}
   \label{fig:trans_ideal}
\end{figure}

In Figure~\ref{fig:trans_ideal}, we plot (a) the trajectory of an ion in the $(x,y)$ plane and (b) its corresponding trajectory in $(v_x,v_y)$ velocity space in the shock-rest frame, where the colors indicate the corresponding segments of the trajectory.
In the upstream region at $x>0$ (black), the black circle centered about the upstream $\mvec{E} \times \mvec{B}$ velocity (black star) corresponds to the Larmor orbit of the ion about the upstream inflow velocity in the $(v_x,v_y)$ plane. 
Upon first crossing the magnetic discontinuity to $x<0$, the ion changes to a Larmor gyration in the $(v_x,v_y)$ plane (blue) about the downstream $\mvec{E} \times \mvec{B}$ velocity (green star).  
In the larger amplitude downstream perpendicular magnetic field, the radius of the Larmor motion in the $(x,y)$ plane in the shock-rest frame is reduced (blue).

Depending on the ion's gyrophase when the ion crosses the magnetic discontinuity, the ion passes back upstream to $x>0$, and once again undergoes a Larmor orbit in the $(v_x,v_y)$ plane (red) about the upstream $\mvec{E} \times \mvec{B}$ velocity (black star).  
In this segment of the trajectory (red), the ion gains perpendicular energy in the shock-rest frame, graphically represented by the distance in velocity space of the ion from the origin of the $(v_x,v_y)$ plane.
Finally, the ion will eventually cross back into the downstream region to $x<0$ (green), resuming its Larmor orbit in the $(v_x,v_y)$ plane (green) about the downstream $\mvec{E} \times \mvec{B}$ velocity (green star).  
Without any additional crossings of the magnetic discontinuity, the ion will simply $\mvec{E} \times \mvec{B}$ drift downstream.

In the segment of the trajectory where the ion can gain energy, it is the motional electric field, $E_y$, that is doing positive work on the ion, exactly like in our self-consistent simulation.
We note that this ion's dynamics---the reflection due to the magnetic gradient and energy gain from its traversal upstream and alignment with the motional electric field---is the well-known single-particle picture of shock-drift acceleration.
In fact, this picture in velocity space of where a single ion gains energy via this reflection by a magnetic gradient has been previously noted \citep{Gedalin:1996a}.
We wish now to connect this Lagrangian perspective on how a single ion gains energy from this reflection off a magnetic gradient to the Eulerian point-of-view we have from the FPC.


\subsection{Velocity-Space Signature of Shock Drift Acceleration}\label{sec:SDAFPC}
\begin{figure}
 \begin{center}
     \hskip 0.05in
     \includegraphics[width=0.45\textwidth]{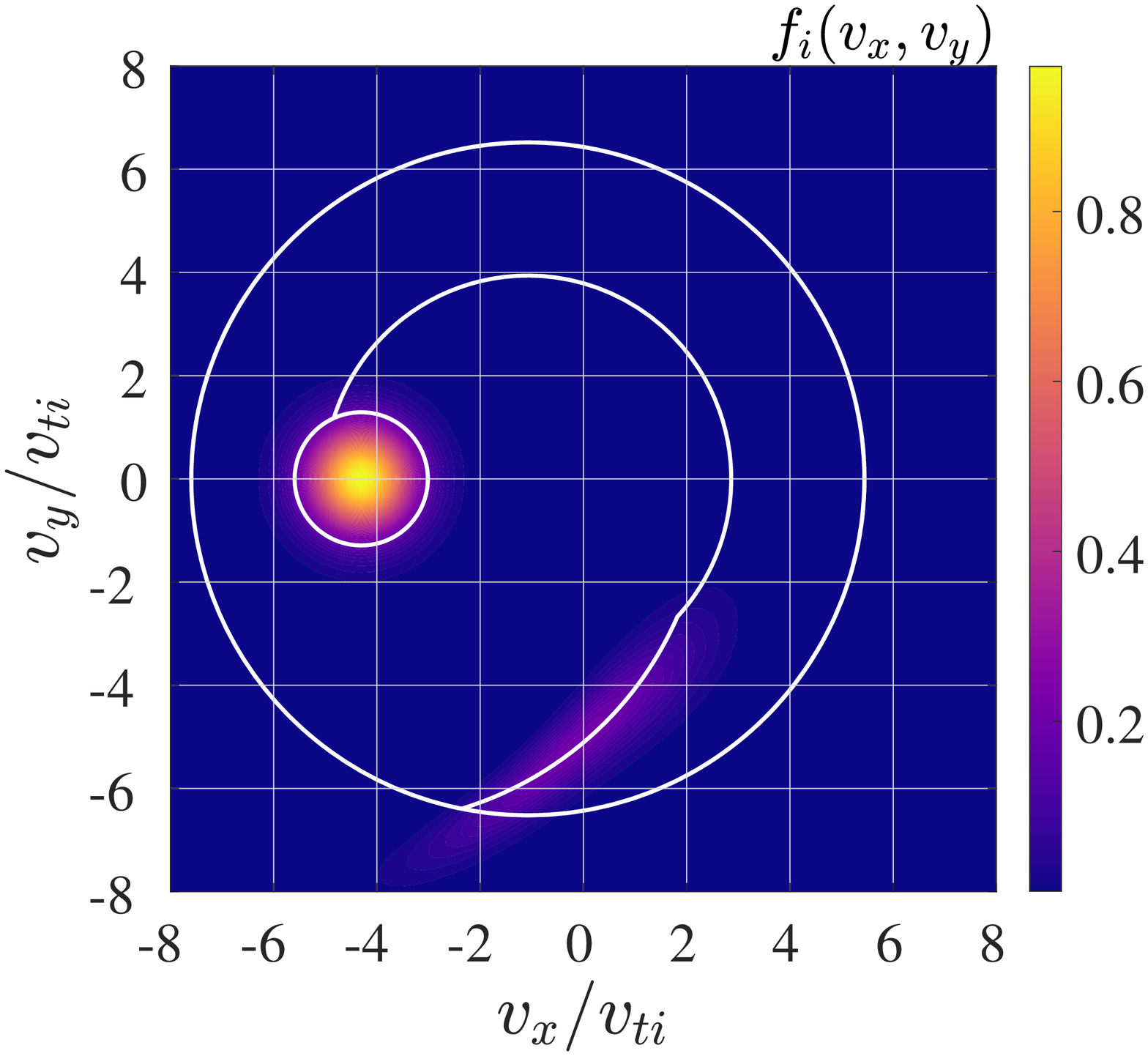}  \hskip 0.25in
    \includegraphics[width=0.48\textwidth]{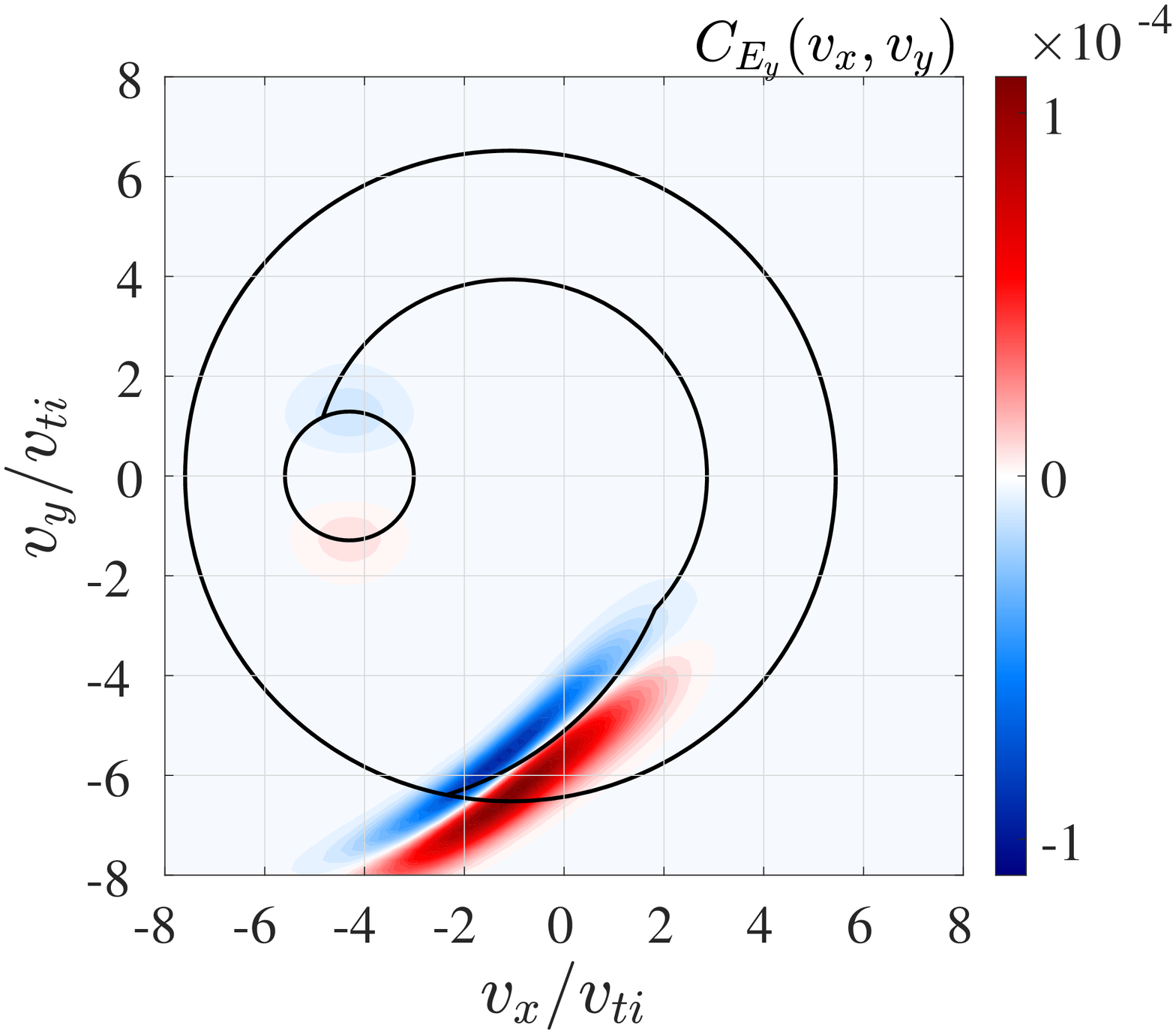}
   \end{center}
 \vskip -2.0in
\hspace*{0.05in} (a)\hspace*{2.45in} (b)
\vskip +1.85in
\begin{center}
     \includegraphics[width=0.49\textwidth]{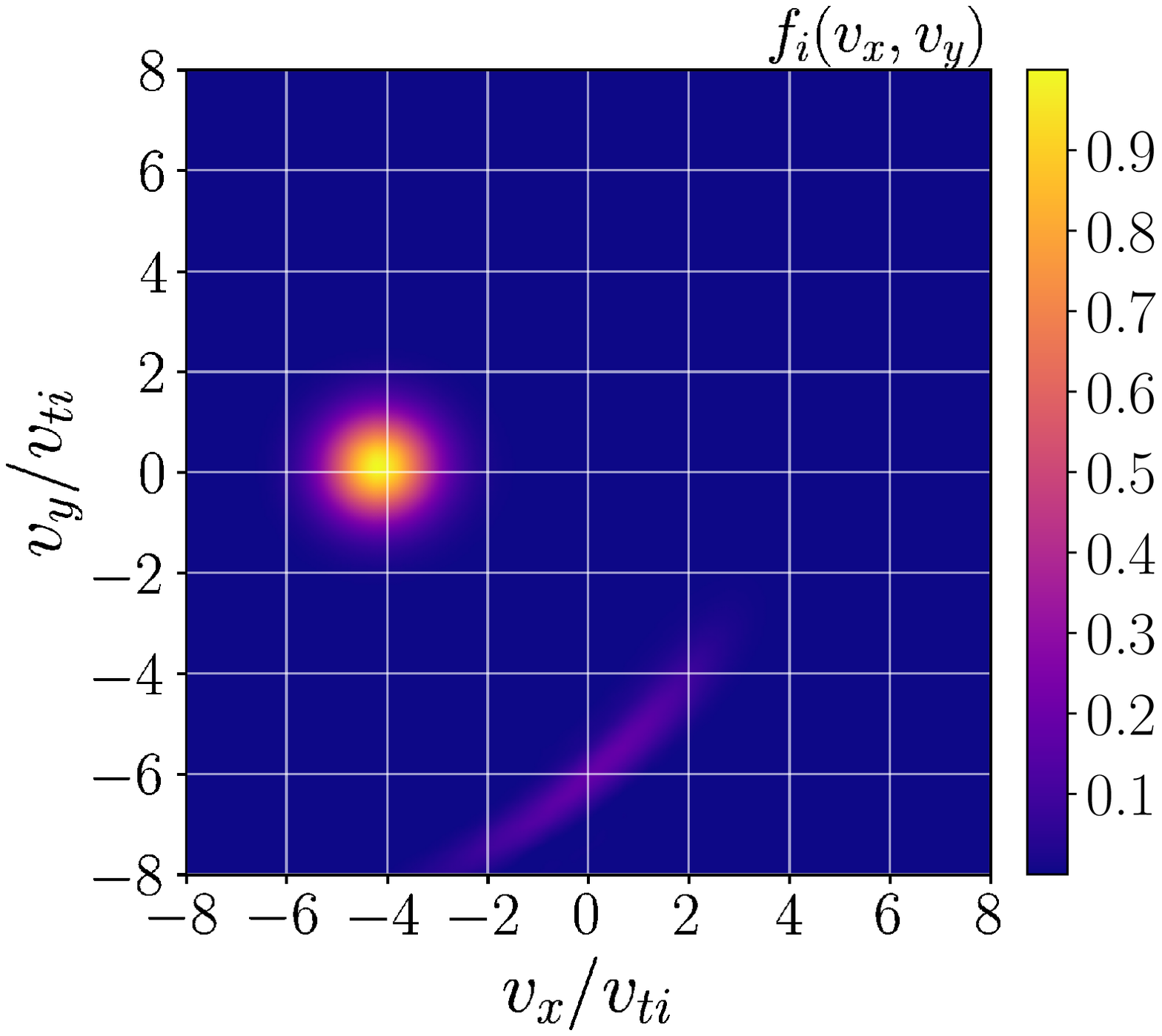}
      \includegraphics[width=0.49\textwidth]{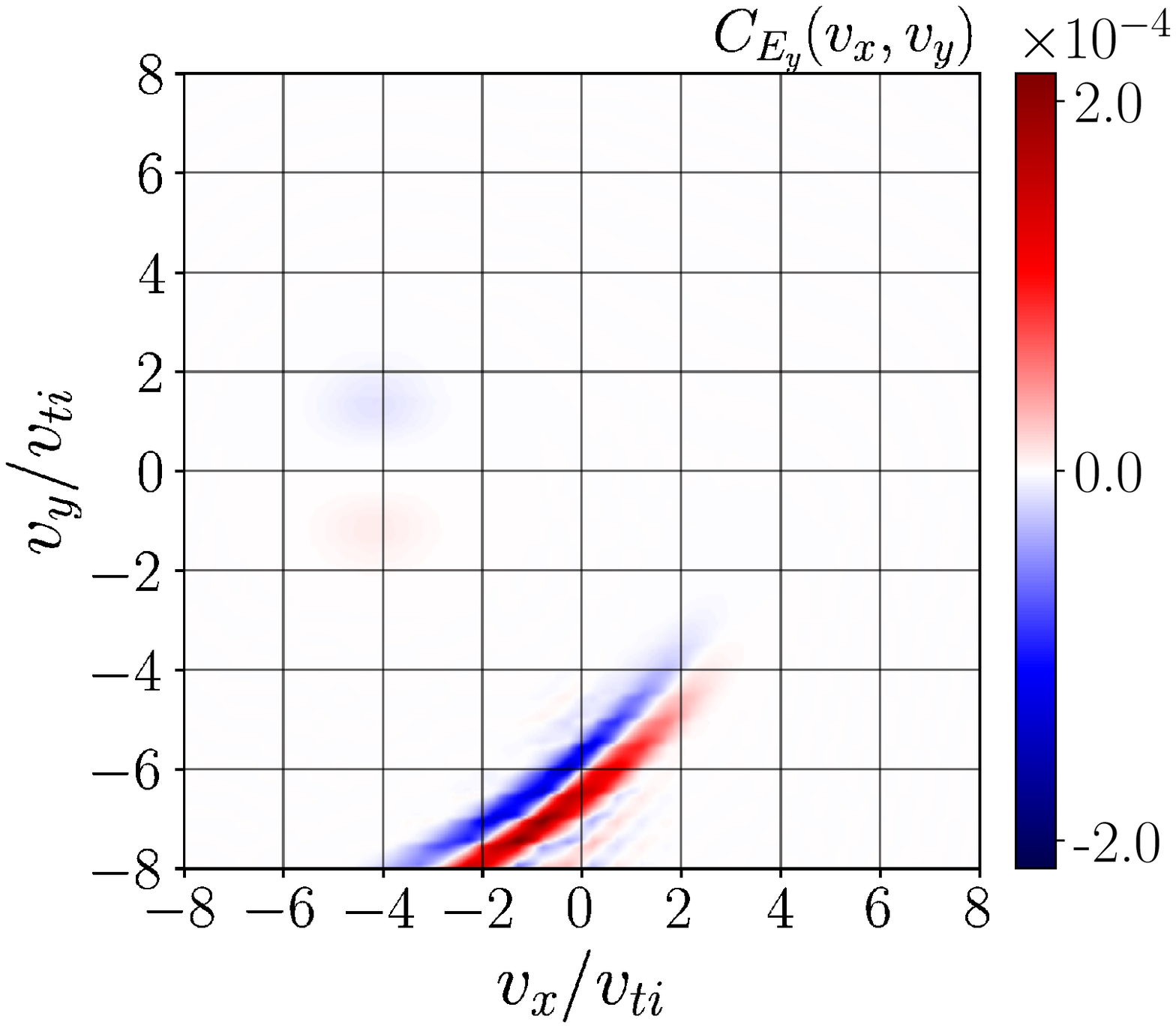}
        \end{center}
 \vskip -2.0in
\hspace*{0.05in} (c)\hspace*{2.45in} (d)
\vskip +1.85in
    \caption{Comparison of the reconstructed ion distribution function (a) and $C_{E_y}$ component of the FPC (b) computed from this reconstruction to the self-consistently produced ion distribution function (c) and $C_{E_y}$ component of the FPC (d) from the \gke~simulation. Using the Vlasov-mapping technique we can connect the single-particle orbits (over-plotted white (a) and black (b) lines) to the distribution function dynamics. $C_{E_y}$ integrated over velocity space is net positive, meaning the observed velocity-space signature corresponds to an energization process. We identify this particular velocity-space signature as the signature of shock-drift acceleration, energization of the reflected ions via the motional electric field in the upstream, via the connection between where in velocity space a single ion is energized and the specific region of velocity space where the strongest energy exchange is occurring.}
    \label{fig:ficey-model-comp}
\end{figure}

To connect the single-particle picture of shock-drift acceleration with how a distribution of ions is energized, we employ the Vlasov-mapping technique \citep{Scudder:1986,Kletzing:1994,Hull:1998,Hull:2000,Hull:2001,Mitchell:2013,Mitchell:2014}, described in \appref{app:vmap}, to determine the velocity distribution function in our simplified model for the electromagnetic fields through the shock.
We show in Figure~\ref{fig:ficey-model-comp} the reconstructed ion distribution $f_i(v_x,v_y)$ (a) at $x = 0.4 d_i$ and the corresponding FPC $C_{E_y}(v_x,v_y)$ (b) computed from the motional electric field, $E_y$, and gradients of this reconstructed distribution function.
In addition, we repeat Figure~\ref{fig:ficexcey}, panels (a) and (c), for reference in comparing the distribution function and generated velocity-space signature between the simplified model and self-consistent simulation.

In the reconstructed distribution function from the idealized model, we identify, in addition to the incoming upstream population centered at the upstream $\mvec{E} \times \mvec{B}$ velocity, a component of reflected particles that have returned upstream, exactly like in the self-consistent simulation.
Overplotted on the ion distribution function and computed FPC from the Vlasov-mapping technique is the trajectory in $(v_x,v_y)$ for the ion analyzed in Figure~\ref{fig:trans_ideal}, showing that this reflected population and velocity-space signature are coincident with the red segment of the trajectory in Figure~\ref{fig:trans_ideal}. 
Integrating this field-particle correlation over velocity space simply yields the net rate of work done by $E_y$, $\int C_{E_y}(v_x,v_y) dv_x dv_y = j_yE_y$, and we find the integration to be positive.
We thus identify the whole population of reflected ions as experiencing net energization, with the velocity-space signature of this energization process, shock-drift acceleration, given by Figure~\ref{fig:ficey-model-comp} (b) and (d).

We have now connected the Lagrangian picture of shock-drift acceleration with the Eulerian picture provided by the FPC technique, and we conclude this section noting that while shock-drift acceleration has been studied extensive theoretically and numerically \citep[\emph{e.g.},][]{Gedalin:1996a,Gedalin:1996b,Gedalin:1997,Gedalin:2000, Park:2013, XGuo:2014a, XGuo:2014b, Park:2015, Gedalin:2018, Xu:2020}, the \emph{velocity-space signature of shock-drift acceleration} provides a new perspective on the energization of the ions in phase space via this process.
In both cases, we understand that a portion of the distribution of ions are reflected via the magnetic field gradient and return upstream, where they can gain energy via the motional electric field.
Although the single-particle trajectory in phase space guides our understanding of where we expect the ions to be gaining energy, using the FPC technique enables us to see clearly the exact region of phase space in which ions are being energized via shock-drift acceleration.

This confirmation of the velocity-space signature of shock-drift acceleration in a self-consistent simulation is a vitally important step for comparison to measured velocity-space signatures of energy exchange using \emph{in situ} spacecraft measurements; however, it is also interesting that the velocity-space signature of shock-drift acceleration is unchanged between the idealized model and a self-consistent simulation given the additional physics of the self-consistent simulation: the finite shock width and cross-shock electric field.
We explore the reasons for the excellent agreement despite these two key differences between the simulation and the idealized model in Appendix~\ref{app:crossShockIon}, where we find the cross-shock electric field assists in reflecting ions, allowing the ions to traverse further back upstream and gain additional energy via shock-drift acceleration.
Thus, while the combination of the finite shock width and cross-shock electric field quantitatively changes the population of ions that are reflected, the qualitative signature of energization in velocity space via shock-drift acceleration remains unchanged.


\section{Field-Particle Correlation Analysis: Electrons}\label{sec:electrons}

\subsection{Velocity-Space Signature of Electron Energization}
\begin{figure}
   \begin{center}
      \includegraphics[width=0.5\textwidth]{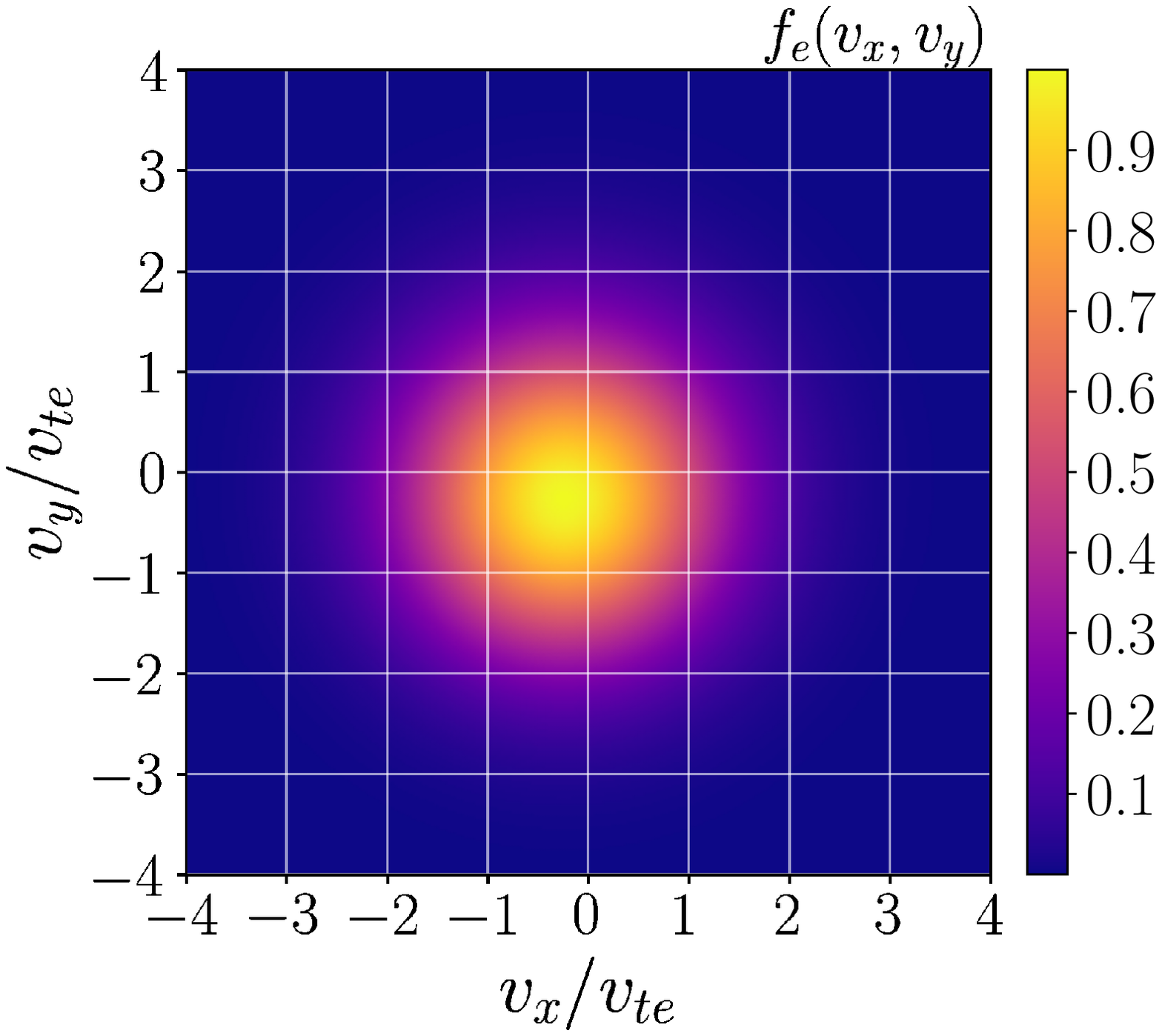}
   \end{center}
\vskip -2.0in
\hspace*{1.35in} (a)
\vskip +1.85in
   \begin{center}
      \includegraphics[width=0.49\textwidth]{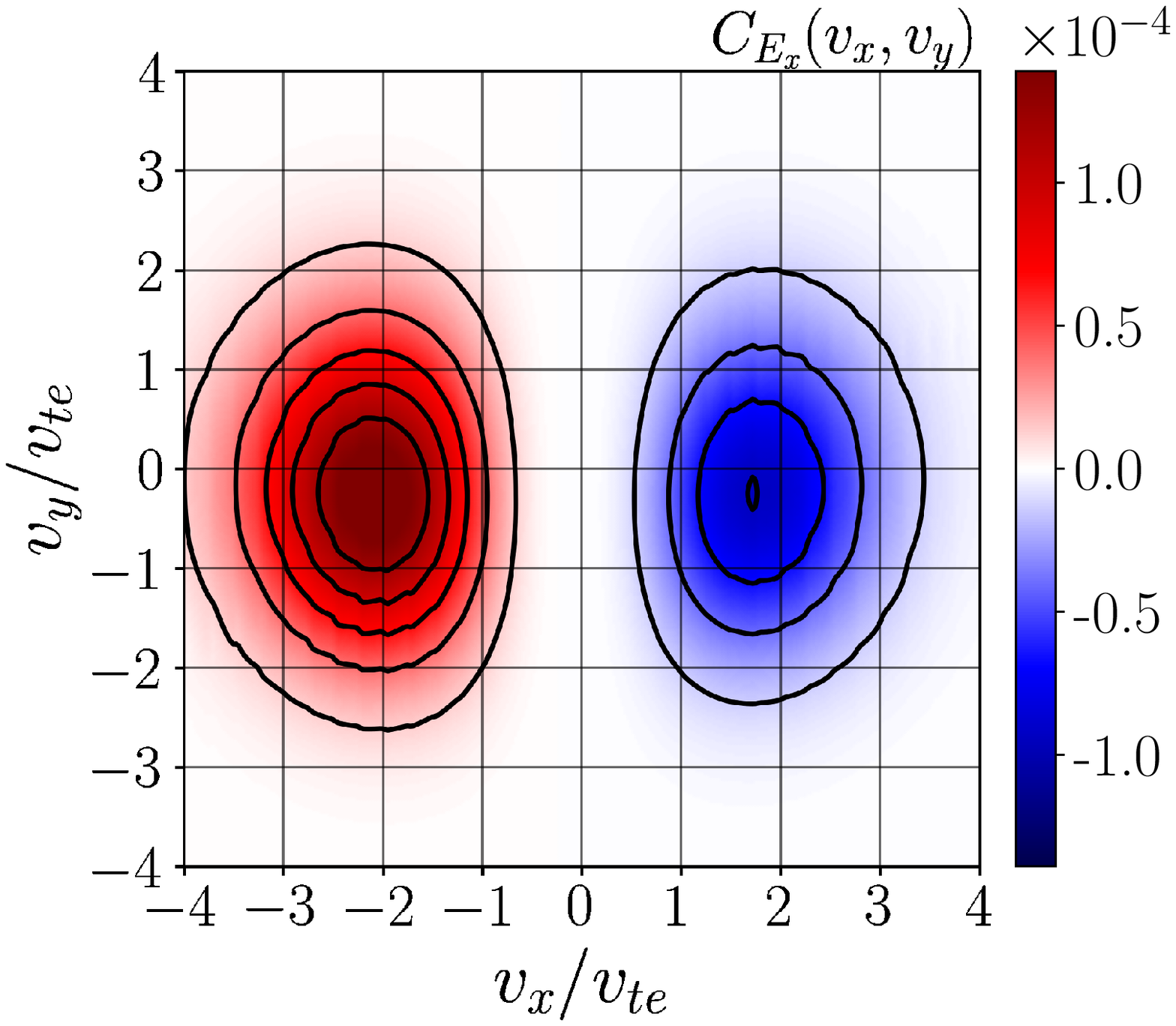}
      \includegraphics[width=0.49\textwidth]{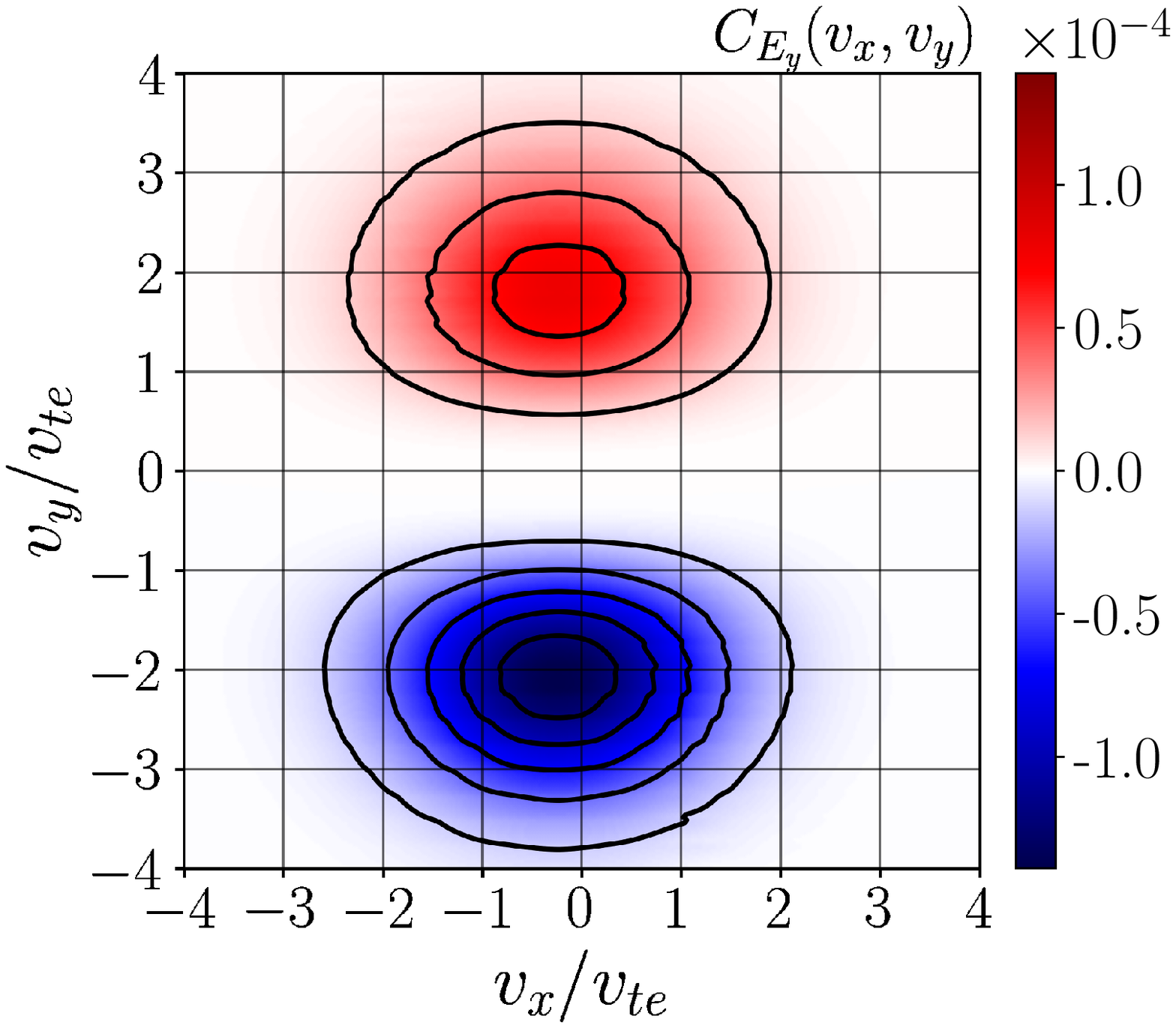}
   \end{center}
\vskip -2.0in
\hspace*{0.05in} (b)\hspace*{2.45in} (c)
\vskip +1.85in
    \caption{The electron distribution function $f_e(v_x,v_y)$ (a), and the $C_{E_x}$ (b) and $C_{E_y}$ (c) components of the FPC, \eqr{\ref{eq:vxCorrelation}} and \eqr{\ref{eq:vyCorrelation}}, computed  at $x_B = 21.8 d_i$ from the self-consistent \gke~simulation. Note that the FPC is computed in the shock-rest frame. The contours on the FPC plots, which are the same in both $C_{E_x}$ and $C_{E_y}$, make clear that $E_y$ leads to a net loss of electron energy, whereas $E_x$ yields a net increase of electron energy.}
    \label{fig:fecexcey}
\end{figure}

We now examine the energization of the electrons by the simulated perpendicular collisionless shock.
Similar to Figure~\ref{fig:ficexcey} for the ions, we show in Figure~\ref{fig:fecexcey}(a) the electron distribution function  $f_e(v_x,v_y)$ and (b) the $C_{E_x}$ and (c) the $C_{E_y}$ components of the FPC, \eqr{\ref{eq:vxCorrelation}} and \eqr{\ref{eq:vyCorrelation}}.
As shown in Figure~\ref{fig:perpShockFields}(e), the electron distribution broadens through the entire shock ramp, so we plot in Figure~\ref{fig:fecexcey} the results of the FPC analysis at $x_B = 21.8 d_i$, where the cross-shock electric field peaks.
In panel (a), the center of the distribution in the shock-rest frame is displaced away from the origin to $v_x<0$ and $v_y<0$ due to the particle drifts in the varying electric and magnetic fields through the shock ramp. 
Because the thermal width of the electron velocity distribution is much larger than the net drift of the distribution in the shock-rest frame, computing the $C_{E_x}$ and $C_{E_y}$ correlations using \eqr{\ref{eq:vxCorrelation}} and \eqr{\ref{eq:vyCorrelation}} leads to the qualitative ``two-lobed" velocity-space signatures observed in Figure~\ref{fig:fecexcey}(b) and (c). 
The small drifts---\emph{i.e.}, $|v_x/v_{te}| \ll 1$---lead to a slight asymmetry of the two-lobed structure, so we over-plot contours of constant $C_{E_{x,y}}$ to make these slight asymmetries more visually apparent. 
Although the gain (red) and loss (blue) of electron energy largely cancels out upon integration over velocity space $(v_x,v_y)$, asymmetries in the two lobes lead to a non-zero net energization.

In Figure~\ref{fig:fecexcey}(b), the asymmetry of the velocity-space signature leads to a net energization of the electrons by the cross-shock component of the electric field $E_x$. 
And, in contrast to the shock-drift acceleration of the ions by the motional electric field $E_y$ seen in Figure~\ref{fig:ficexcey}(c), we see in Figure~\ref{fig:fecexcey}(c) that the electrons experience a net loss of energy due to the $E_y$ component of the electric field. 
As with the ion analysis, we now turn to an idealized model for the electron dynamics through the shock layer to understand these velocity-space signatures for the electron energization.


\subsection{Adiabatic Heating in an Idealized Perpendicular Shock}\label{sec:elc-SPM}
\begin{figure}
   \begin{center}
      \includegraphics[width=0.59\textwidth]{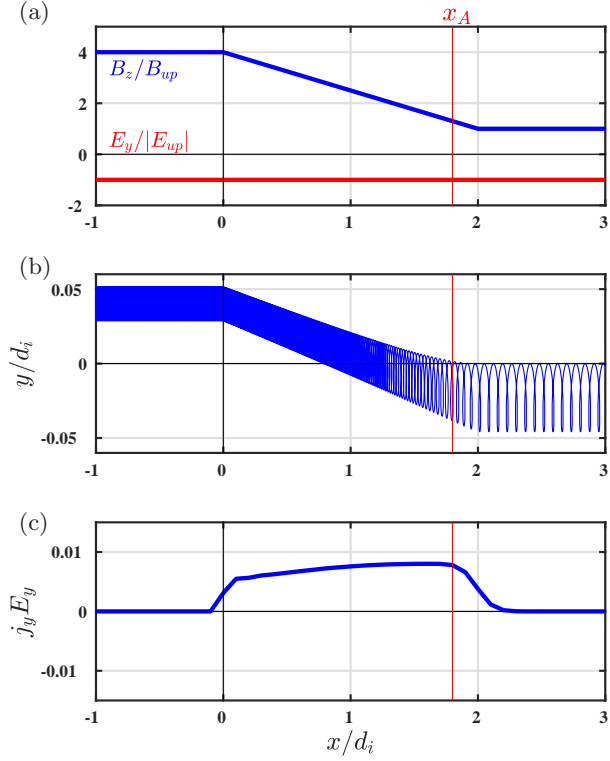}
   \end{center}
\vskip -4.0in
\hspace*{1.1in} (a)
\vskip +1.2in
\hspace*{1.1in} (b)
\vskip +1.2in
\hspace*{1.1in} (c)
\vskip +1.2in
   \caption{(a) Profiles along the shock normal direction of the
    perpendicular magnetic field $B_z$ (blue) and the motional
    electric field $E_y$ (red), (b) trajectory in
    the $(x,y)$ plane of an electron as it traverses the finite-width ramp in the magnetic field, and (c) the rate of work done by the electric
    field on the distribution of particles $j_y E_y$.}
   \label{fig:elec_profile}
\end{figure}

Unlike the ions for which $L_{shock} \lesssim \rho_i$, the electron gyroradius is much smaller than the gradient length scale of the shock, $L_{shock} \gg \rho_e$, and we thus expect the electrons to stay well magnetized through the shock. 
Therefore, we adopt a simplified model for the electron dynamics through the shock by taking a shock ramp with a linearly increasing magnetic field over a length $L= 2 d_i$ and mass ratio $m_i/m_e=1836$, satisfying the limit $L_{shock} \gg \rho_e$.  The other parameters are the same as the simple shock model used for the ion analysis in Section~\ref{sec:ion-SPM}: a magnetic field increase of $B_{d}/B_{u}=4$, a constant and uniform motional electric field $E_y<0$ in the shock-rest frame, with the same ion and electron plasma betas such that the shock velocity is comparable to the self-consistent simulation $M_A = 4.9$ and $M_f = 3.0$.  Note that this idealized model for the electrons has no cross-shock component of the electric field, $E_x=0$; the implications of this choice are discussed at length in the upcoming subsections.

In this model, the increase in the magnetic field magnitude through the ramp leads to a steady decrease in the $\mvec{E} \times \mvec{B}$ velocity as the plasma flows through the shock transition. 
In addition, the gradient of the magnetic field magnitude in the shock-normal direction induces a $\nabla B$ drift in the $+y$ direction.
In Figure~\ref{fig:elec_profile}, we plot (a) the profile of the perpendicular magnetic field $B_z(x)$ (blue) and the motional electric field $E_y(x)$ (red) along the shock normal direction, as well as (b) the trajectory of an electron in the $(x,y)$ plane as it flows through the shock ramp over $0 \le x/d_i \le 2$.  
The trajectory plot shows clearly the $\nabla B$ drift in the $+y$ direction.
We note that, as the electrons flow through the shock ramp over $0 \le x/d_i \le 2$ and undergo a $\nabla B$ drift in the $+y$ direction, (c) the net energization for a distribution of electrons $j_y E_y$ is positive.

In  the region where the perpendicular magnetic field changes magnitude, $0 \le x/d_i \le 2$, the $\nabla B$ drift is anti-aligned with the motional electric field, thus leading to net energization of electrons.
In fact, the rate of energization of the electrons by the $\nabla B$ drift in the motional electric field is precisely that needed to conserve the first adiabatic invariant of the electron, i.e., the magnetic moment $\mu = m_e v_\perp^2/2 B_z$.
This energization via conservation of the electron's adiabatic invariant is thus often referred to as adiabatic heating.

We can show the relationship between the energization via the $\nabla B$ drift and the conservation of the electron's magnetic moment by considering the change in the perpendicular kinetic energy of the electrons,
\begin{align}
    \frac{d m_e v_\perp^2/2}{dt}=q_e u_{\nabla B} E_y
\label{eq:elecmu}
\end{align}
where the magnitude of the $\nabla B$ drift in the $+y$ direction is given by
\begin{align}
u_{\nabla B}= \frac{m_e v_\perp^2}{2 q_e B_z} 
\left(\frac{1}{B_z} \frac{\partial B_z}{\partial x} \right).
\label{eq:gradbdrift}
\end{align}
For the static fields in this model, the total time derivative is dominated by the $\mvec{E} \times \mvec{B}$ velocity, $d/dt = \partial /\partial t + u_x \partial/\partial x = u_{E \times B} \partial/\partial x$. 
Substituting $u_{E \times B}=E_y/B_z$, we can manipulate \eqref{eq:elecmu} to obtain
\begin{align}
\frac{\partial }{\partial x} \frac{m_e v_\perp^2}{2B_z}= \frac{\partial \mu }{\partial x}=0,
\label{eq:mucons}
\end{align}
proving that the electron's magnetic moment $\mu$ is conserved.
As before, we now wish to connect this single-particle, Lagrangian picture of adiabatic heating with the Eulerian picture provided by the FPC technique.


\subsection{Cross-Shock Electric Field Impact on Velocity-Space Signature of Adiabatic Heating}\label{sec:elcFPC}
\begin{figure}
 \begin{center}
     \hskip 0.05in \includegraphics[width=0.45\textwidth]{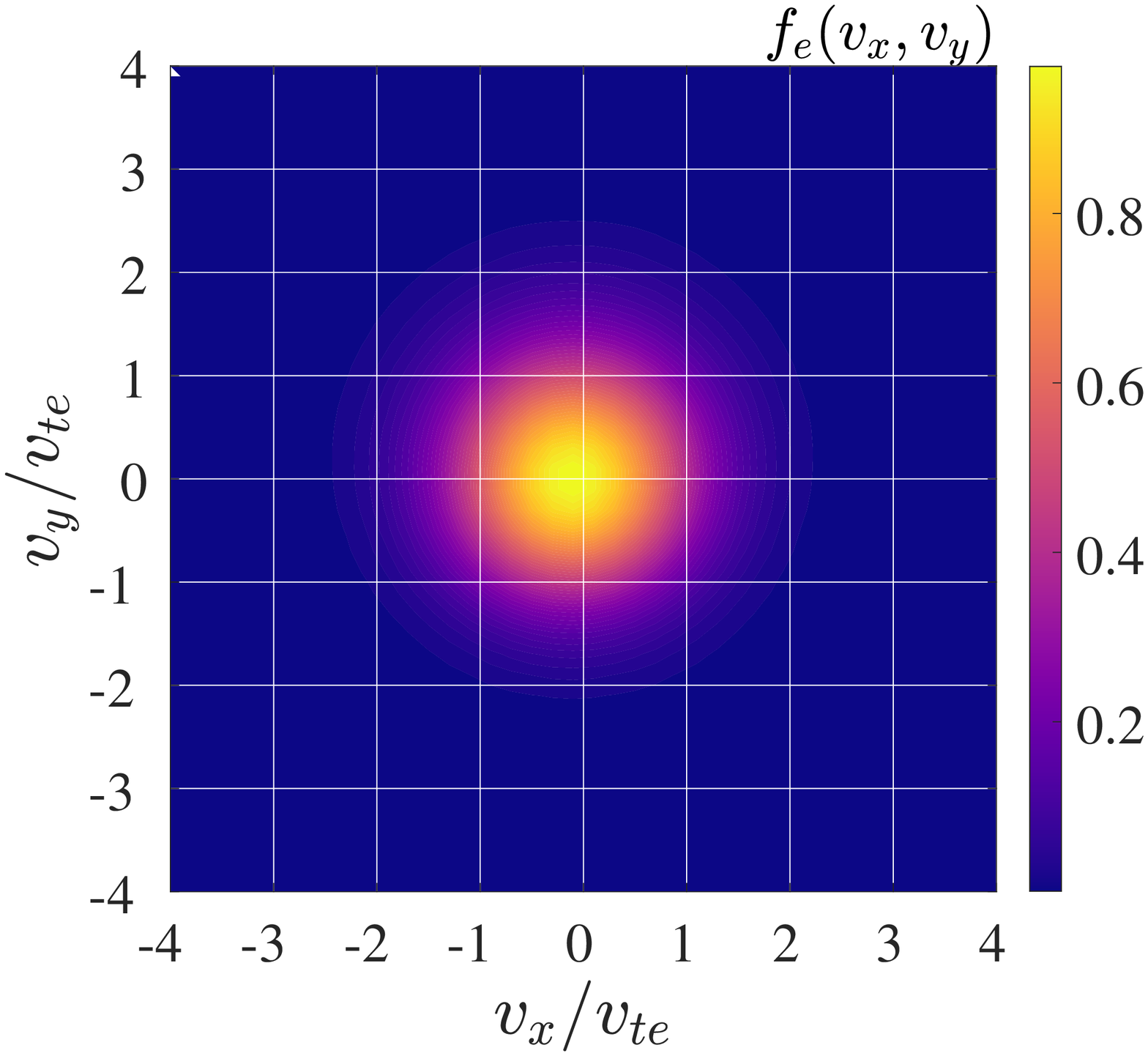} \hskip 0.25in
    \includegraphics[width=0.48\textwidth]{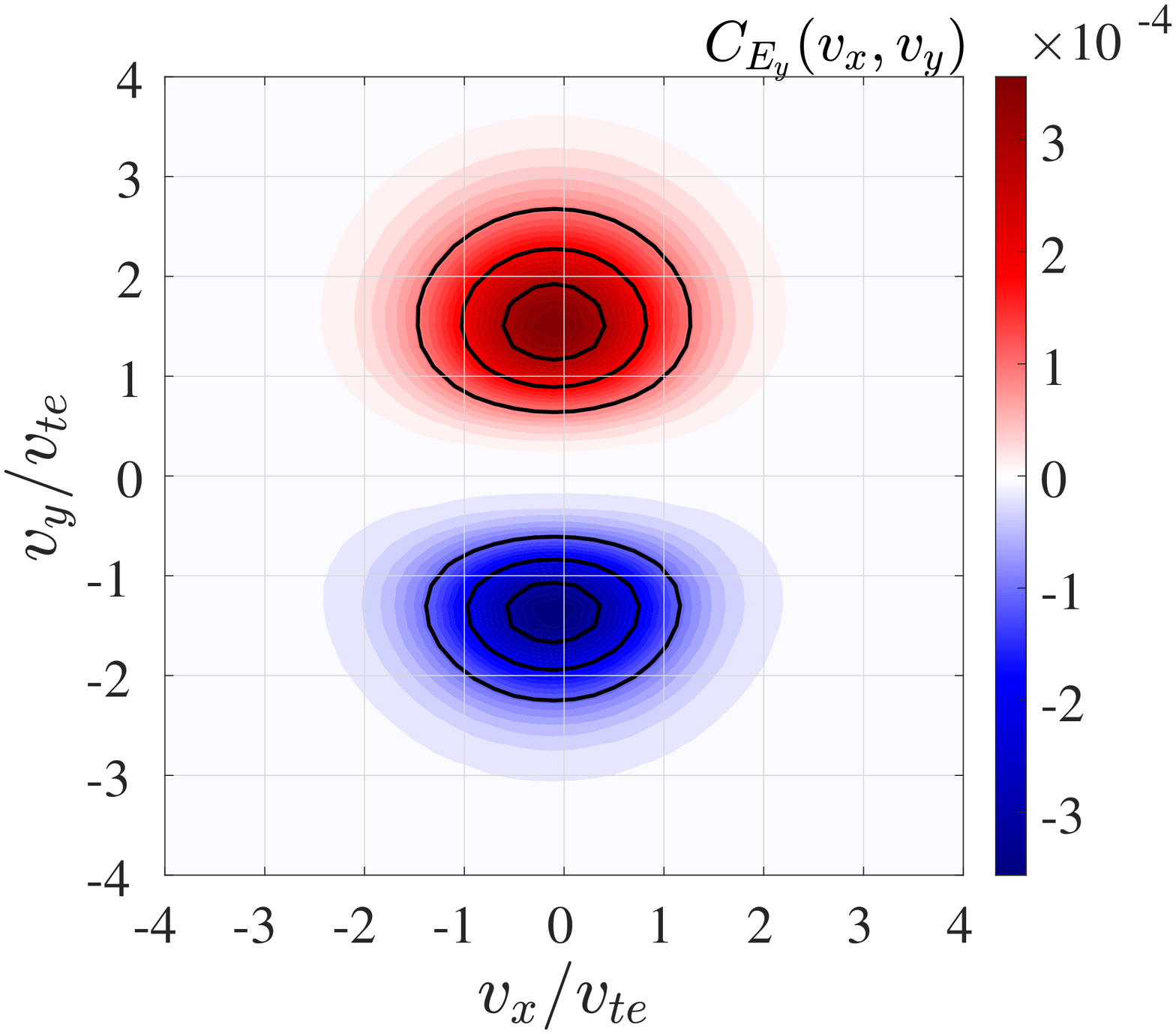}
   \end{center}
 \vskip -2.0in
\hspace*{0.05in} (a)\hspace*{2.45in} (b)
\vskip +1.85in
\begin{center}
     \includegraphics[width=0.49\textwidth]{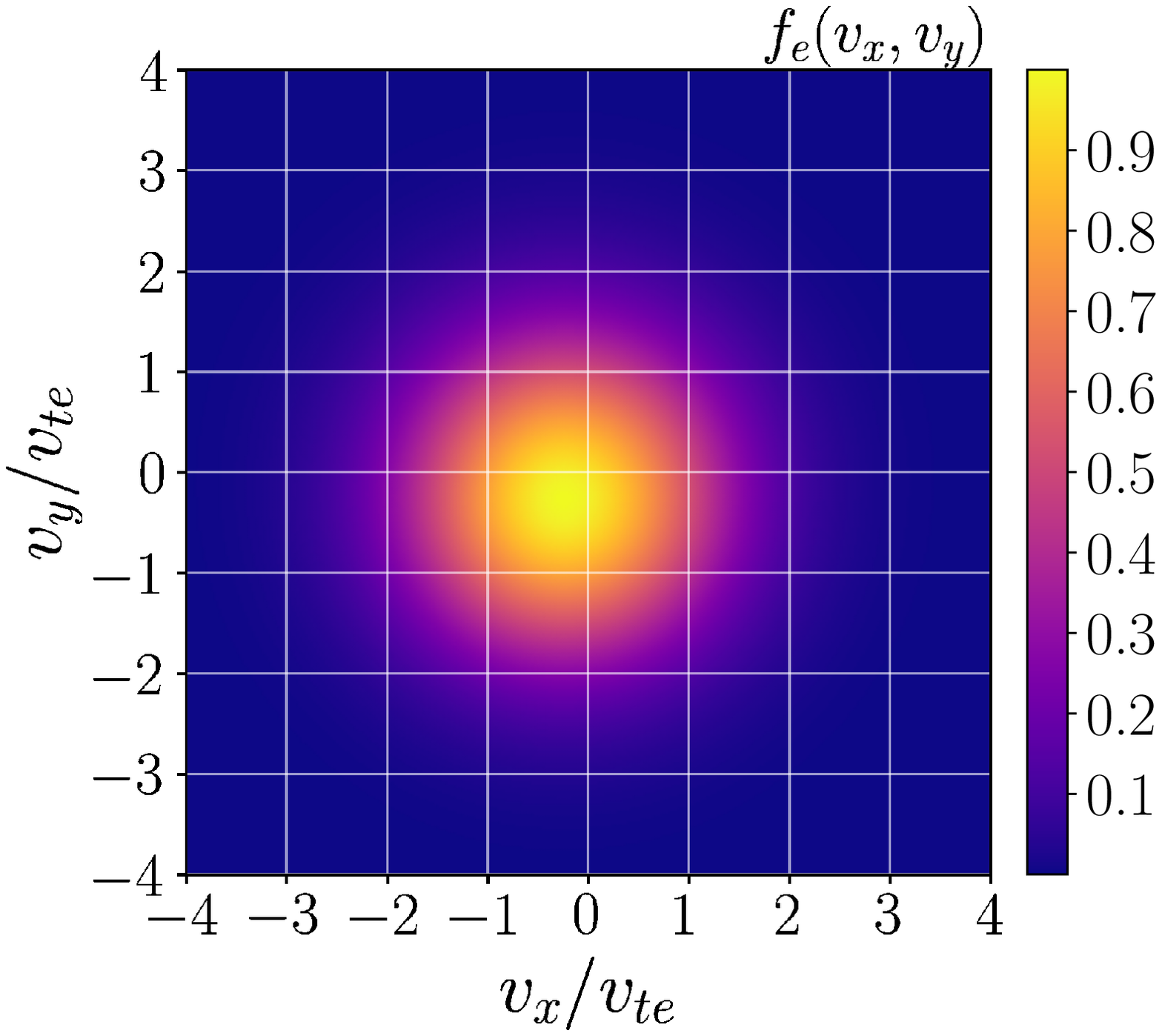}
      \includegraphics[width=0.49\textwidth]{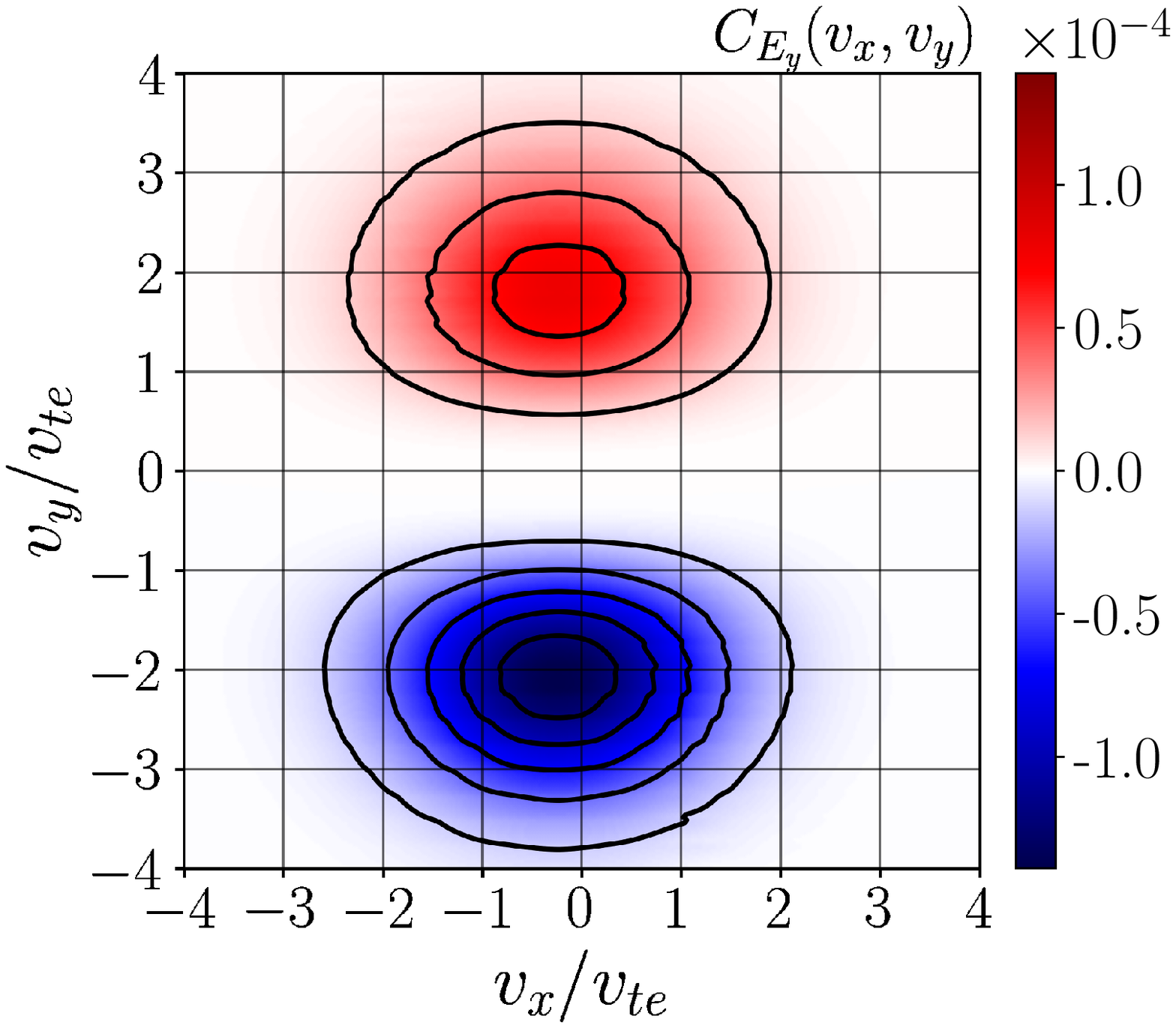}
        \end{center}
 \vskip -2.0in
\hspace*{0.05in} (c)\hspace*{2.45in} (d)
\vskip +1.85in
    \caption{Comparison of the reconstructed electron distribution function (a) and $C_{E_y}$ component of the FPC (b) computed from this reconstruction to the self-consistently produced electron distribution function (c) and $C_{E_y}$ component of the FPC (d) from the \gke~simulation. Unlike with the ion comparison presented in Figure~\ref{fig:ficey-model-comp}, the model $C_{E_y}$ displays the opposite asymmetry from the $C_{E_y}$ computed from the self-consistent simulation. Even though the signatures are qualitatively similar, this first computation of $C_{E_y}$ suggests that $E_y$ is responsible for a net loss of energy through the shock.}
    \label{fig:fecey-model-comp}
\end{figure}

We again employ the Vlasov-mapping technique, as in Section~\ref{sec:SDAFPC}, to reconstruct the electron distribution function through the idealized shock model and the resulting FPC velocity-space signatures.  In Figure~\ref{fig:fecey-model-comp}, we present (a) the resulting electron distribution function $f_e(v_x,v_y)$  and (b) the $C_{E_y}$ correlation at position $x_A = 1.8 d_i$ in the model, where the velocity-integrated energy transfer rate $j_y E_y$ is positive, as seen in Figure~\ref{fig:elec_profile}(c).  Note that the small shift (relative to $v_{te}$) of the distribution to $v_y>0$ due to the $\nabla B$ drift leads to an asymmetry in the velocity-space signature because of the $v_y^2$ weighting in \eqr{\ref{eq:vyCorrelation}} for $C_{E_y}$ \footnote{Note that this shift of $f_e$ and resulting asymmetry in $C_{E_y}$ is somewhat difficult to discern visually in Figure~\ref{fig:fecey-model-comp}(a) and (b) because we have taken a realistic mass ratio $m_p/m_e=1836$ in the model and the $\nabla B$ drift for the electrons in \eqr{\ref{eq:gradbdrift}} is proportional to $m_e$.}. Although a large part of the energy transfer rate represented by this two-lobed velocity-space signature cancels out upon integration over $v_y$, the slight asymmetry leads to a net positive energization of the electrons, yielding  $j_y E_y>0$, as plotted in Figure~\ref{fig:elec_profile} (c).

\begin{figure}
    \centering
    \includegraphics[width=\textwidth]{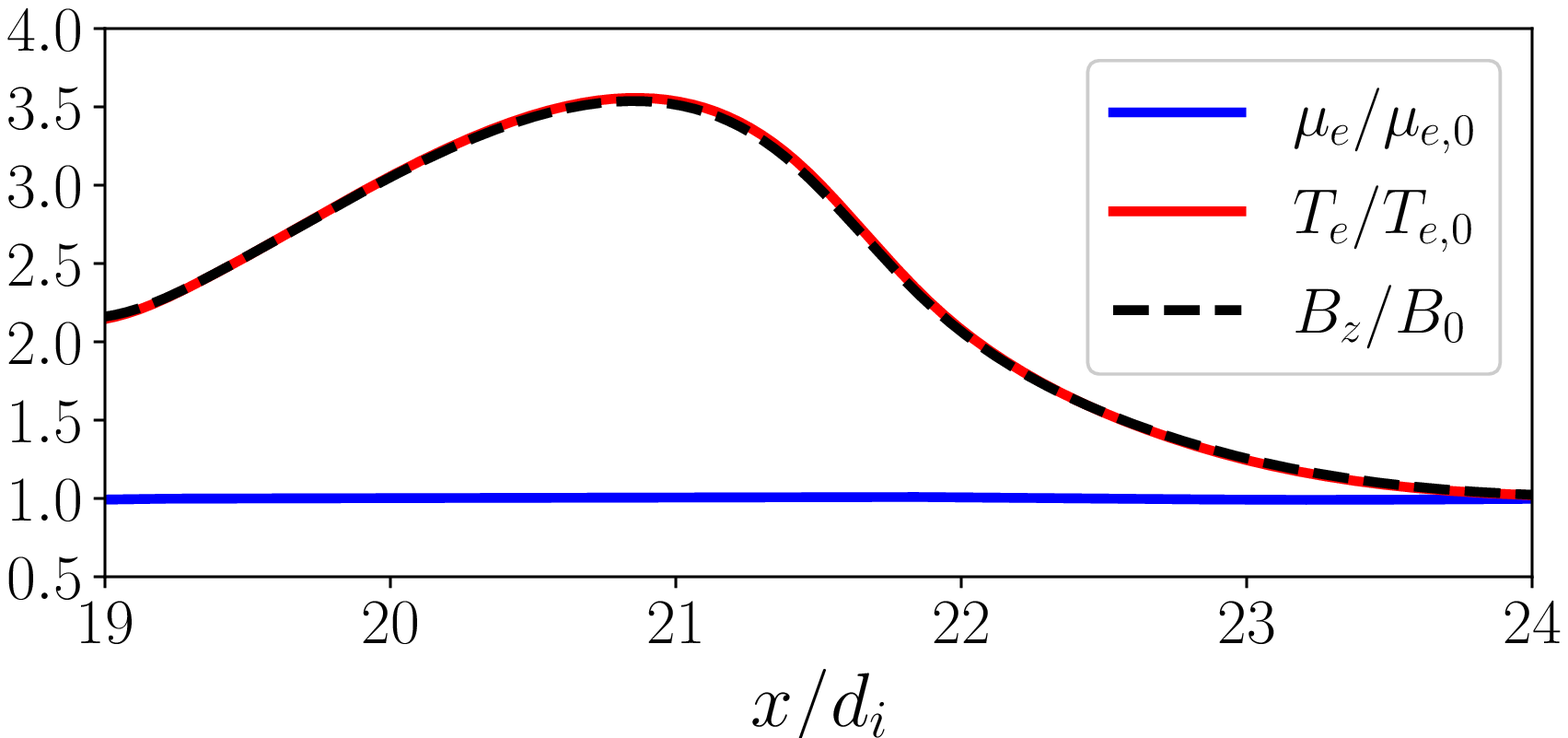}
      \caption{The electron adiabatic invariant, $\mu_e = T_\perp/B_z$ (blue solid), the electron temperature (red solid), and the magnetic field (black dashed), normalized to their value upstream and plotted through the shock. The electron temperature rises commensurate with the compression of the magnetic field such that the electron $\mu_e$ is well conserved through the shock.}
    \label{fig:electronMu}
\end{figure}

We compare the predicted velocity-space signature of electron adiabatic heating from the idealized shock model in Figure~\ref{fig:fecey-model-comp}(b) to the \gke~simulation results, where we plot in 
Figure~\ref{fig:fecey-model-comp}(c) the $f_e(v_x,v_y)$ and (d) the $C_{E_y}$ from the simulation at position $x_B=21.8 d_i$. 
While the correlation $C_{E_y}$ from the simulation has the same qualitative, two-lobed structure indicative of drift energization of the electrons, the net drift of the electron velocity distribution has $v_y<0$, and so therefore the resulting asymmetry in $C_{E_y}$ has the opposite overall sign, leading to a net loss of energy for the electrons due to $E_y$.  How can we reconcile these apparently contradictory results for the electron energization by the motional electric field $E_y$, particularly given that we have shown in \eqr{\ref{eq:mucons}} that the $\nabla B$ drift in the $+y$ direction leads to adiabatic heating of the electrons?

We can check if the adiabatic invariant of a distribution of electrons, 
\begin{align}
    \mu_e= \frac{1}{\int f_e \thinspace d\mvec{v}} \left \{ \int \frac{m_e \left [ (v_x - u_x)^2 + (v_y - u_y)^2 \right]}{2 B_z} f_e \thinspace d\mvec{v} \label{eq:distMu} \right \}
\end{align}
is constant through the shock, and whether our intuition about how the electron temperature should increase through a magnetic field gradient has merit.
Note that in the 1D-2V geometry of the simulated perpendicular shock where both velocity coordinates are perpendicular to the magnetic field, the adiabatic invariant of the distribution of electrons can be simplified to
\begin{align}
    \mu_e = \frac{T_{\perp,e}}{B_z},
\end{align}
where $T_{\perp,e}$ is the temperature of the electrons perpendicular to the magnetic field.
Indeed, as shown in Figure~\ref{fig:electronMu}, the adiabatic invariant for the distribution of electrons, $\mu_e$, is well conserved through the shock transition.

\begin{figure}
\begin{center}
      \includegraphics[width=0.8\textwidth]{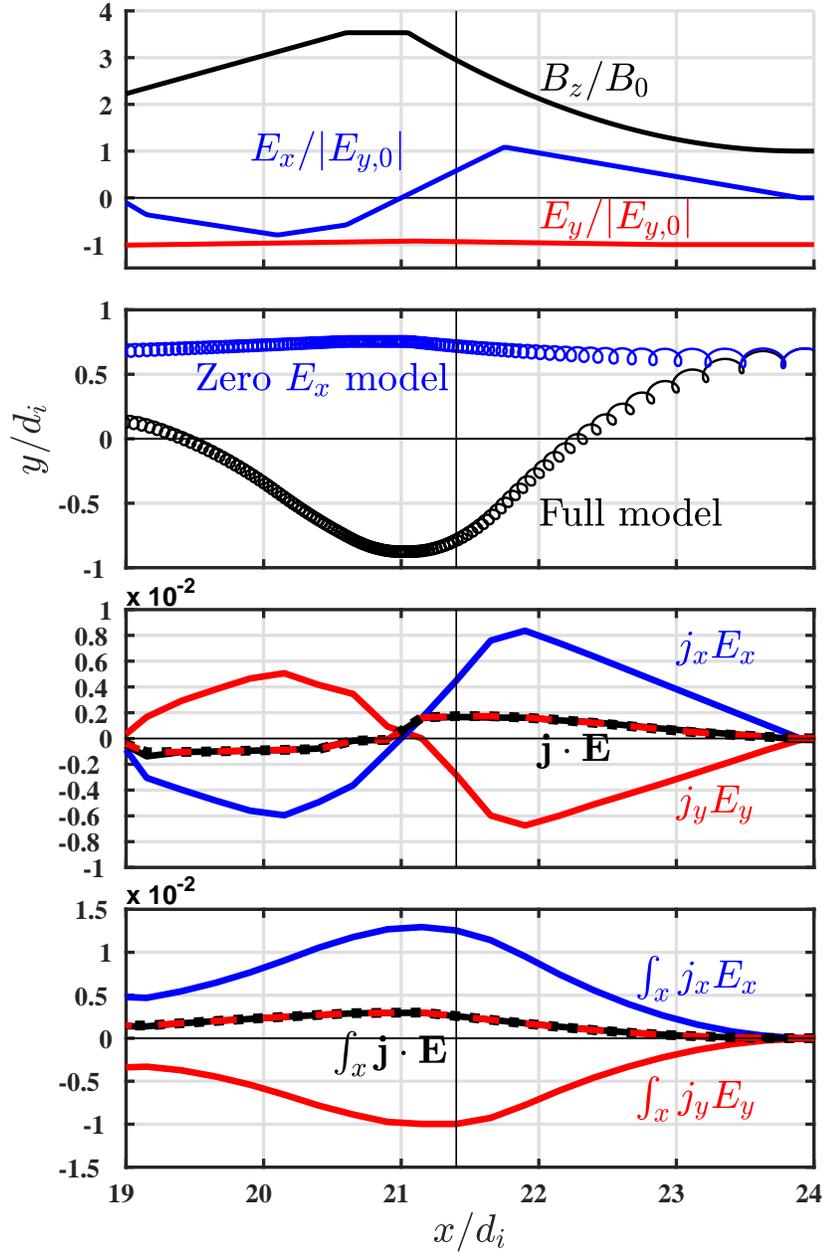}
      \end{center}
    \caption{(a) Electromagnetic fields approximated from  the self-consistent \gke~simulation. (b) Example electron trajectories for full model (black) and zero $E_x$ model (blue), showing qualitatively different drifts in the $y$ direction. (c) Rate of work done by the components of the electric field, $j_yE_y$ (red) and $j_xE_x$ (blue) for the full model (solid) and zero $E_x$ model (dashed), along with total 
    $\mvec{j} \cdot \mvec{E}$ (black). Note that the total energization (black, solid and dashed) is the same for both cases. (d) Cumulative work done integrated from upstream $\int_x \mvec{j} \cdot \mvec{E}$. The electrons experience adiabatic heating in both cases, although the detailed mechanisms of energization involve qualitatively different drifts.}
    \label{fig:spm_e_comparison}
\end{figure}

This result suggests that the cross-shock electric field is complicating the electron dynamics and energy exchange through the shock.
Here we can leverage the Vlasov-mapping model to explore the effect of the cross-shock electric field on the energetics of the electrons.  
In Figure~\ref{fig:spm_e_comparison}, we present a comparison of (b) the electron trajectories, (c) the rate of energization of the electrons by the electric field $\mvec{j}_e \cdot \mvec{E}$, and (d) the cumulative electron energization integrated from upstream $\int_{x_{up}}^x dx \ \mvec{j}_e \cdot \mvec{E}$ for two models: (i) a ``full model'' (solid) which integrates electron trajectories in the self-consistently produced electromagnetic fields in Figure~\ref{fig:spm_e_comparison}(a); and (ii) a ``zero $E_x$ model'' (dashed), in which we artificially set the cross-shock electric field to zero. 

The trajectories in Figure~\ref{fig:spm_e_comparison}(b) show a clear qualitative difference between the zero $E_x$ model and the full model: in the zero $E_x$ model (blue), the transverse drift through the shock ramp is relatively weak and in the $+y$ direction, whereas the full model (black) yields a much larger amplitude drift in the opposite direction.  
This qualitative difference in the transverse drift direction explains the stark differences in $C_{E_y}$ between the idealized model in Figure~\ref{fig:fecey-model-comp} (b) and the simulation in (d). 

Looking at the rate of electron energization by the electric field in Figure~\ref{fig:spm_e_comparison}(c), we indeed see that $j_{ye} E_y$ (red dashed) is positive for the zero $E_x$ model (and is the only means of energization of the electrons since $E_x=0$), but $j_{ye} E_y$ is negative for the full model (red solid). 
However, when the energization by $j_{xe} E_x$ (blue solid) is combined with $j_{ye} E_y$ (red solid), the net rate of energization $\mvec{j}_e \cdot \mvec{E}$ of the two models is exactly the same (black solid and red dashed overlap).  
Therefore, although the dynamics of the electrons differ qualitatively in the presence or absence of the cross-shock electric field, their energization is the same in either case.  
To explain this puzzling finding, we exploit the limit  $L_{shock} \gg \rho_e$ to execute a guiding-center drift analysis of the electron energization.


\subsection{Guiding-Center Drift Analysis of Electron Energization}

In the idealized model presented in Section~\ref{sec:elc-SPM}, there are only two drifts: an $\mvec{E} \times \mvec{B}$ drift in the $x$ direction due a constant $E_y$ through the magnetic field ramp and a $\nabla B$ drift in the $y$ direction due to the linearly increasing magnetic field.
The introduction of a cross-shock electric field, along with the transition from a single-particle picture to a distribution of particles, adds several new drifts to the full list of potentially dynamically important drifts.
We now not only have an $\mvec{E} \times \mvec{B}$ drift that has a component in the $x$ direction due to the motional electric field $E_y$ supporting the incoming supersonic flow, but also a new $y$ component due to the cross-shock electric field $E_x$.
For a distribution of electrons the $\nabla B$ drift in the $y$ direction is modified from its single particle form,
\begin{align}
    u_{\nabla B}^{SP} = \frac{m_e v_{\perp,e}^2}{2 q_e B^2_z} \pfrac{B_z}{x} \hat{\mvec{y}}
\end{align}
to
\begin{align}
    u_{\nabla B} = \frac{1}{q_e n_e} \frac{p_{\perp,e}}{B^2_z} \pfrac{B_z}{x} \hat{\mvec{y}};, \label{eq:gradB1x2v} 
\end{align}
where $p_{\perp,e}$ is the electron perpendicular pressure,
\begin{align}
    p_{\perp,e} = \frac{1}{2} m_e \int (\mvec{v}_\perp - \mvec{u}_{e \perp})^2 f_e \thinspace d\mvec{v}.
\end{align}
We now must also consider a polarization drift in the $x$ direction, and we note that the total time derivative as the electrons flow through the shock is dominated by the convective contribution $d/dt = \partial/\partial t + \mvec{U} \cdot \nabla \simeq U_x \partial/\partial x$, giving
\begin{align}
    u_{dE/dt} = \frac{1}{\Omega_{cs} B_z} \frac{d\mvec{E}}{dt} \sim \frac{1}{\Omega_{cs} B_z}  U_x \frac{\partial E_x}{\partial x} \hat{\mvec{x}}.
\end{align}
Finally, we have the magnetization drift, $\nabla \times \mvec{M}$, a bulk drift in the $y$ direction due to the increasing density through the shock ramp, 
\begin{align}
 u_{\nabla \times \mvec{M}} = \frac{\nabla \times \mvec{M}}{q_e n_e} =\frac{1}{q_e n_e}  \nabla \times \left (- \frac{p_{\perp, e}}{B_z} \right ) =  \frac{1}{q_e n_e} \pfrac{}{x}  \left (\frac{p_{\perp,e}}{B_z} \right )\hat{\mvec{y}}, \label{eq:magnetizationDrift}
\end{align}
where the magnetization vector $\mvec{M}$ \citep{Hazeltine:1998} is given by
\begin{align}
    \mvec{M} = -p_\perp \frac{\mvec{B}}{|\mvec{B}|^2}. \label{eq:magnetization1x2v}
\end{align}
Although there can, in principle, be a polarization drift in the $y$ direction due to the variation in $E_y$ through the shock, in the shock rest frame $E_y$ changes very little, so this drift is negligible.
We note that the $\nabla B$ drift and magnetization drift can be combined to form the diamagnetic drift,
\begin{align}
   u_{\mbox{diamag}} & =  u_{\nabla B} + u_{\nabla \times \mvec{M}}  = \frac{1}{q_e n_e} \left [ \frac{p_{e \perp}}{B^2_z} \pfrac{B_z}{x} +  \pfrac{}{x}  \left (\frac{p_{e \perp}}{B_z} \right ) \right ], \notag \\
    & =  \frac{1}{q_e n_e} \frac{1}{B_z} \pfrac{p_{e \perp}}{x} =  -\frac{1}{q_e n_e} \frac{\nabla p_{e \perp} \times \mvec{B}}{|\mvec{B}|^2}.
    \label{eq:diamag}
\end{align}

This generalization from the single-particle picture to a distribution of particles is somewhat subtle and often dubbed Spitzer's paradox from the early work on plasma equilibria in a fusion context \citep{Spitzer:1952, Spitzer:1962, Qin:2000}.
While the concept of pressure is ill-defined for a single particle, magnetic field gradients must inevitably be balanced by pressure gradients in equilibrium because the $\nabla B$ drift depends on the particles' velocity, and thus different parts of the distribution of particles will experience different $\nabla B$ drifts.
As a consequence, the pressure of the plasma must change through the magnetic field gradient, presuming of course that the plasma is magnetized.
To understand this generalization from the single-particle picture to a distribution of particles, a detailed derivation of these drifts from the first moment of the Vlasov equation can be found in Appendix~\ref{app:GCBulkDrifts}.

\begin{figure}
 \begin{center}
      \includegraphics[width=0.49\textwidth]{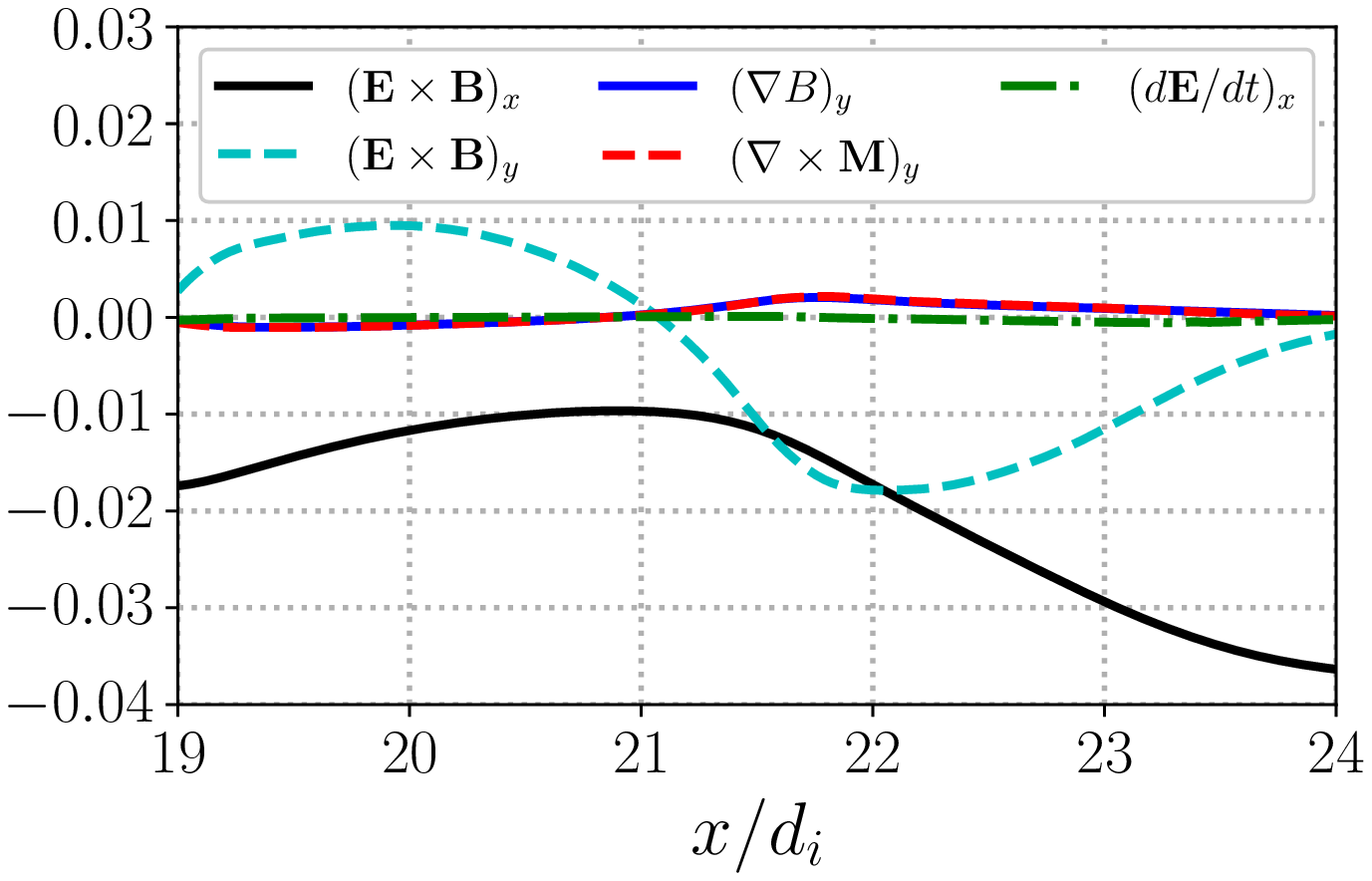}
      \includegraphics[width=0.49\textwidth]{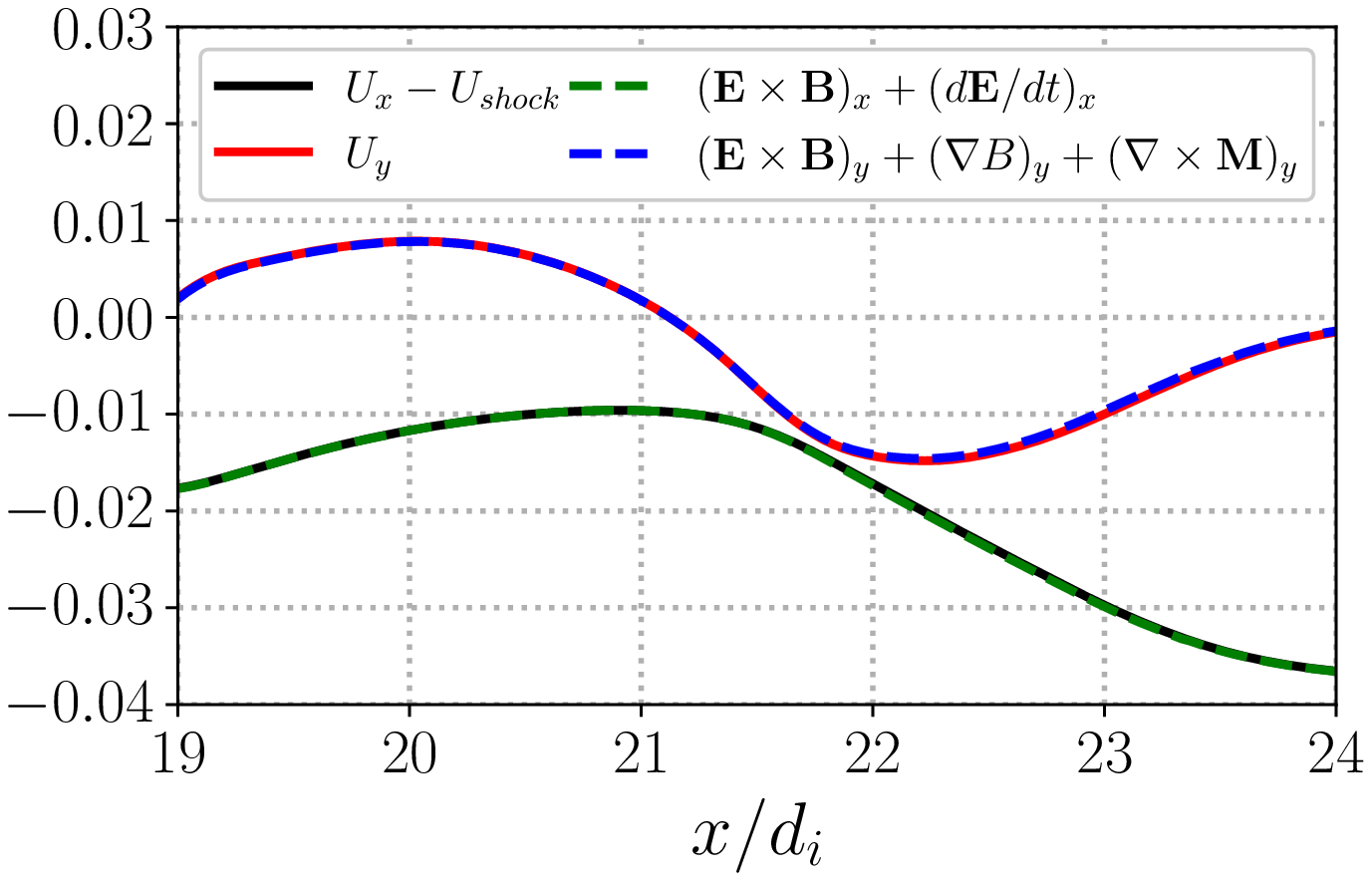}
   \end{center}
       \vskip -1.8in
\hspace*{0.35in} (a)\hspace*{2.15in} (b)
\vskip +1.7in
 \begin{center}
    \includegraphics[width=0.49\textwidth]{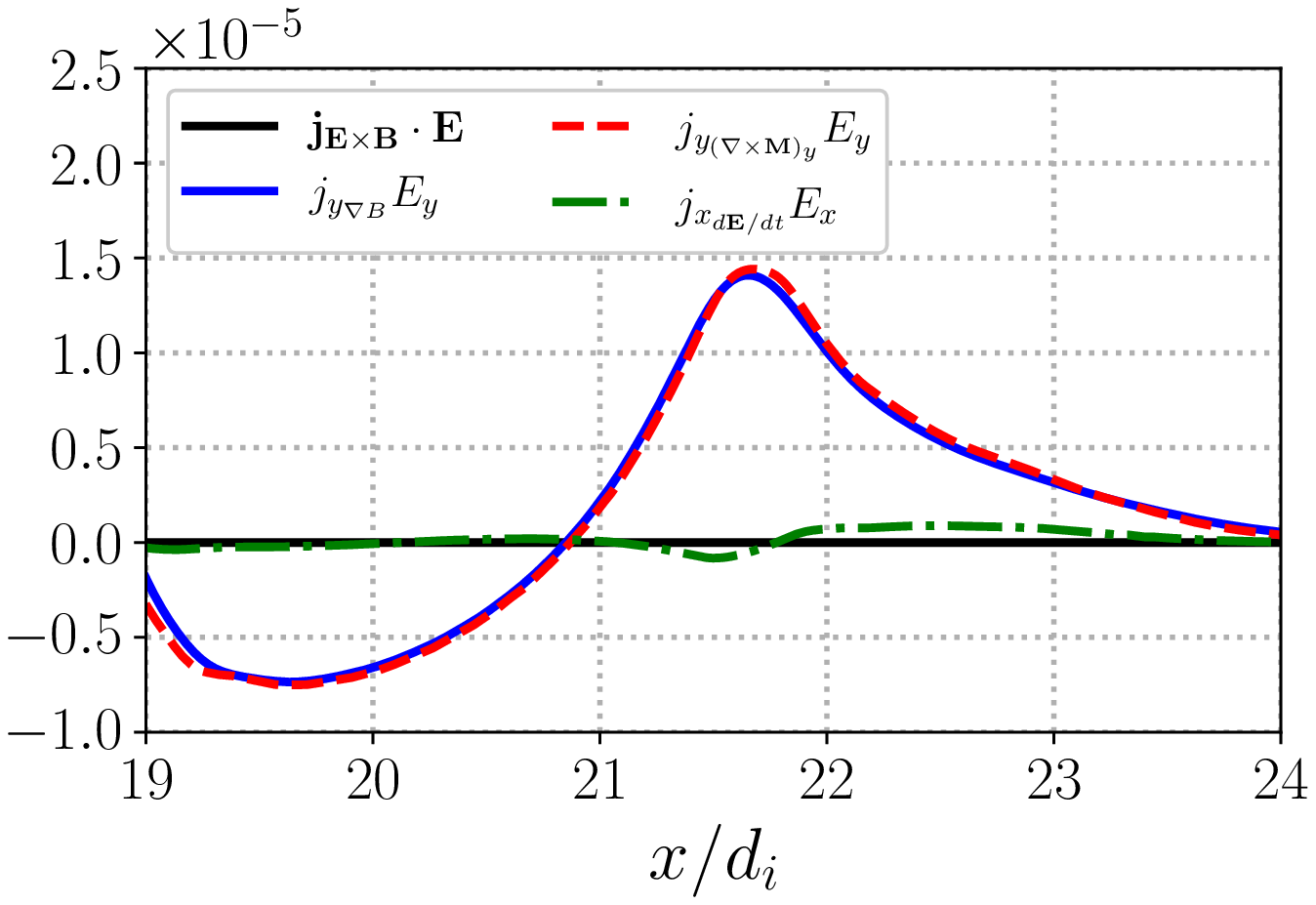}
    \includegraphics[width=0.49\textwidth]{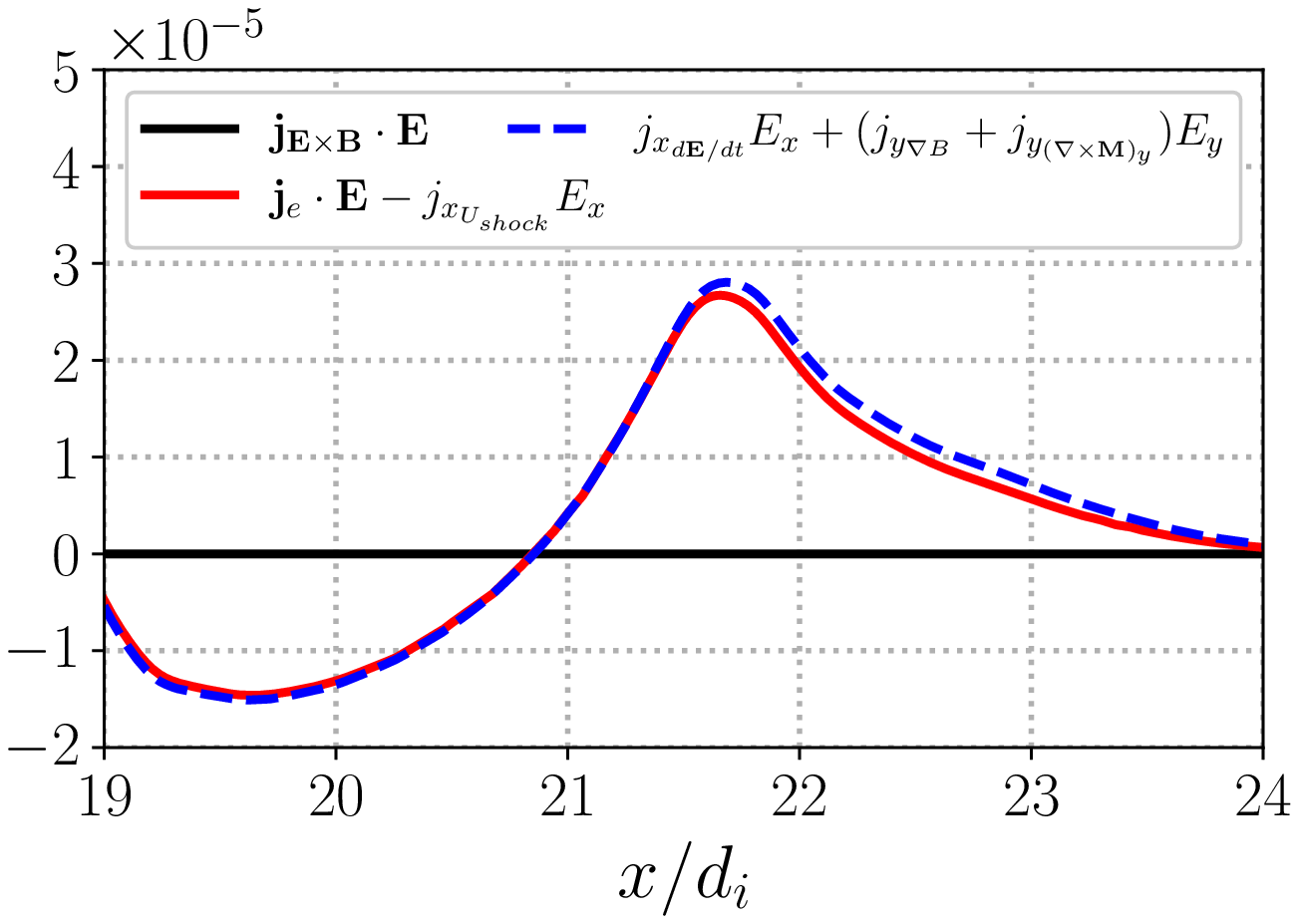}
    \end{center}
       \vskip -1.8in
\hspace*{0.35in} (c)\hspace*{2.15in} (d)
\vskip +1.7in
      \caption{(a) A comparison of the major drifts through the shock evaluated in the shock-rest frame of reference: (i) $\mvec{E} \times \mvec{B}$ drift in $x$ (black) and $y$ (green dashed), (ii) the $\nabla B$ drift in $y$ (blue), (iii) the magnetization drift in $y$ (red dashed), and (iv) the polarization drift in $x$ (magenta dashed-dotted). We check that these drifts sum to the total first moment computed from the electron distribution function (b) as well as determine how each of these drifts contributes to the overall energy exchange, $\mvec{j}_e \cdot \mvec{E}$ (c) and compare the $\mvec{j}_e \cdot \mvec{E}$ computed from these drifts to the total $\mvec{j}_e \cdot \mvec{E}$ computed from moments of the electron distribution function. Note that in comparing how each of these drifts contributes to $\mvec{j}_e \cdot \mvec{E}$, we plot the polarization drift multiplied by $E_x$ (green dashed-dotted), the $\nabla B$ and magnetization drifts multiplied by $E_y$ in the shock-rest frame (blue and red dashed respectively), and the total $\mvec{j}_e \cdot \mvec{E}$ arising from both components of the $\mvec{E} \times \mvec{B}$ flow (black), as we expect the total energization due to the $\mvec{E} \times \mvec{B}$ flow to be zero.}
    \label{fig:drift-comp}
\end{figure}

We plot these drifts in the shock-rest frame in Figure~\ref{fig:drift-comp}(a) as a function of the distance through the shock, showing that there is a clear ordering of the magnitude of these drifts: both components of the  $\mvec{E} \times \mvec{B}$ drift are dominant, the $\nabla B$  and magnetization $\nabla \times \mvec{M}$ drifts provide equal contributions at a smaller amplitude, and the polarization drift is yet smaller.  
To demonstrate that the sum of these drifts indeed completely describes the bulk flow of the plasma in the shock-rest frame, we show in (b) that the first moment of the electron distribution agrees well with the sum of the drifts for both $x$ and $y$ components.

The rate of energization of the electrons in the drift approximation is equal to the rate of work done by the electric field on the drifting electrons, given by 
\begin{align}
    \mvec{j}_{d,e} \cdot \mvec{E} = q_e n_e \mvec{U}_d \cdot \mvec{E}
\end{align}
where $\mvec{U}_d$ is the total drift motion, which can be decomposed into the contributions by the individual drifts identified above in, e.g., Eqns.\thinspace(\ref{eq:gradB1x2v})--(\ref{eq:magnetization1x2v}).
This analysis follows in the vein of studies of electron energization in reconnection using the guiding center approximation \citep{Dahlin:2014, Dahlin:2015, Dahlin:2016, Dahlin:2017, Drake:2019}.
However, we emphasize again that we have generalized from a single-particle picture to a distribution function picture as part of our goal to understand the energization of the electrons in phase space, and thus our equivalent guiding center energization equation has additional terms such as the energization of the plasma via the magnetization drift \eqr{\ref{eq:magnetizationDrift}}, which do not appear in the guiding center energy equation of a single particle, \emph{e.g.}, Eq.~(1) in \citet{Dahlin:2014} and \citet{Dahlin:2016}\footnote{In fact, this generalization from a single-particle picture to a distribution function picture is also pointed out, at least implicitly, by \citet{Dahlin:2014} (and subsequent studies) when summing over particles, thus permitting the definition of the pressure of the plasma, and measuring the total heating of the plasma. While the curvature drift term of interest in these studies is proportional to $v_\parallel^2$ for a single particle, the evolution of the energy density, or integrated energy density as in Eq.~(5) in \citet{Dahlin:2014}, transforms this term to be proportional to $p_\parallel$ because particles of different parallel velocities experience different curvature drifts.}.
We reiterate that the perspective provided by the Lagrangian point-of-view, wherein one considers the energization of individual particles, has significant merit, but that the Eulerian perspective has its own advantages, and to obtain the Eulerian point-of-view we must generalize from single particles to distributions of particles.

In Figure~\ref{fig:drift-comp}(c), we plot the rate of electron energization by each of these four drifts in the shock-rest frame.
The first key takeaway from this figure is that, although the two components of the  $\mvec{E} \times \mvec{B}$ drift dominate the total drift motion, there is no net work done by the $\mvec{E} \times \mvec{B}$ drift as we expect because $(\mvec{E} \times \mvec{B}) \cdot \mvec{E}=0$.
So, while the energization due to one component of the $\mvec{E} \times \mvec{B}$ drift may appear large, summed over all components the energization must be zero, as shown in Figure~\ref{fig:drift-comp}(c).
In this regard, separately plotting the contributions to the rate of electron energization $\mvec{j}_e \cdot \mvec{E}$ by the different components of the $\mvec{E} \times \mvec{B}$ drift can be somewhat misleading, as both components of the $\mvec{E} \times \mvec{B}$ flow are much larger than the other drifts.

That the full $\mvec{E} \times \mvec{B}$ drift leads to zero net energization of the particles also explains the puzzling finding in our single-particle modeling of the electron energization shown in Figure~\ref{fig:spm_e_comparison}(c). 
Although the  $j_{e_x} E_x$ and  $j_{e_y} E_y$ from the full model were larger than and significantly different from the zero $E_x$ model, when summed they yielded the same net rate of energization of the electrons as the zero $E_x$ model. 
This cancellation is exactly the result of the two components of ``energization'' from the $\mvec{E} \times \mvec{B}$ flow cancelling, as we know they must. 
The remaining net energization is then solely from the other drifts and their alignment with the motional electric field $E_y$.

In Figure~\ref{fig:drift-comp}(d), we check that the total rate of energization of the electrons by the electric field in the shock-rest frame, $\mvec{j}_e \cdot \mvec{E}$, agrees with the sum of the energization by the $\nabla B$, magnetization, and polarization drifts, finding good agreement.  
Note that for this comparison, we are computing the electron current in the shock-rest frame from the electron distribution function, i.e.,
\begin{align}
    \mvec{j}_e = \int q_e (\mvec{v}' - U_{shock} \hat{\mvec{x}}) f_e d\mvec{v}'.
\end{align}
After the comparison between the full model and zero-$E_x$ model in Figure~\ref{fig:spm_e_comparison}(c) and (d) revealed that the total $\mvec{j}_e \cdot \mvec{E}$ was roughly equivalent between the two models, when the zero-$E_x$ model only had energy exchange due to the alignment of the $\nabla B$ drift with the motional electric field, we might have anticipated that the only energy gain was due to this same adiabatic heating process from the idealized model in Section~\ref{sec:elc-SPM}.
Importantly though, we see from the drift analysis in Figure~\ref{fig:drift-comp}(c) that the model in Section~\ref{sec:elc-SPM} must be generalized to the case when a distribution of particles is drifting.

While the energy gain by a single electron $\nabla B$ drifting in the model fields is exactly the energy gain required for that single electron's magnetic moment $\mu$ to be conserved, in the self-consistent simulation it is not only the $\nabla B$ drift that ensures the electron distribution's adiabatic invariant $\mu_e$ is well conserved in the shock.
We also have an equal contribution to the energization from the magnetization drift.  
Together, as shown by \eqr{\ref{eq:diamag}}, the $\nabla B$ and magnetization drift are equivalent to the diamagnetic drift, $ u_{\mbox{diamag}} =  u_{\nabla B} + u_{\nabla \times \mvec{M}}$, so another perspective on the electron energization via adiabatic heating is the adiabatic invariant of a distribution of electrons, $\mu_e$ in \eqr{\ref{eq:distMu}}, is conserved due to the alignment of the diamagnetic drift and the motional electric field, $E_y$.

Two important questions remain: (i) what is responsible for the slight disagreement between the energy gain due to the $\nabla B$ and magnetization drifts and $\mvec{j}_e \cdot \mvec{E}$ from velocity moments of the electron distribution function in Figure~\ref{fig:drift-comp}(d)?; and (ii) even if the physics of adiabatic heating is the same with the electrons gaining energy via the alignment of drifts with the motional electric field, is the velocity-space signature of adiabatic heating the same as that predicted by the Vlasov-mapping model in Figure~\ref{fig:spm_e_comparison}(b)?
To answer the first question, we have performed a more realistic mass ratio simulation, $m_i/m_e = 400$, in \appref{app:massRatio400} where we find better agreement between the energy gain due  to the $\nabla B$ and magnetization drifts and $\mvec{j}_e \cdot \mvec{E}$ from velocity moments of the electron distribution function.
Thus, the small disagreement between these methods of measuring the electron's energy gain in the $m_i/m_e = 100$ simulation is simply due to the fact that the electron gyroradius is not asymptotically smaller than the shock's extent.
To answer the second question, we seek a means of eliminating the $\mvec{E} \times \mvec{B}$ component of the energy exchange in the FPC.


\subsection{Velocity-Space Signature of Adiabatic Electron Heating}

If the large $\mvec{E} \times \mvec{B}$ flows are polluting the analysis of the overall exchange of the energy, when fundamentally the electron heating is principally due to the alignment of the $\nabla B$ and magnetization drifts with the motional electric field, we return to the velocity-space signature plotted in Figure~\ref{fig:fecexcey}(c) and Figure~\ref{fig:fecey-model-comp}(d) to determine how we might remove the contribution from the large $\mvec{E} \times \mvec{B}$ flows to the FPC signal.
Guided by the fact that the transverse electric field component $E_y$ governs the adiabatic heating of the electrons, we seek to eliminate the large contribution to the rate of energization associated with the $y$-component of the  $\mvec{E} \times \mvec{B}$ drift (which is ultimately canceled by energization associated with its $x$-component, as we show in \appref{app:exb}).  
Therefore, we transform to a frame of reference moving in the transverse direction at the same velocity as the $y$-component of the  $\mvec{E} \times \mvec{B}$ drift, $\mvec{U}_{td}= -E_x/B_z \yhat$.
We define the \emph{transverse drift frame} of reference: (i) in the $x$ direction, or shock-normal direction, the shock is at rest; (ii) in the $y$ direction, or transverse direction, the frame moves at a velocity equal to the $y$-component of the local  $\mvec{E} \times \mvec{B}$ drift in the shock-rest frame.

Critically, the electric field transforms from the shock-rest frame ($sf$, unprimed) to the transverse drift frame ($td$, double primed), as
\begin{align}
    \mvec{E}''= \mvec{E} + \mvec{U}_{td} \times \mvec{B} = \mvec{E} + (-\frac{E_x}{B_z}\yhat) \times B_z \zhat = \mvec{E} -E_x \xhat = E_y \yhat 
\end{align}
where we have assumed $E_z=0$ in this 1D-2V perpendicular shock.
Therefore, in the transverse drift frame, the cross-shock electric field is zero, $E_x''=0$, and the motional electric field is unchanged from the shock-rest frame, $E_y''=E_y$.
The $y$-component of the velocity coordinate in the transverse drift frame is
\begin{align}
    v_y''= v_y - U_{td,y} = v_y + \frac{E_x}{B_z}.
\end{align}
For completeness, the $x$ velocity coordinate and magnetic field are unchanged from the shock-rest frame, $v_x''=v_x$ and $\mvec{B}''=\mvec{B}$.

Although the transverse drift frame changes with position through the shock as the cross-shock electric field changes along the normal direction, one can determine this frame of reference at any position from a local, single-point measurement of the electric field. 
In contrast, other drifts generally depend on gradients, and therefore cannot be uniquely specified using only single-point measurements. 
The benefit of the transverse drift frame of reference is not only that the rate of energization associated with the total  $\mvec{E} \times \mvec{B}$ drift is equal to zero, which is true in any frame of reference, but also that the rates of energization associated with each component of the $\mvec{E} \times \mvec{B}$ drift are separately zero. \footnote{Note that the transverse drift frame is not the only frame of reference in which the contributions to the energization due to each component of the $\mvec{E} \times \mvec{B}$ drift are zero.  One could also define a normal drift frame moving at the normal component of the local $\mvec{E} \times \mvec{B}$ drift, which would yield $E_y''=0$.  This could be useful in determining the energization associated with the polarization drift in the $x$ direction, but since this is a subdominant contribution, we do not pursue that line of investigation here.}

\begin{figure}
 \begin{center}
    \includegraphics[width=0.49\textwidth]{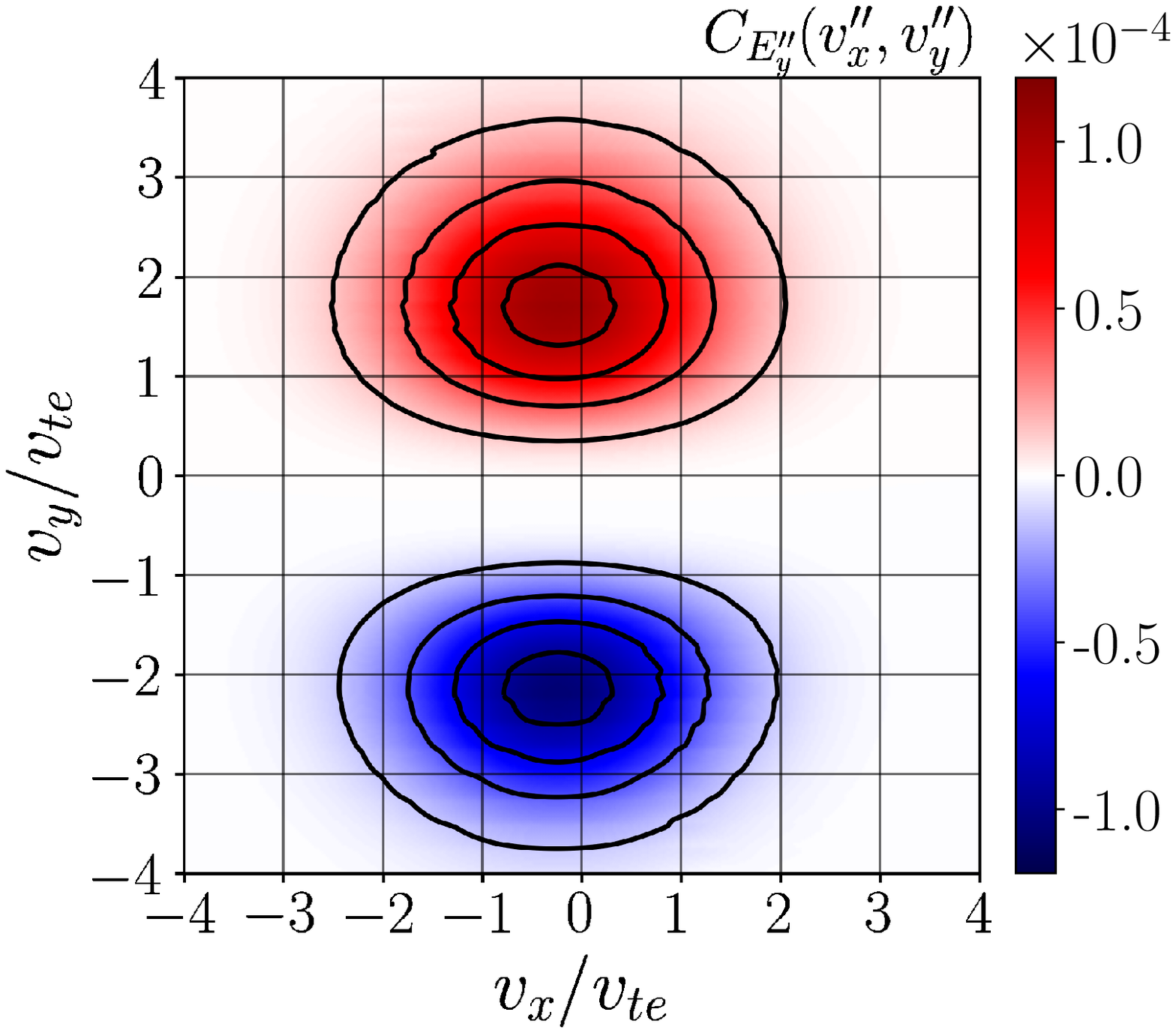}
    \includegraphics[width=0.49\textwidth]{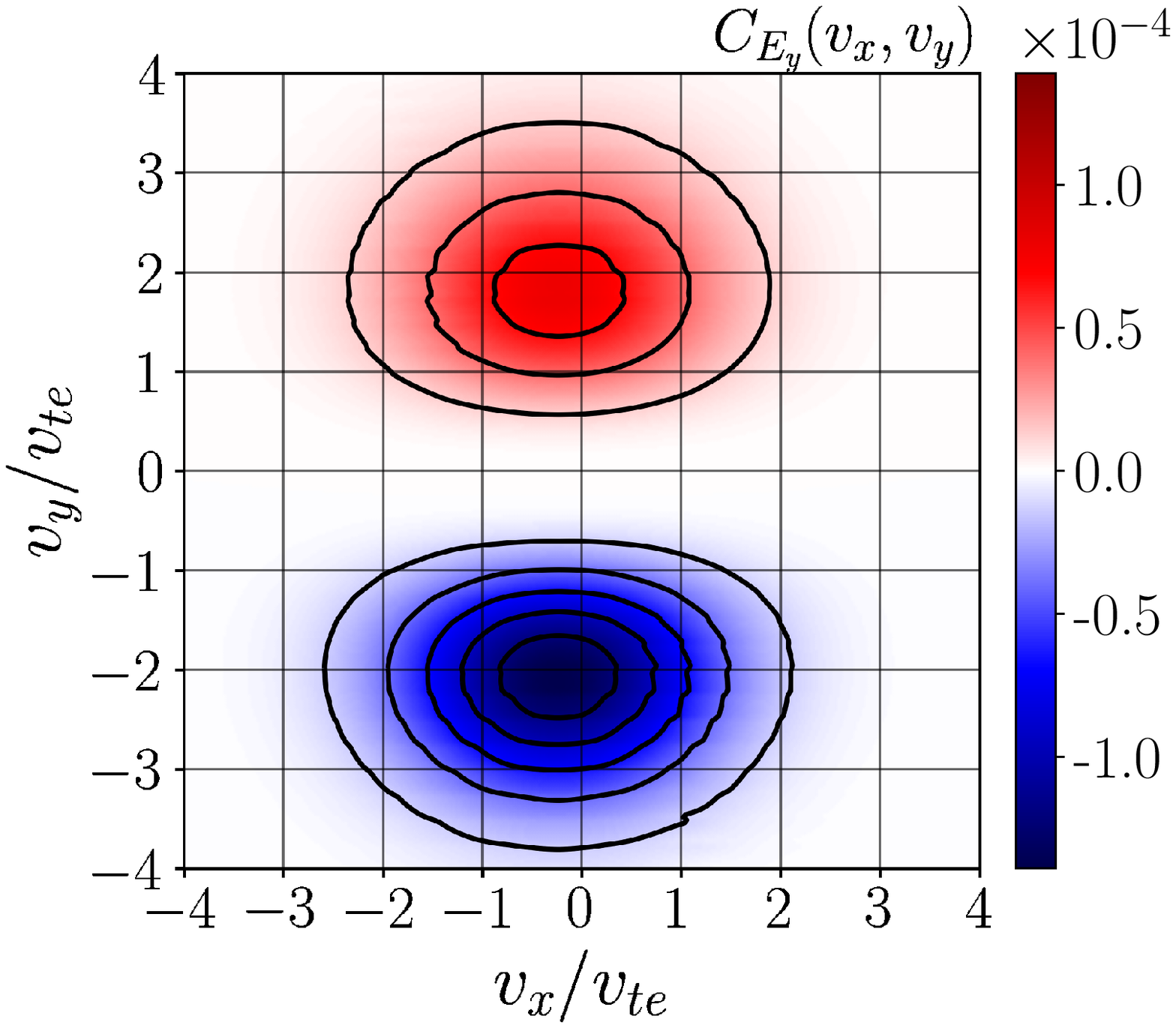}
   \end{center}
 \vskip -2.0in
\hspace*{0.05in} (a)\hspace*{2.45in} (b)
\vskip +1.85in
    \caption{A comparison of the FPC from $E_y$ where $v_x$ and $v_y$ are shifted to the shock-rest frame and local $\mvec{E} \times \mvec{B}$ frame respectively, i.e. the transverse drift frame, $C_{E''_y}(v''_x, v''_y)$ (a) to our previous computation of the FPC from $E_y$ using only a frame transformation in $v_x$ to the shock-rest frame, $C_{E_y}(v_x,v_y)$ (b). Note that panel (b) is a repeat of panel (c) of Figure~\ref{fig:fecexcey} and panel (d) of Figure~\ref{fig:fecey-model-comp}. While the previous correlation, $C_{E_y}$, suggested the electrons were losing energy in this degree of freedom, the newly transformed correlation, $C_{E''_y}$, demonstrates that the electrons are in fact gaining energy once the $\mvec{E} \times \mvec{B}$ motion in this degree of freedom is subtracted. This energization mechanism, caused by an alignment of the $\nabla B$ and magnetization drifts with the motional electric field, $E_y$, is the same mechanism responsible for energizing the electrons in the idealized model, and the velocity-space signature for this adiabatic heating now exactly matches the results of the idealized model presented in Sections~\ref{sec:elc-SPM} and \ref{sec:elcFPC}.}
    \label{fig:transformedCey}
\end{figure}

Since $E_x''=0$ in the transverse drift frame, we need only examine the $y$ correlation, $C_{E_y''}(v_x'',v_y'')$ to explore the energization of the electrons.
We plot in Figure~\ref{fig:transformedCey}(a) the correlation $C_{E_y''}(v_x'',v_y'')$ in the transverse drift frame at position $x_B=21.8 d_i$ compared to (b) the corresponding correlation $C_{E_y}(v_x,v_y)$ in the shock-rest frame (a repeat of Figure~\ref{fig:fecexcey}(c)). 

We note two things immediately from Figure~\ref{fig:transformedCey}. 
First, transformation to the transverse drift frame has switched the sign of the net rate of electron energization (from integrating the correlation over velocity space) relative to the case in the shock-rest frame, made clear by counting the over-plotted equally space contours in the blue-red, two-lobe structure of each case.
Second, the correlation  $C_{E_y''}(v_x'',v_y'')$  in the transverse drift frame is strikingly similar to the correlation found in Figure~\ref{fig:fecey-model-comp}(b) computed from the reconstructed distribution function from the idealized model.
Both points are perhaps unsurprising, as this transformation has removed the energy exchange in the $v_y$ degree of freedom due to the $\mvec{E} \times \mvec{B}$ flow, and we are left with a similar energization mechanism found in the idealized model: alignment of the $\nabla B$ and magnetization drifts with the motional electric field, $E_y$.
The alignment of these two combined drifts (constituting the diamagnetic drift) with the motional electric field is exactly what is required for the electron distribution's adiabatic invariant $\mu_e$ to be well-conserved and for the electrons to gain energy through the increasing magnetic field of the perpendicular shock.  
Figure~\ref{fig:transformedCey}(a) contains the second key result of this paper, the velocity-space signature of adiabatic electron heating.

It is worth emphasizing that transformation to the transverse drift frame not only revealed the same velocity-space signature for adiabatic heating as in the idealized model analyzed in Sections~\ref{sec:elc-SPM} and \ref{sec:elcFPC}, but also allowed the extraction of an energization signature which was buried in the large background energy exchange from the $\mvec{E} \times \mvec{B}$ flow.
Ultimately, the adiabatic heating via alignment of the $\nabla B$ and magnetization drifts with the motional electric field was masked by the large oscillation of energy between the $v_x$ and $v_y$ degrees of freedom due to the two components of the $\mvec{E} \times \mvec{B}$ motion. 
The need to carefully consider the frame of reference in which the energization analysis is performed, especially due to the impact the frame of reference choice has on the cross-shock electric field, has been pointed out in previous studies \citep{Goodrich:1984}.

One must therefore exercise extreme caution in attributing energization to a particular drift when separately considering the work done by the different components of the electric field.  
In Figure~\ref{fig:spm_e_comparison}(c), for example, it would be easy to attribute erroneously the energization of the electrons to the cross-shock electric field $E_x$.
Rather, the energization due to the $x$-component of the  $\mvec{E} \times \mvec{B}$ drift and $E_x$ is exactly canceled by a loss of energy due to the $y$-component of the  $\mvec{E} \times \mvec{B}$ drift and  $E_y$. 
The net result is a much smaller positive rate of energization due to the motional electric field $E_y$ and the remaining drifts in the transverse direction.

That the addition of the magnetization drift did not qualitatively alter the velocity-space signature of adiabatic electron heating can be understood by considering the general qualitative features of drift energization. 
In the drift approximation for a warm electron distribution relevant to most heliospheric plasmas, it is often true that the magnitude of the drift (not including the dominant $\mvec{E} \times \mvec{B}$ drift, which cannot energize particles) is much less than the thermal velocity of the electrons, $U_{d,e} \ll v_{te}$.  
In this case, the center of the drifting velocity distribution is offset from the origin of velocity space, but this offset will be much smaller than the thermal width of the distribution.
When the correlation is computed by taking the velocity derivatives of $f_e$ and multiplying by the appropriate component of the electric field weighted by $|\mvec{v}|^2$, the result will generally produce a two-lobed velocity-space signature, qualitatively similar to that shown in  Figure~\ref{fig:transformedCey}(a).

A careful consideration of the drifts is essential to interpret properly the mechanism responsible for the adiabatic heating. 
But much of the interest in collisionless shocks is focused not on adiabatic heating and instead on identifying mechanisms of non-adiabatic heating. 
In fact, because the observed temperature increase in the electrons is entirely in the perpendicular temperature to conserve the electron distribution's adiabatic invariant, even if the electron response is initially adiabatic, this anisotropy will itself be a source of instabilities, such as the whistler anisotropy instability \citep{Gary:1993, Gary:2011, Wilson:2013a, Wilson:2020}, that will further complicate the energy exchange.

Likewise, the transverse drift frame relies on the electrons being magnetized so that their $\mvec{E} \times \mvec{B}$ motion is well-defined.
While the transformation is sensible here because $L_{shock} \gg \rho_e$, there are many observations of collisionless shocks where the shock ramp is not asymptotically larger than the electron gyroradius \citep{Hobara:2010} and thus would warrant caution in the application of the transverse drift frame to reveal any masked energization signatures such as the velocity-space signature of adiabatic heating in Figure~\ref{fig:transformedCey}(a).
In addition, the transverse-drift frame is a non-inertial frame because the $\mvec{E} \times \mvec{B}$ velocity is changing through the shock.
While the transformation to the instantaneous transverse-drift frame at a single point in configuration space in a simulation is a perfectly reasonable frame transformation, care will be required applying this technique to analysis of spacecraft data, which inevitably average over a small volume of configuration space.

Nevertheless, the velocity-space signatures of the mechanisms that govern non-adiabatic heating are likely to be entirely distinct from the simple two-lobed appearance of adiabatic heating, and so will be easy to distinguish using a field-particle correlation analysis. 
In addition, it is useful to first characterize the ``background'' signature of adiabatic heating, as represented by the typical velocity-space signature shown in Figure~\ref{fig:transformedCey}(a).
In cases where the adiabatic response produces a temperature anisotropy that drives instabilities, the energetic response, and thus the electron's velocity-space signature through the shock, is likely to be characterized by a combination of both adiabatic and non-adiabatic signatures. 


\section{Summary and Future Outlook}\label{sec:conclusions}

This paper presents the first attempt to characterize the energy exchange in a collisionless shock using the field-particle correlation technique \citep{Klein:2016, Howes:2017}. 
We have examined a self-consistent perpendicular collisionless shock using the continuum Vlasov-Maxwell solver in the \gke~simulation framework and our results can be summarized as follows:
\begin{enumerate}
\item[(i)] We have identified the velocity-space signatures of shock-drift acceleration of ions in Figure~\ref{fig:ficey-model-comp}(d) and adiabatic heating of electrons in Figure~\ref{fig:transformedCey}(a).
\item[(ii)] Using simplified models of single-particle motion through idealized models of the electromagnetic fields through the shock transition, we identified the conditions under which we expect to observe these velocity-space signatures for these energization processes. 
\begin{enumerate}
\item We determined that the velocity-space signature of shock-drift acceleration can be seen clearly in a reflected ion population and is robust to the presence of the finite shock width and cross-shock electric field that arise in the  self-consistent simulation.
\item For the electrons, we determined it was critical to eliminate the energization associated with the separate components of the $\mvec{E} \times \mvec{B}$ drift by transforming to the transverse drift frame of reference, as the large $\mvec{E} \times \mvec{B}$ drifts in both $x$ and $y$ due to the incoming supersonic flow and cross-shock electric field significantly obscured the effect of the energization of the electrons via other drifts. 
\end{enumerate}
\item[(iii)] Finally, we observed a general strength of this method of analysis using the field-particle correlation technique, going beyond identification of  where $\mvec{j} \cdot \mvec{E}$ is positive and which components are positive, and furthermore captured the subtleties of a generalization from a single-particle picture to a distribution of particles. 
\begin{enumerate}
\item In the case of the ions, a cursory glance at the $x$ component of $\mvec{j} \cdot \mvec{E}$, which is overall negative due to the slowing down of the bulk distribution, obscures the role the cross-shock electric field plays in increasing the reflected fraction of ions---see \appref{app:crossShockIon} for further details. 
\item In the case of electrons, just using the components of $\mvec{j} \cdot \mvec{E}$ would completely miss the actual source of energization, as one might expect that the positive $x$ component of $\mvec{j} \cdot \mvec{E}$ corresponds to net energization via the cross-shock electric field. In fact, it is not the cross-shock component of the electric field $E_x$ which leads to the observed energization of the electrons through the shock, but rather the motional electric field $E_y$. And while in the single-particle picture the energization is simply the alignment of said motional electric field with the $\nabla B$ drift, the generalization to a distribution of electrons introduces an additional drift, the magnetization drift, which when combined with the $\nabla B$ drift, forms the diamagnetic drift. When aligned with the motional electric field, these two drifts provide the necessary energization for the adiabatic invariant of the electron distribution to be well conserved through the shock.
\end{enumerate}
\end{enumerate}

The work presented here is only the beginning of a program of study to determine, in general, how we may be able to leverage the full information contained in the particle velocity distribution function to ascertain the details of the energy exchange in a collisionless shock.
While historically the Lagrangian perspective of examining how individual particles gain and lose energy has led to enormous improvements in our understanding of the dynamics and energetics of collisionless shocks, this complementary Eulerian approach, directly analyzing the energy exchange in phase space using the field-particle correlation technique, has significant value for interpreting both simulation and spacecraft data.
Especially when advances in spacecraft instrumentation provide ever higher resolution and higher cadence particle velocity distribution measurements of collisionless shocks \citep{ChenLJ:2018, Goodrich:2018}, the time is ripe for fully exploiting the information contained in phase space to provide a deeper of understanding of the mechanisms of particle energization at a collisionless shock.

Further studies of higher dimensional, higher magnetosonic Mach number, and more general geometry collisionless shocks are of the utmost importance. 
As reviewed in the introduction, there is a large variety of processes not considered in this study that have been studied previously as potential energization mechanisms, from shock surfing acceleration to diffusive shock acceleration.
As with the body of work utilizing the field-particle correlation technique for analyzing dissipation via resonant processes, we will require a systematic study of all of these processes if we have any hopes of distinguishing their velocity-space signatures.
We may expect certain energization processes such as diffusive shock acceleration and the Fast Fermi process, which also rely on particle reflection, may have qualitatively similar velocity-space signatures to shock-drift acceleration but still contain all the requisite information to identify the particular process locally occurring, just as we can utilize information such as the velocity around which a resonant wave process is identified to characterize the particular waves which are resonantly energizing the plasma.

We expect as we move to higher dimensionality, higher magnetosonic Mach number, and more general shock geometry, our analysis will also be further complicated by upstream kinetic instabilities such as those observed in the Earth's bow shock.
The velocity-space signatures of these instabilities are of equal importance to characterize and study using the field-particle correlation technique.
While the instantaneous field-particle correlation technique employed in this study was well-suited to the impulsive ion energization via shock-drift acceleration and the steady electron energization via adiabatic heating, we may require finite time correlations to characterize the energy exchange within the upstream fluctuations and turbulence of the shocks in exact analogy with previous field-particle correlation studies \citep{Klein:2017b, Klein:2020, Horvath:2020}.
Nevertheless, an exciting frontier awaits in applying the field-particle correlation technique to the distribution functions produced by more realistic collisionless shock simulations and classifying the observed velocity-space signatures of particle energization.

\section{Acknowledgements}
The authors thank L.J. Chen and S. Wang for enlightening discussions on collisionless shocks and the entire \gke~team, especially N. Mandell, M. Francisquez, T. Bernard, and P. Cagas, for all of their insights.
This work used the Extreme Science and Engineering Discovery Environment (XSEDE), which is supported by National Science Foundation grant number ACI-1548562.
J. Juno was supported by a NSF Atmospheric and Geospace Science Postdoctoral Fellowship (Grant No. AGS-2019828) and a NASA Earth and Space Science Fellowship (Grant No. 80NSSC17K0428).
G. G. Howes, J. M. TenBarge, D. Caprioli, and A. Spitkovsky were supported by NASA grant 80NSSC20K1273.
G. G. Howes was also supported by NASA grants 80NSSC18K1366, 80NSSC18K1217, 80NSSC18K1371, and 80NSSC18K0643.
J. M. TenBarge was also supported by DOE grant DE-SC0020049.
A. Spitkovsky was also supported by NSF grants PHY-1804048 and AST-1814708.
D. Caprioli was partially supported by NASA (grant NNX17AG30G, 80NSSC18K1218, and 80NSSC18K1726) and NSF (grants AST-1714658, AST-1909778, PHY-1748958, PHY-2010240).
K. G. Klein was supported by DOE grant DE-SC0020132 and NASA grant 80NSSC19K0912.
J. Juno, G. G. Howes, D. Caprioli, and A. Spitkovsky were partially supported by the ``Multiscale Phenomena in Plasma Astrophysics''  program at the Kavli Institute for Theoretical Physics (NSF grant PHY-1748958).
The work was also supported by the International Space Science Institute's (ISSI) International Teams programme.

\appendix

\section{Component-wise Separation of the Field-Particle Correlation Technique}\label{app:v2FPC}
\begin{figure}
   \begin{center}
      \includegraphics[width=0.49\textwidth]{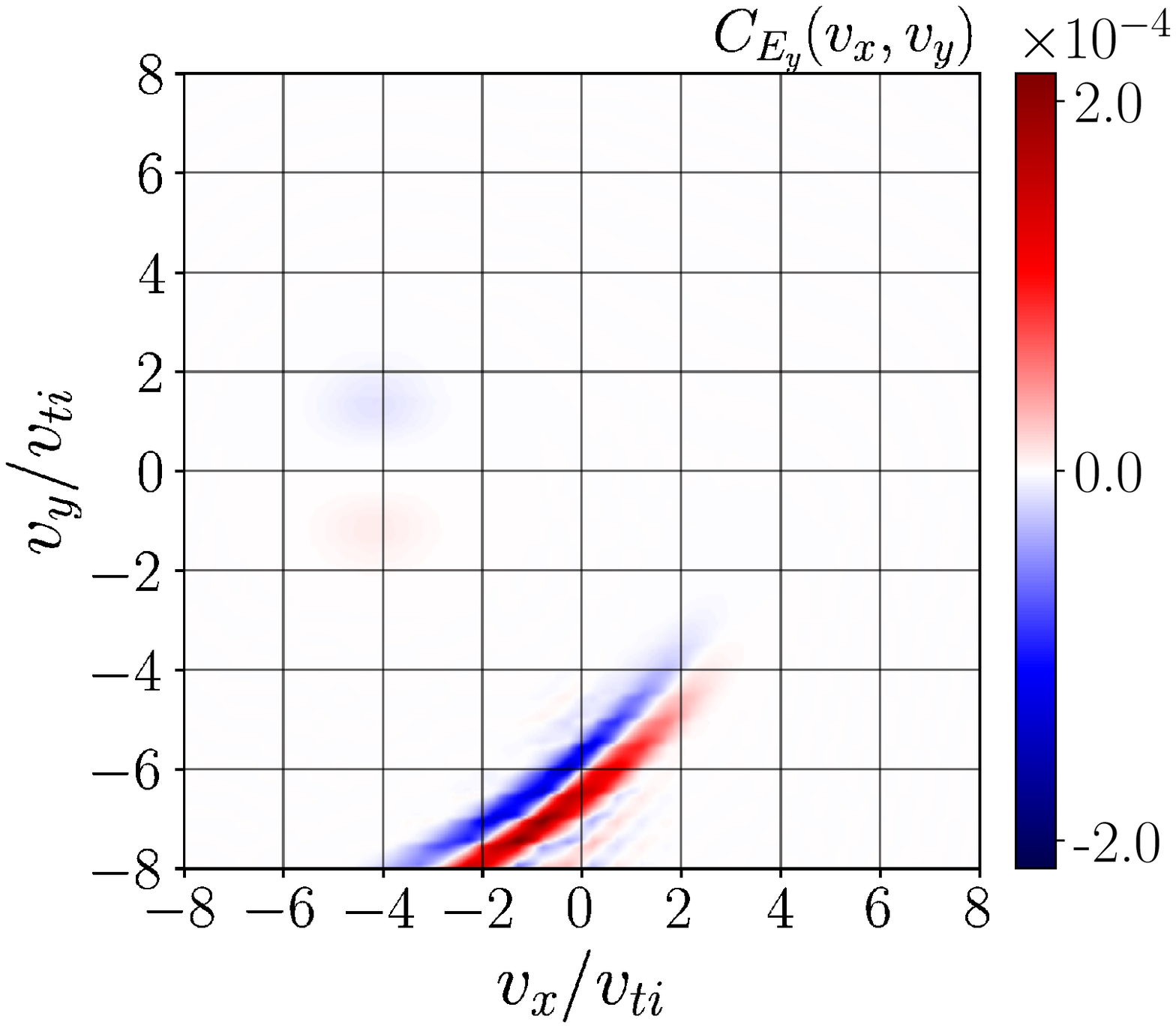}
      \includegraphics[width=0.49\textwidth]{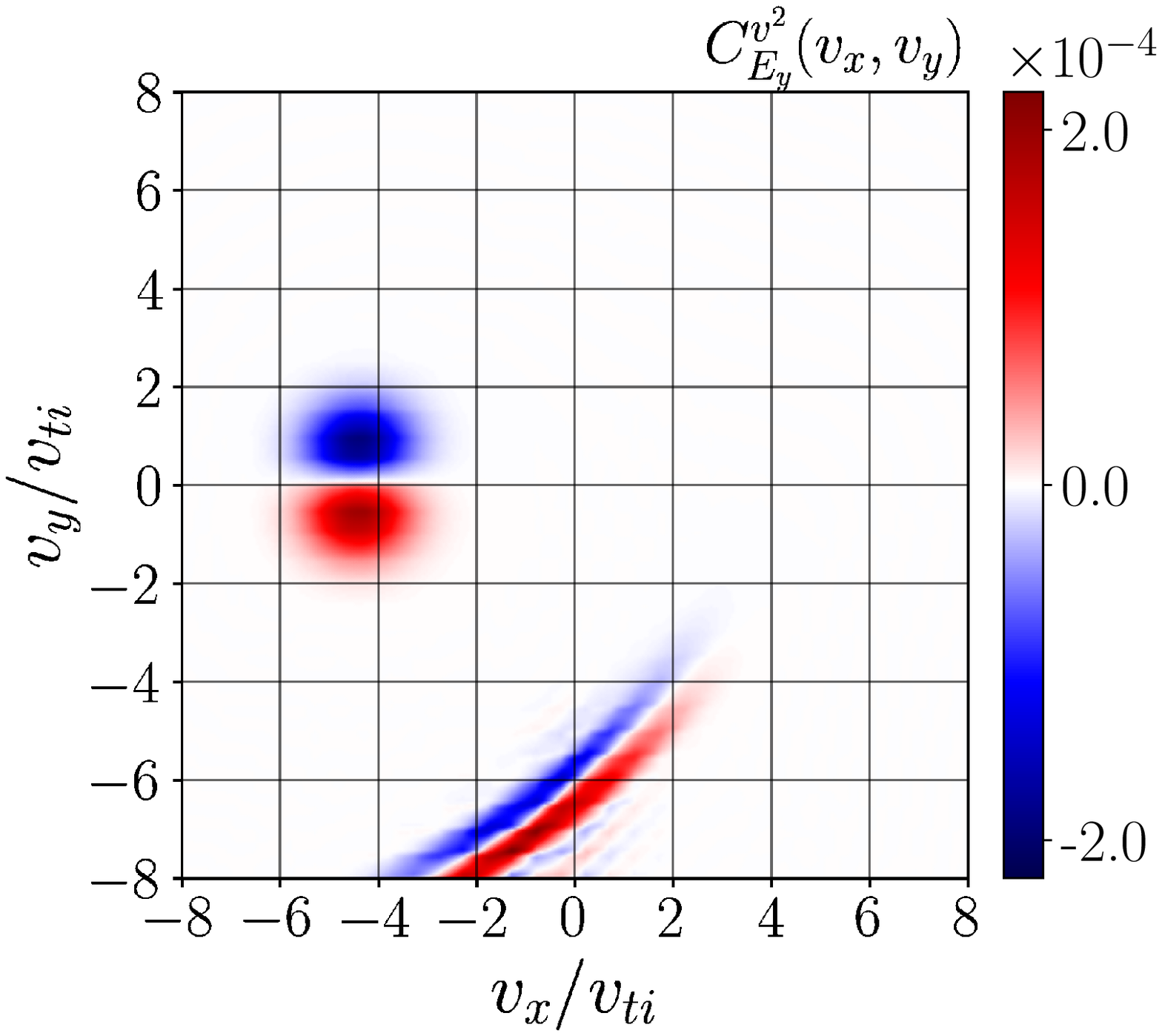}
   \end{center}
\vskip -2.0in
\hspace*{0.05in} (a)\hspace*{2.45in} (b)
\vskip +1.85in
   \caption{(a) The field particle correlation $C_{E_y}(v_x,v_y)$ from \eqr{\ref{eq:vyCorrelation}} using weighting $v_y^2$   vs.~(b) the correlation $C^{(v^2)}_{E_x}(v_x,v_y)$ using the full 
  $v^2$ weighting, 
   both computed from the ion distribution function of the self-consistent \gke~simulation.}
   \label{fig:vyvsfull} 
\end{figure}

In \eqr{\ref{eq:FPC-full}} derived in Section~\ref{sec:fpc}, the field-particle correlation was expressed as a dot product between the electric field and the velocity-space gradient of the particle distribution function.
While this gives the total energy exchange, it is often useful to identify components of the energy exchange by decomposing the field-particle correlation technique like so,
\begin{align}
  C^{(v^2)}_{E_x}(\mvec{v},t,\tau) & = -q_s\frac{v^2}{2} E_x(\mvec{x}_0,t) \cdot \frac{\partial f_s(\mvec{x}_0,\mvec{v},t)}{\partial v_x}, \\    
  C^{(v^2)}_{E_y}(\mvec{v},t,\tau) & = -q_s\frac{v^2}{2} E_y(\mvec{x}_0,t) \cdot \frac{\partial f_s(\mvec{x}_0,\mvec{v},t)}{\partial v_y}, \label{eq:ceyfullv2}
\end{align}
similar to \eqr{\ref{eq:vxCorrelation}} and \eqr{\ref{eq:vyCorrelation}}, but without the substitution of the components of $v^2$.
We now justify this additional substitution to obtain the form of the field-particle correlation technique employed throughout this manuscript.

Although this substitution alters the rate of change of phase-space energy density as a function of velocity space $(v_x,v_y)$, the difference in these two forms vanishes upon integration of the field-particle  correlation over velocity space. 
In other words, the change does not alter the net rate of particle energization at a given spatial position $\mvec{x}_0$. 
That this replacement does not alter the net rate of energization is easily seen by examining the $x$ contribution to the dot product in the second term of \eqr{\ref{eq:dws}} when integrated over $(v_x,v_y)$ velocity space,
\begin{eqnarray}
\lefteqn{\int_{-\infty}^{-\infty}dv_x \int_{-\infty}^{-\infty}dv_y 
\left( - q_s\frac{(v_x^2+v_y^2)}{2}  E_x \frac{\partial f_s}{\partial v_x} \right)} & & \nonumber \\
& = &-\frac{q_s}{2}  E_x \int_{-\infty}^{-\infty}dv_y \left[
\int_{-\infty}^{-\infty}dv_x  v_x^2 \frac{\partial f_s}{\partial v_x}
+ v_y^2 \cancelto{0}{\int_{-\infty}^{-\infty}dv_x \frac{\partial f_s}{\partial v_x} }
\right] \nonumber \\
& = & \int_{-\infty}^{-\infty}dv_x \int_{-\infty}^{-\infty}dv_y 
\left( - q_s\frac{v_x^2}{2}  E_x \frac{\partial f_s}{\partial v_x}   \right)
\end{eqnarray}
where the $v_x$ integral of the factor with $v_y^2$ is a perfect differential, and thus contributes nothing assuming appropriate velocity-space boundary conditions, $\lim_{v_x \to \pm \infty} f(v_x,v_y)=0$.

The primary motivation for this substitution of the components of $v^2$ is to highlight the regions  in velocity space that contribute to the net energy transfer between the particles and fields. 
Let us compare the velocity-space signature obtained with \eqr{\ref{eq:ceyfullv2}} to \eqr{\ref{eq:vyCorrelation}} for the case considered in Section~\ref{sec:ions}.
In Figure~\ref{fig:vyvsfull}, we plot the two forms of the field-particle correlation (a) \eqr{\ref{eq:vyCorrelation}}, $C_{E_y}$, and (b) \eqr{\ref{eq:ceyfullv2}}, $C_{E_y}^{(v^2)}$ for the same case shown in Figure~\ref{fig:ficexcey}, panel(e).
Using the alternative  form in Figure~\ref{fig:vyvsfull}(b) $C_{E_y}^{(v^2)}$ given by \eqr{\ref{eq:ceyfullv2}}, we see that there is a large feature in the velocity-space signature of the ion energization associated with the incoming ion flow, but that significant feature leads to zero net ion energization. 
In fact, apparent energy transfer associated with $E_y$ in this form is actually canceled exactly by the magnetic field term $(\mvec{v} \times \mvec{B})_x \partial f/\partial v_x$  in the Lorentz force, as discussed in Appendix \ref{app:exb}.

Using the preferred form in Figure~\ref{fig:vyvsfull}(a) $C_{E_y}$, this net zero energy transfer associated with the incoming ion beam does not appear. 
Only the energy transfer associated with the reflected ions appears when using the form in \eqr{\ref{eq:vyCorrelation}}. 
Therefore, although using only the  $v_y^2$ contribution does not capture the full energy flow in velocity space, it does capture the energy transfer associated with the net rate of energization, and so this form is preferable for the study of particle energization.


\section{Calculation of Field-Particle Correlation for the $\mvec{E}\times \mvec{B}$ Drift}
\label{app:exb}

Here we calculate the field-particle correlation for the rate of change of phase-space energy density of a plasma undergoing uniform $\mvec{E} \times \mvec{B}$ motion. 
Consider the case, relevant to the particular transverse magnetized shock problem addressed here, of a constant transverse
magnetic field $\mvec{B}=B_{z0} \zhat$ and a constant electric field $\mvec{E} = -E_{y0} \yhat$ where $E_{y0}>0$, giving an upstream $\mvec{E}\times \mvec{B}$ velocity of $\mvec{u}_{E\times B} = -(E_{y0}/B_{z0}) \xhat$.
The 2V Maxwellian distribution drifting with this $\mvec{u}_{E\times B}$ velocity is given by
\begin{align}
f_s(v_x,v_y) =\frac{n_0}{\pi v_{ts}^2} e^{-[(v_x-u_{E\times B})^2 + v_y^2]/v_{ts}^2},
\end{align}
where we have assumed no spatial variation, in analogy with the upstream region of the perpendicular shock studied here.

With no spatial variation, the rate of change of phase-space energy density $w_s(x,v_x,v_y,t) \equiv m_s v^2 f_s(x,v_x,v_y,t)/2$ is given by
\begin{align}
   \frac{\partial w_s(\mvec{x},\mvec{v},t)}{\partial t} = -q_s\frac{v^2}{2}  \left( \mvec{E} +
   \mvec{v} \times \mvec{B} \right)
   \cdot \frac{\partial f_s}{\partial \mvec{v}}.
   \label{eq:dwsexb}
\end{align}
Substituting in the fields and the velocity-space derivatives
\begin{align}
   \frac{\partial f_s}{\partial v_x} = \frac{-2(v_x-u_{E\times B})}{v_{ts}^2} f_s ,
\end{align}
\begin{align}
   \frac{\partial f_s}{\partial v_y}  =\frac{-2 v_y}{v_{ts}^2} f_s,
\end{align}
we obtain the following result
\begin{align}
   \frac{\partial w_s(\mvec{x},\mvec{v},t)}{\partial t} =
   -q_s\frac{v^2}{2}  \left( -v_y E_{y0}
   -\cancel{v_x v_y B_{z0}} + \cancel{v_y v_x B_{z0}} -  v_y u_{E\times B} B_{z0} 
   \right) \frac{2 f_s}{v_{ts}^2},
   \label{eq:dwsexb2}
\end{align}
where we see that the term from $ \left(\mvec{v} \times \mvec{B}\right)_y (\partial f_s/\partial v_y)$ cancels with the contribution from $ \left(\mvec{v} \times \mvec{B}\right)_x (\partial f_s/\partial v_x)$ that is not associated with the $\mvec{E}\times \mvec{B}$ flow. 
Now, if we substitute for the $\mvec{E}\times \mvec{B}$  velocity, where $u_{E\times B} = -(E_{y0}/B_{z0})$, we obtain
\begin{align}
   \frac{\partial w_s(\mvec{x},\mvec{v},t)}{\partial t} =
   -q_s\frac{v^2}{2}  \left( -\cancel{v_y E_{y0}} -  \cancel{v_y \left[-\frac{E_{y0}}{B_{z0}}\right] B_{z0}}
   \right) \frac{2 f_s}{v_{ts}^2}=0,
   \label{eq:dwsexb3}
\end{align}
where the change of phase-space energy density due to the electric field in the first term is canceled by the change of phase-space energy density due to the magnetic field acting on the $\mvec{E}\times \mvec{B}$ flow.

Importantly, \eqr{\ref{eq:dwsexb3}} demonstrates that the instantaneous rate of change of the phase-space energy density, at every point in velocity space, is zero for a Maxwellian plasma simply undergoing uniform $\mvec{E} \times \mvec{B}$ motion. 
Of course, we expect that an $\mvec{E} \times \mvec{B}$ flow produces no net energization.
But, in combination with our intuition that $\mvec{E} \times \mvec{B}$ flows produce no net energization, the result presented here more strongly motivates the form of the field-particle correlation technique presented in \appref{app:v2FPC}, and other transformations employed throughout this study to eliminate the contribution of $\mvec{E} \times \mvec{B}$ flows to individual components of the energization, such as the transformation to the transverse-drift frame in Section~\ref{sec:electrons}.
Using these transformations, we can then gain further insight into the energy exchange in phase space without having to sum over components, as is required in \eqr{\ref{eq:dwsexb3}} to completely cancel the $\mvec{E} \times \mvec{B}$ contribution to the rate of change of the phase-space energy density.


\section{Energy Conversion vs.~Energization}\label{app:energizationVsEnergyConversion}
\begin{figure}
   \begin{center}
      \includegraphics[width=0.59\textwidth]{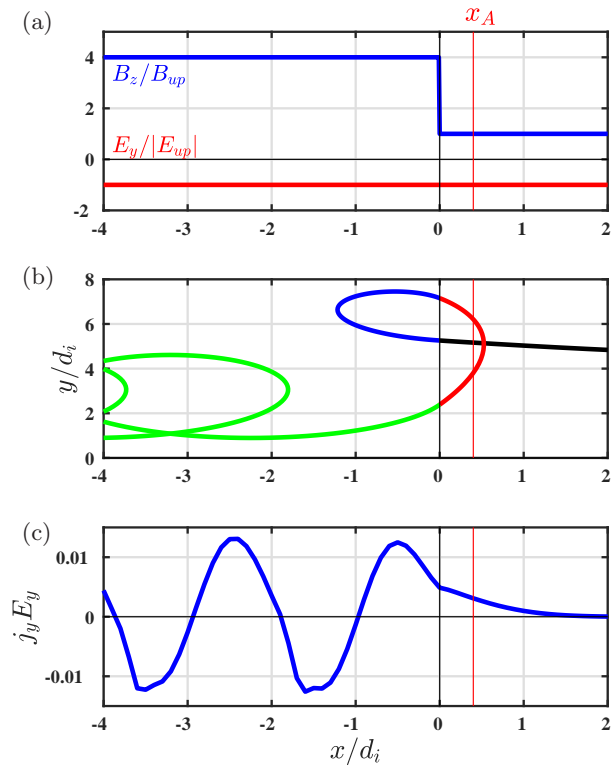}
   \end{center}
\vskip -4.0in
\hspace*{1.1in} (a)
\vskip +1.2in
\hspace*{1.1in} (b)
\vskip +1.2in
\hspace*{1.1in} (c)
\vskip +1.2in
   \caption{(a) Profiles along the shock normal direction of the transverse magnetic field $B_z$ (blue) and the motional electric field $E_y$ (red), (b) trajectory of a reflected ion in the $(x,y)$ plane, and (c) the rate of work done by the electric field on the distribution of particles $j_y E_y$. }
   \label{fig:ion_profile}
\end{figure}

The purpose of this appendix is to clarify terminology on energy conversion within the plasma versus energization of the plasma via the processes present in this collisionless shock.
In the analysis of the ion energization in the \gke~simulation presented in Section~\ref{sec:ions}, the narrow upstream velocity distribution broadens as it passes into the downstream region, as can be seen in Figure~\ref{fig:distThroughShock}(d).
We wish to distinguish between this broadening of the ion distribution, which we identify as energy conversion from bulk kinetic to internal energy, and energization of the ion distribution, i.e., energy transfer from the electromagnetic fields to the ions.

To understand this distinction, we return to the simplified model for the ion dynamics through the perpendicular shock presented in Section~\ref{sec:ions}. 
For the reader's benefit, we re-plot the single-particle motion and fields in the simplified model, along with the net rate of work done on the full ion particle velocity distribution, $j_y E_y$, in Figure~\ref{fig:ion_profile}.
In addition, we have marked with $x_A$ the point that the field-particle correlation was calculated in Section~\ref{sec:ions}.

\begin{figure}
   \begin{center}
      \includegraphics[width=0.47\textwidth]{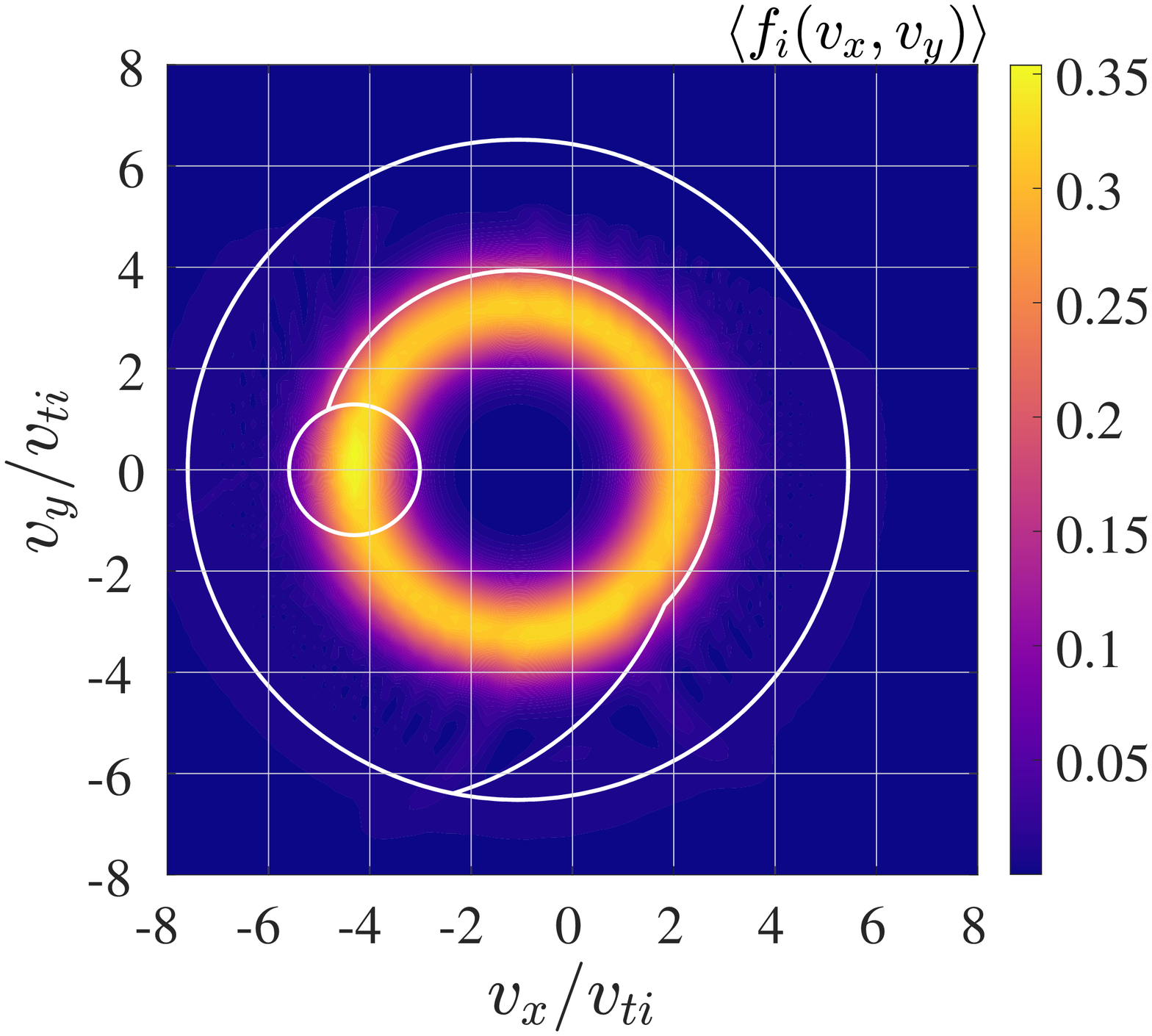}
      \includegraphics[width=0.49\textwidth]{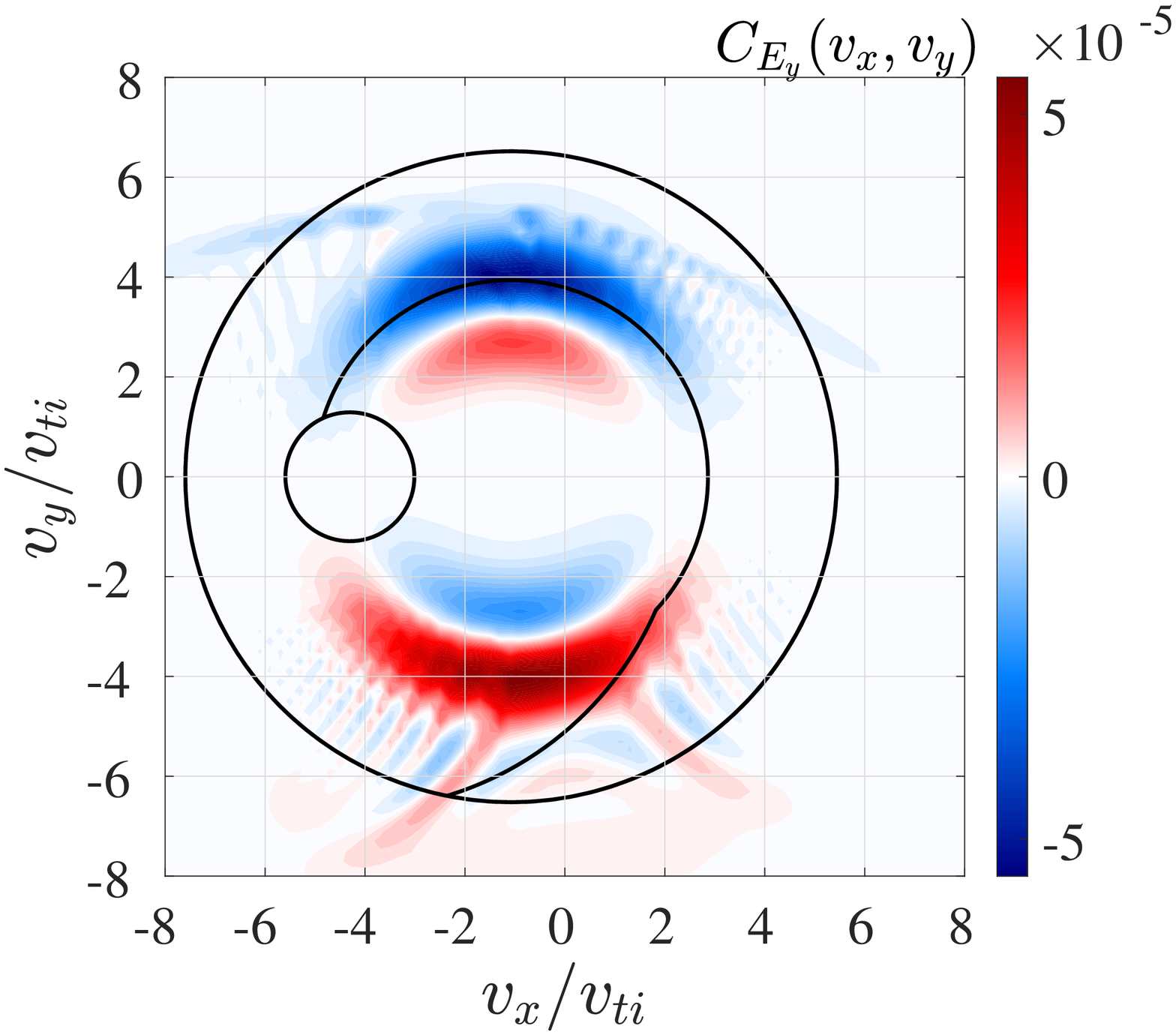}
   \end{center}
       \vskip -2.0in
\hspace*{0.in} (a)\hspace*{2.4in} (b)
\vskip +1.85in
   \caption{Averaging over the downstream region $-4 \le x/d_i \le -2 $ in the idealized shock model yields (a) the distribution function $\langle f_i(v_x,v_y) \rangle$ and (b) the field-particle correlation $C_{E_y}$ of the averaged distribution. The averaged field-particle correlation is approximately symmetric, corresponding to zero net energization, in agreement with a spatial average of $j_y E_y$ in Figure~\ref{fig:ion_profile} (c), which suggests that the ions experience no further energization once they have crossed downstream.}
   \label{fig:ion_average}
\end{figure}

We now plot in Figure~\ref{fig:ion_average} that spatially averaged downstream distribution function and field-particle correlation, averaging over  $-4 \le x/d_i \le -2$.
We observe a broadened ring distribution and an approximately anti-symmetric velocity-space signature, corresponding to zero net energization over this spatial interval.
Importantly, if we transform to the downstream frame of reference and compare the distance to the origin of the upstream distribution (white circle) to the distance to the origin of the ring distribution, we find that both distributions, upstream and spatially-averaged downstream, are roughly equidistant to the origin.
In other words, the energy of these two distributions is roughly equivalent.
In the upstream, the energy is dominantly bulk kinetic, while in the downstream the energy is mostly nonthermal, an increase in the effective perpendicular temperature of the distribution.
But, this energy conversion is conservative, not changing the total net energy of the ions. The only process in the simplified model which changes the total microscopic kinetic energy of the ions is the energization via shock-drift acceleration.

It is worth expanding upon this subtlety of energy conversion versus energization by considering a more generic idealized model in which we vary the amplitude of the magnetic field increase at the magnetic discontinuity. 
In general, as an ion $\mvec{E} \times \mvec{B}$ drifts through a magnetic discontinuity, the perpendicular velocity in the local bulk-flow frame of reference increases at the expense of the diminished bulk-flow $\mvec{E} \times \mvec{B}$ velocity.
The downstream perpendicular velocity relative to the upstream bulk-flow velocity $v_{\perp d}/U_u$ is determined by three dimensionless parameters for this idealized problem: (i) the ratio of the downstream to the upstream magnetic field magnitude
$B_d/B_u$; (ii) the ratio of the upstream perpendicular velocity to the upstream bulk-flow velocity $v_{\perp u}/U_u$; and (iii) the gyrophase $\theta$ of the ion\footnote{Note that, since the inflow velocity is in the $-x$ direction for the model configurations considered here, we define gyrophase $\theta$ as the angle measured \emph{clockwise} from the $-x$ direction. 
Therefore, $\theta=0$ corresponds to a perpendicular velocity that increases the magnitude of the inflow velocity, and $\theta=180^\circ$ decreases the magnitude of the inflow velocity.}
when it first reaches the magnetic discontinuity.

In Figure~\ref{fig:ion_profile}(b), the specific ion trajectory plotted returns upstream (red segment) due to the increased magnetic field downstream of the discontinuity. 
If the ion does \emph{not} return upstream, then one can compute the downstream perpendicular velocity $v_{\perp d}/U_u$ as the difference between the velocity upon crossing the discontinuity and the downstream $\mvec{E} \times \mvec{B}$ velocity, yielding
\begin{align}
    \frac{v_{\perp d,th}}{U_u} = \left\{ \left[\frac{v_{\perp u}}{U_u}\cos \theta + \left(1- \frac{B_u}{B_d} \right) \right]^2 
    + \left[\frac{v_{\perp u}}{U_u}\sin \theta  \right]^2  \right\}^{1/2}.
    \label{eq:theory}
\end{align}
Note that, although the perpendicular energy relative to the local (upstream or downstream) bulk-flow velocity generally increases, this increase comes at the expense of the kinetic energy of the incoming bulk flow, and the total kinetic energy of each ion is conserved in this process. 
This statement can be proven for a ring of ions with perpendicular  velocity $v_{\perp u}$ about upstream velocity $U_u$ by squaring \eqr{\ref{eq:theory}}, substituting $B_d/B_u=U_u/U_d$, integrating the gyrophase $\theta$ over $2\pi$, and multiplying by $m_i/2$, to obtain the expression
\begin{align}
  \frac{1}{2} m_iv^2_{\perp d,th}= \frac{1}{2} m_iv^2_{\perp u} +  \frac{1}{2} m_i
  (U_u-U_d)^2.
    \label{eq:energy_cons}
\end{align}
The conservation of energy is obvious when evaluated in the downstream frame ($U_d=0$), where  \eqr{\ref{eq:energy_cons}} proves that the downstream perpendicular energy of the ring of ions is simply the sum of the upstream perpendicular energy plus the ``bulk" kinetic energy of the ring distribution moving at $U_u$.

\begin{figure}
\begin{center}
    \includegraphics[width=0.5\textwidth]{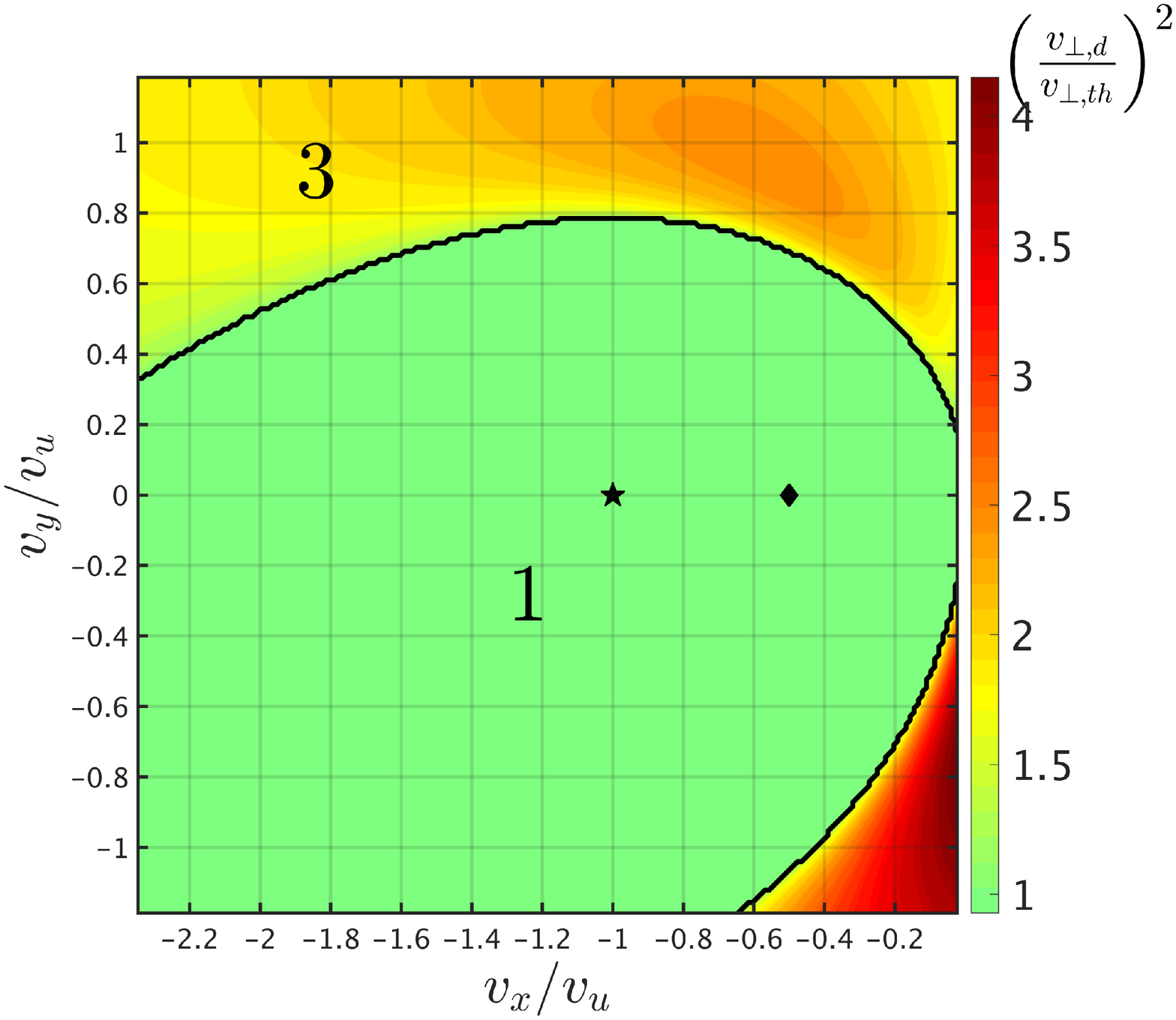}    
\end{center}
\vskip -2.0in
\hspace*{1.35in} (a)
\vskip +1.85in
\begin{center}
    \includegraphics[width=0.495\textwidth]{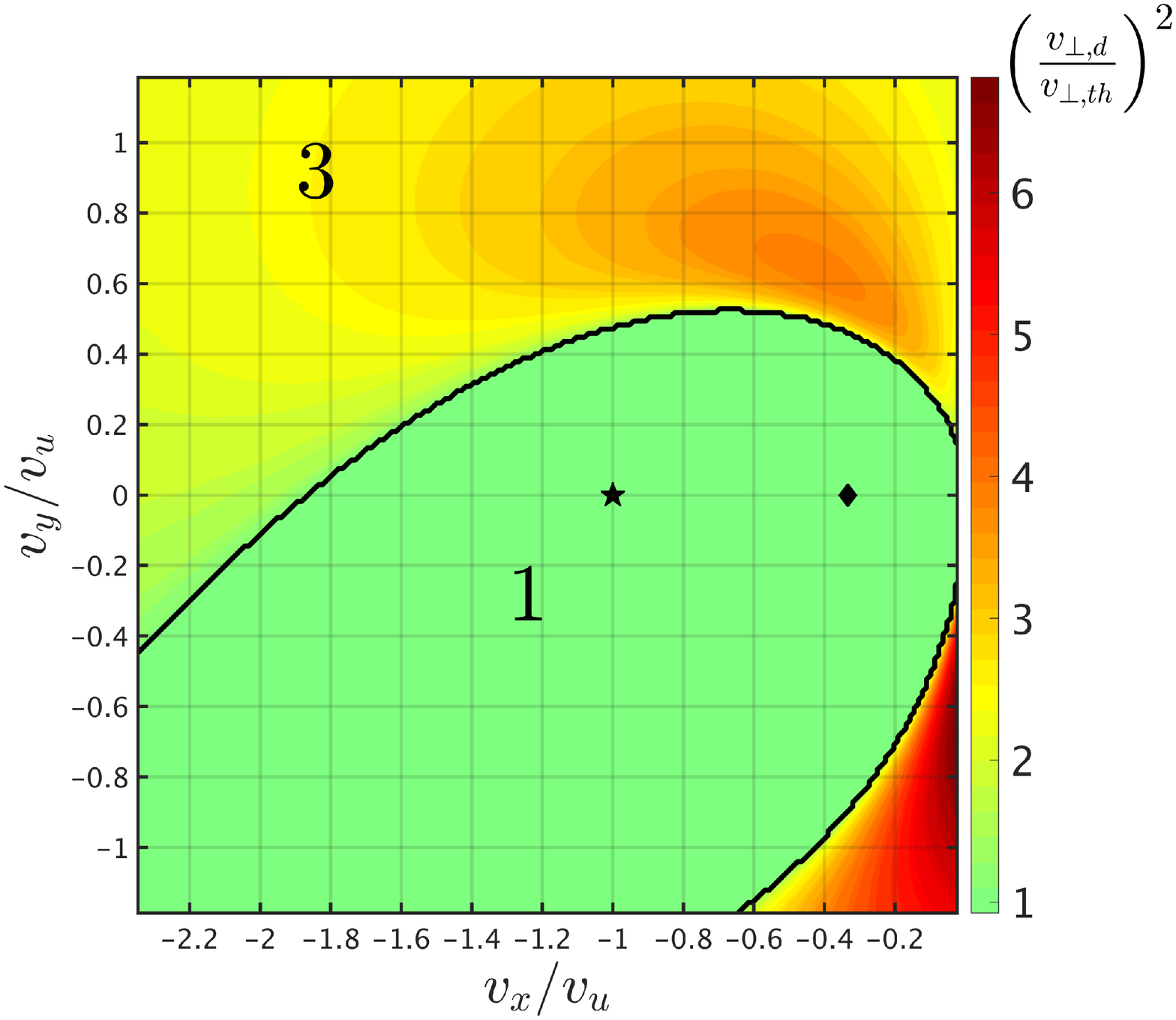}
    \includegraphics[width=0.495\textwidth]{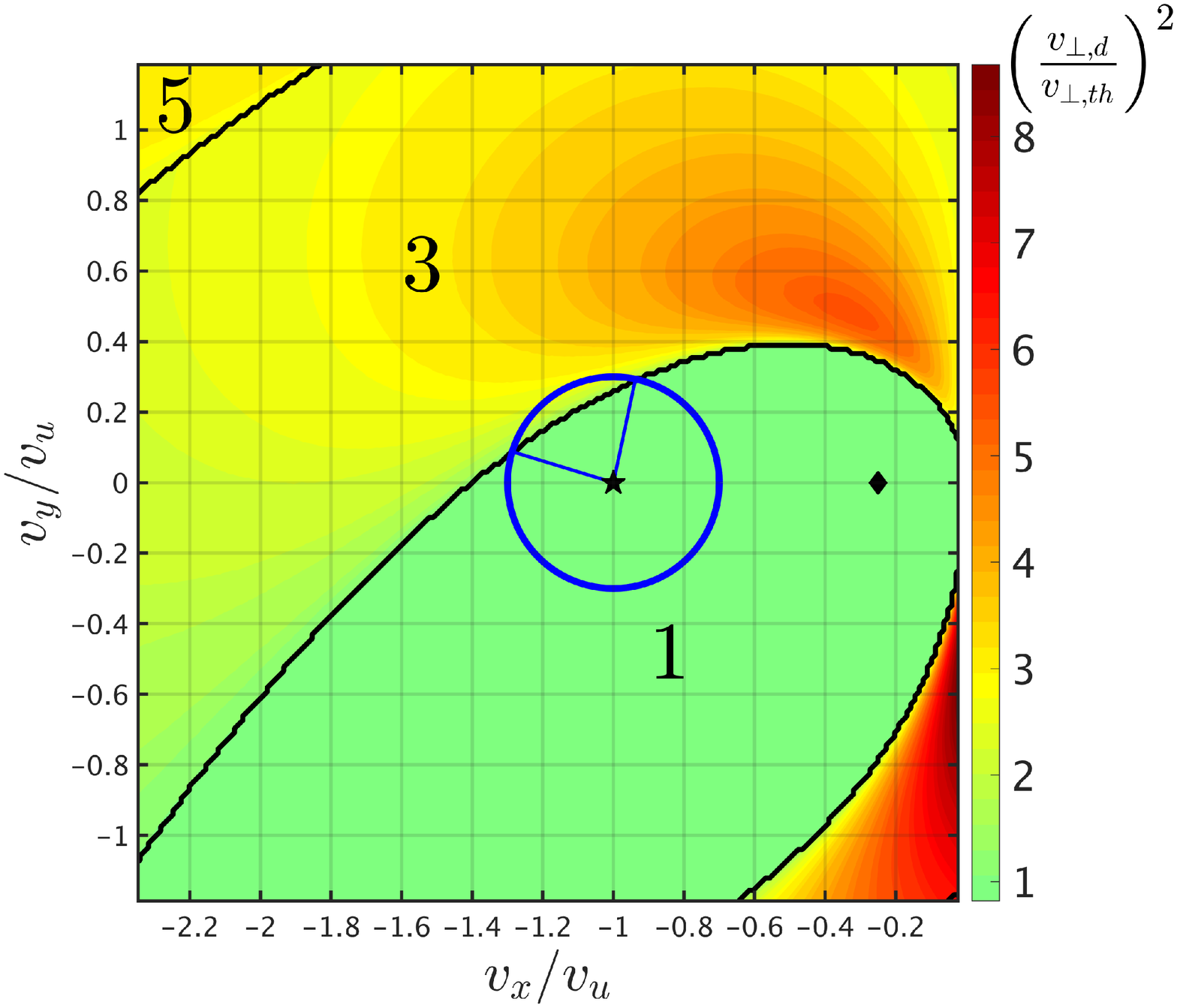}    
\end{center}
\vskip -2.0in
\hspace*{0.05in} (b)\hspace*{2.45in} (c)
\vskip +1.85in
    \caption{Ion energization as a function of  $v_{\perp u}/U_u$
and $\theta$ on the $(v_x,v_y)$ plane for $B_d/B_u= $ (a) 2, (b) 3, and (c) 4.  
The upstream bulk velocity is given by the star, the downstream bulk velocity is given by the diamond.  The blue circle in (c) represents particles with a specific upstream perpendicular velocity $v_\perp$, where only particles with gyrophases $\theta$ within the indicated range undergo reflection.}
    \label{fig:ion-theory}
\end{figure}

Of course, if an ion does return upstream, it can gain energy by the process of shock-drift acceleration via the alignment of transverse (to the shock normal) component of the Larmor velocity and motional electric field that supports the inflow at the $\mvec{E} \times \mvec{B}$ velocity.
Only if an ion returns upstream is there any net energy exchange between the electromagnetic fields and the particles.
We demonstrate this energy gain for the idealized model in Figure~\ref{fig:ion-theory}, which plots the gain of perpendicular energy $\left(v_{\perp d}/v_{\perp d, th}\right)^2$ as a function of $(v_x,v_y)$ for $B_d/B_u=2,3,4$ in panels (a,b,c). 

In this figure, the black contours separate regions with different numbers of crossings of the magnetic discontinuity ($x=0$ in Figure~\ref{fig:ion_profile}), where every ion must cross the magnetic discontinuity an odd number of times to eventually cross downstream, given by the large numbers on the plot. 
The increase of the perpendicular energy due to shock drift acceleration is given by the colorbar.
In the self-consistent simulation, this energy gain comes at the expense of the field energy, while for the idealized model, this energy gain via shock-drift acceleration is not conservative. 
Nevertheless, the idealized model helpfully illustrates where in phase space we observe merely energy conversion versus where we expect to see actual energization due to the electromagnetic fields.
All ions that cross the magnetic discontinuity only once conserve their energy, leading to the increase of the perpendicular energy (relative to the downstream frame) predicted by \eqr{\ref{eq:theory}} (green color).

We conclude this appendix by making a few final notes on this distinction between energy conversion and energization.
The movement of particles from one position in velocity space to another requires acceleration, so this energy conversion is still mediated by forces in the plasma: the $\mvec{v} \times \mvec{B}$ force in the single-particle picture, which combines with the $\mvec{v} \cdot \nabla_{\mvec{x}}$ streaming term in the distribution function picture.
The latter term corresponds to the traditional picture of pressure work, as once we have a distribution of particles, the pressure can participate in this energy conversion.

It is important though to distinguish pressure work, which simply converts one form of energy into another, and a pressure supported electric field, which requires gradients in the pressure and can energize the plasma.
For example, the cross-shock electric field which arises in the self-consistent simulation is a result of the electron pressure gradient.
This pressure-supported cross-shock electric field both increases the reflection of ions---see \appref{app:crossShockIon} for further details---and is a critical component to the increase in $T_\perp$ of the electrons via adiabatic heating, where the pressure gradient provides the relevant drifts for the electron distribution's adiabatic invariant to be conserved through the magnetic field gradient.

Fundamentally, we seek to be as precise as possible in what we are diagnosing with the field-particle correlation technique by focusing exclusively on the electric field component of the evolution of the phase-space energy density.
While the energy conversion that occurs in collisionless shocks is a component of the overall increase in the temperature of, e.g., the ions, this process of energy conversion is distinct from the energization processes that occur, such as shock-drift acceleration.
And it is these energization processes that we seek the velocity-space signatures of, as we may then be able to leverage this same toolkit for understanding the processes present in spacecraft observations of collisionless shocks.


\section{Vlasov Mapping Technique to Determine Full Particle Velocity Distributions}
\label{app:vmap}

We can explore the evolution of the particle velocity distributions in our idealized perpendicular shock models by a technique that we denote \emph{Vlasov mapping} \citep{Scudder:1986,Kletzing:1994,Hull:1998,Hull:2000,Hull:2001,Mitchell:2013,Mitchell:2014}.
At the physical point $\mvec{x}_{\mbox{obs}}$ at which we want to ``observe'' the velocity distribution, we repeat the single-particle-motion analysis for every point $(v_{x,\mbox{init}},v_{y,\mbox{init}})$ in the velocity space, integrating backwards in time until we reach a point $\mvec{x}_{\mbox{up}}$ upstream in the unperturbed, in-flowing plasma.
This backwards integration yields a final position in velocity space $(v_{x,\mbox{fin}},v_{y,\mbox{fin}})$ by following along the ion trajectory through phase space. 
Since the velocity distribution upstream is known, we know the phase-space density at this final point in velocity space $(v_{x,\mbox{fin}},v_{y,\mbox{fin}})$. 
For a collisionless plasma, Liouville's theorem dictates that the phase-space density is invariant along the particle trajectories through 3D-3V phase space, so we may set the phase-space energy density at $(v_{x,\mbox{init}},v_{y,\mbox{init}})$ at the point of observation equal to the phase-space density upstream at $(v_{x,\mbox{fin}},v_{y,\mbox{fin}})$, giving
\begin{align}
f_s(\mvec{x}_{\mbox{obs}},v_{x,\mbox{init}},v_{y,\mbox{init}})
= f_s(\mvec{x}_{\mbox{up}},v_{x,\mbox{fin}},v_{y,\mbox{fin}}).
\end{align}

This Vlasov mapping technique is, of course, not self-consistent with respect to how the evolving particle velocity distributions may become unstable and generate electromagnetic field fluctuations through kinetic instabilities. 
It is essentially an extension of the single-particle-motion analysis, computing the evolution of the full velocity distribution due to known electromagnetic fields. 
Furthermore, it is  possible in general that regions of phase space downstream do not connect to any  position upstream, leading to voids in the downstream phase space, but for the  perpendicular collisionless shock evaluated here, this potential difficulty is not encountered.


\section{Effect of Finite Ramp Width and Cross-Shock Electric Field on Ion Energization}\label{app:crossShockIon}
\begin{figure}
\begin{center}
      \includegraphics[width=0.8\textwidth]{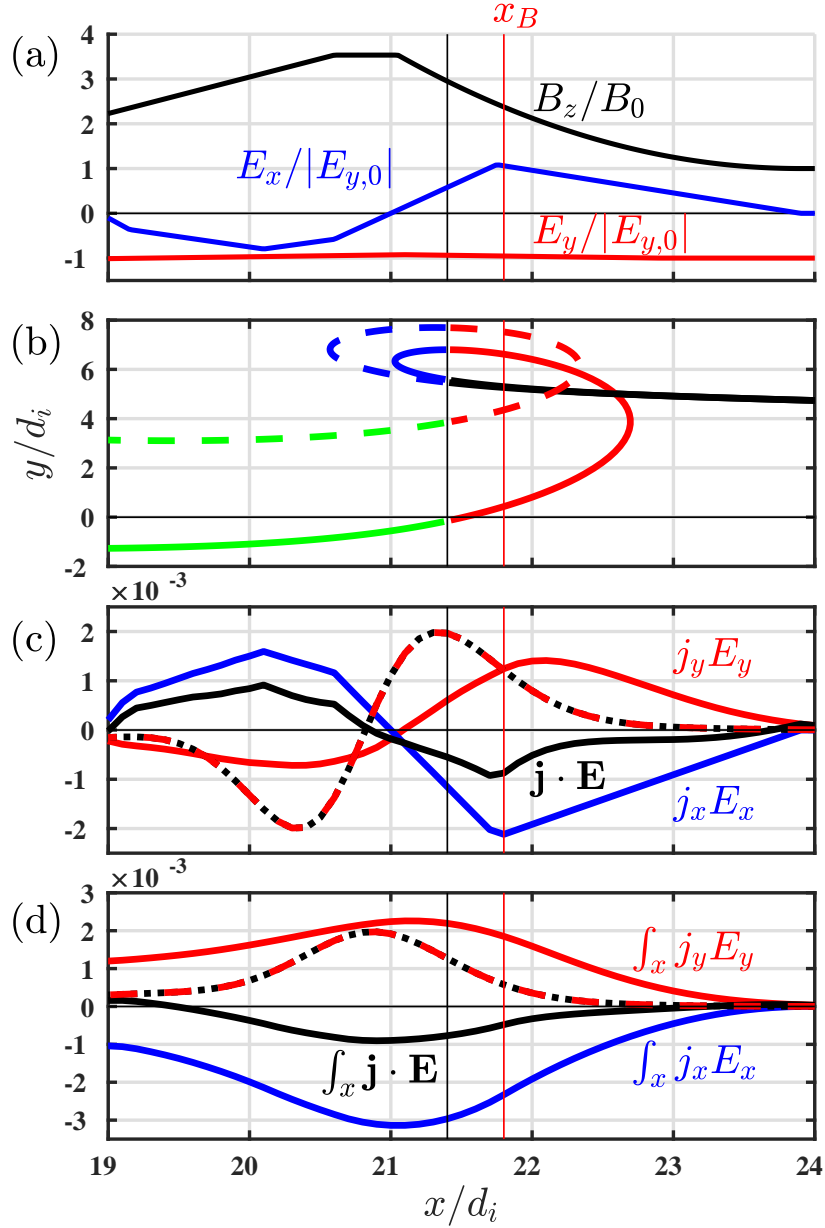}
      \end{center}
    \caption{(a) Electromagnetic fields approximated from  the self-consistent \gke~simulation. (b) Example ion trajectories for the full model (solid) and zero $E_x$ model (dashed). (c) Rate of work done by the components of the electric field, $j_yE_y$ (red) and $j_xE_x$ (blue) for the full model (solid) and zero $E_x$ model (dashed), along with total 
    $\mvec{j} \cdot \mvec{E}$ (black). (d) Cumulative work done integrated from the upstream $\int_x \mvec{j} \cdot \mvec{E}$. Inclusion of the cross-shock electric field enhances ion reflection, thereby achieving a larger energy gain due to the motional electric field $E_y$ through shock-drift acceleration.}
    \label{fig:no-cross-shock}
\end{figure}

To understand the effects of the finite shock width and the cross-shock electric field on the ion energization, we first use the Vlasov mapping model to separate out the effects of different components of the electric field on the ion trajectories, similar to the analysis of the electrons presented in Section~\ref{sec:elcFPC}.
In Figure~\ref{fig:no-cross-shock}(b), we compare the ion trajectories between two Vlasov-mapping models: (i) the ``full model'' (solid) which computes the ion trajectories and evolution of the ion velocity distribution using the full electromagnetic fields from the \gke~simulation; and (ii) the ``zero $E_x$ model'' (dashed), in which we artificially set the cross-shock electric field to zero. 
In Figure~\ref{fig:no-cross-shock}(c), we plot the rates of energization of the ion distribution by the electric field, $j_xE_x$ (blue) and $j_yE_y$ (red), along with the total energization, $\mvec{j} \cdot \mvec{E}$ (black), for the two models. 
We also show the cumulative total energization of the ion distribution by integrating from upstream $\int_{x_{up}}^x \mvec{j} \cdot \mvec{E} dx$, along with the separate contributions from each component of the electric field for both models in  Figure~\ref{fig:no-cross-shock}(d).

First, we adopt a Lagrangian approach to examine the ion energization along its trajectory.
The comparison of the example reflected ion trajectories in Figure~\ref{fig:no-cross-shock}(b) illustrates how the cross-shock electric field alters the ion trajectory.  
Because $E_x$ opposes the flow of ions into the shock, the ion penetrates less deeply into the shock before turning back upstream (solid blue segment of trajectory) than for the zero $E_x$ model (blue dashed).  
For the full model including $E_x$, the ion returns further upstream (red solid), where the lower magnitude of the magnetic field leads to a larger Larmor radius of its orbit. 
This return further upstream in the full model is particularly important when the shock ramp has a finite width.
The enhancement of the ion reflection by $E_x$ significantly affects energization of these reflected ions by $E_y$ through the shock drift acceleration mechanism, where Figure~\ref{fig:no-cross-shock}(c) shows that the rate of ion energization $j_yE_y$ (red) in the foot and ramp region, $22 \le x/d_i \le 24$, is much larger for the full model (solid) than for the zero $E_x$ model (dashed).
This increased energization is a direct result of the larger distance the full model ion traverses in $y$ upstream as its gyroradius is increased by the combination of acceleration by $E_x$ and the decreased magnetic field amplitude upstream.

\begin{figure}
 \begin{center}
      \includegraphics[width=0.49\textwidth]{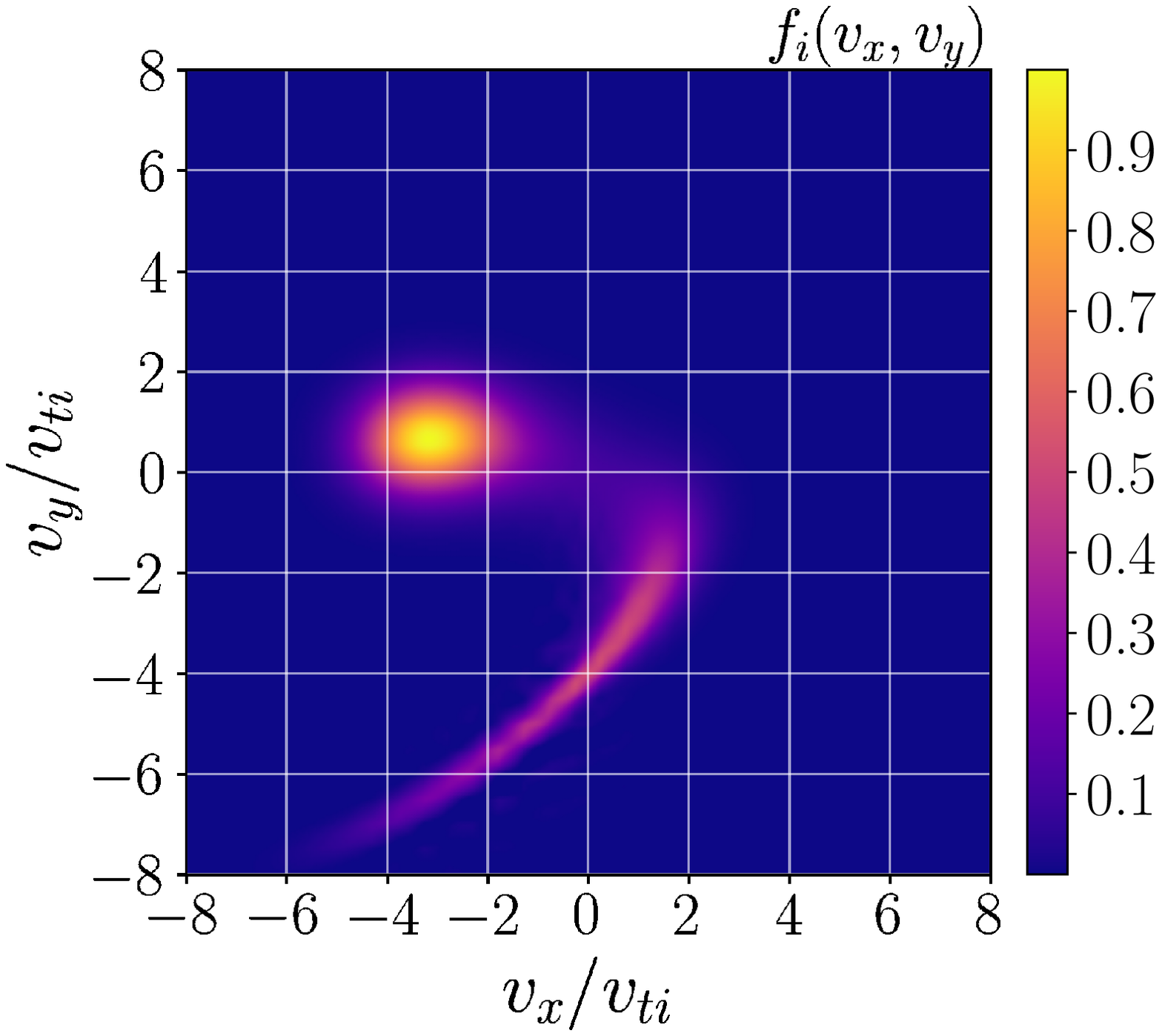}
      \includegraphics[width=0.49\textwidth]{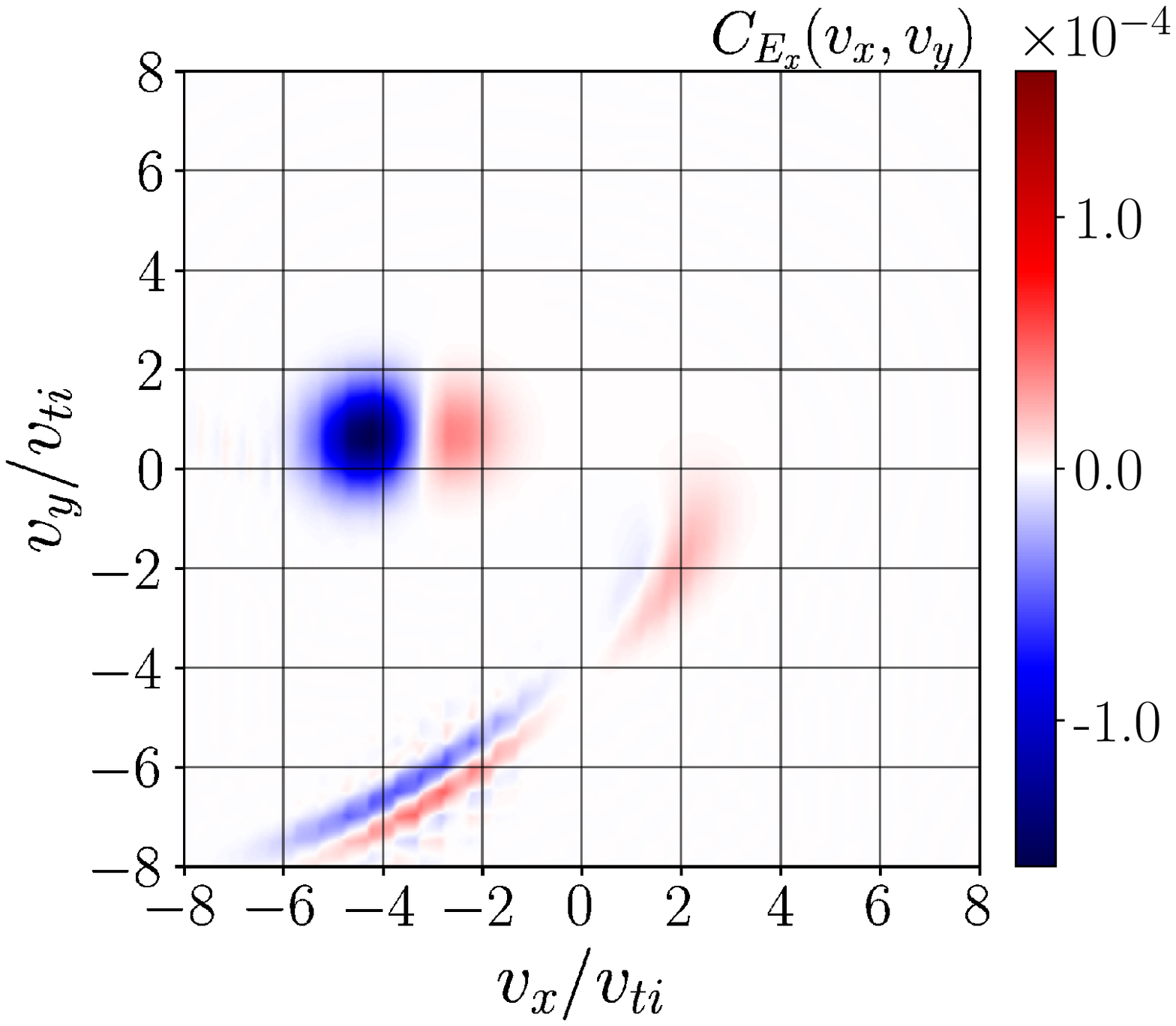}
   \end{center}
       \vskip -2.0in
\hspace*{0.05in} (a)\hspace*{2.45in} (b)
\vskip +1.85in
      \caption{The ion distribution function from the \gke~simulation (left), and $C_{E_x}$ computed from the \gke~simulation (right) plotted at $x_B = 21.8 d_i$, near the peak of the cross-shock electric field. We note two features in the velocity-space signature found from computing $C_{E_x}$: the strong negative correlation coincident with the incoming beam, denoting the deceleration of the incoming flow and transfer of energy from the bulk upstream kinetic energy to electromagnetic energy, and the modest positive correlation at $v_y < 0, v_x > 0$ where particles can now be accelerated by the cross-shock electric field and pushed back upstream. This acceleration of ions of particular velocities is the principal reason for the increased efficiency of shock-drift acceleration despite the finite shock width, as the cross-shock electric field assists in increasing the phase-space density of reflected ions that can gain energy along the motional electric field upstream.}
    \label{fig:ion-Ex-correlation}
\end{figure}

Another way to view the effect of the cross-shock electric field in increasing the efficiency of shock drift acceleration is to employ a complementary Eulerian point of view to examine the energization as a function of velocity space $(v_x,v_y)$ at a single point in configuration space.
Following this approach, we explore the enhanced reflection due to the cross-shock electric field by examining $C_{E_x}$ to understand how $E_x$ accelerates or decelerates ions in different regions of velocity space. 
In Figure~\ref{fig:ion-Ex-correlation}, we plot (a) the ion distribution function $f_i(v_x,v_y)$ and (b) the correlation with the cross-shock electric field $C_{E_x}(v_x,v_y)$ from the simulation at the position $x_B=21.8 d_i$ (vertical red line in Figure~\ref{fig:no-cross-shock} and the same point where the electron analysis in Section~\ref{sec:electrons} was performed) where the cross-shock electric field peaks.

The ion distribution at this position is dominated by the incoming beam, with a small fraction of reflected ions forming a ``boomerang'' shaped distribution. 
The dominant effect is that $E_x$ decelerates the incoming ion beam.  
But, the population of reflected ions with $v_x>0$ at $x_B$---which corresponds to the upper crossing of the red segment of the trajectory with the vertical line at $x_B$ in Figure~\ref{fig:no-cross-shock}(b)---is being accelerated by $E_x$, causing these ions both to return further upstream and to increase their perpendicular velocity, thereby leading to a larger Larmor radius. 
These two effects $E_x$ has on the reflected ions with $v_x>0$ reinforce the enhanced reflection and increased energization of these ions by the shock drift acceleration mechanism. 

In this regard, we reiterate a powerful feature of the FPC: the velocity-space signatures produced by the FPC reveal how electric fields energize different components of the ion distribution in qualitatively different ways.
The cross-shock electric field decelerates the incoming beam while accelerating the reflected population with $v_x>0$, as shown by the blue and red signatures respectively in these regions of phase space. 
The separation of the energization of different populations of the ion velocity distribution from an Eulerian perspective, provided by the FPC method, enables a deeper understanding of the underlying mechanisms of ion acceleration at the shock.
We emphasize that by looking only at the velocity-integrated rate of energization by $E_x$---given by $j_x E_x$ in Figure~\ref{fig:no-cross-shock}(c) at $x_B$---one sees just the net loss of ion energy due to $E_x$, masking the important effect that the cross-shock electric field plays to enhance the ion reflection. 
While the role of the cross-shock electric field in enhancing the reflection of the ions has been previously theorized to be an important component of energizing the reflected ion population \citep{Cohen:2019}, the Eulerian perspective provided by the FPC makes the physics of the cross-shock electric field especially clear by illustrating where the ions are gaining and losing energy in phase space.

\section{Bulk Guiding Center Drifts}\label{app:GCBulkDrifts}

To derive the bulk drifts, we begin with the momentum equation, obtained via taking the first velocity moment of the Vlasov equation of plasma species $s$,
\begin{align}
    \frac{d\mvec{u}_s}{dt} + \frac{1}{m_s n_s}\nabla\cdot\mvec{P}_s = \frac{q_s}{m_s}(\mvec{E} + \mvec{u}_s\times\mvec{B}), \label{eq:momentumEq}
\end{align}
where
\begin{align}
    \frac{d}{dt} \equiv \frac{\partial}{\partial t} + \mvec{u}_s\cdot\nabla,
\end{align}
and $\mvec{P}_s$ is the pressure tensor,
\begin{align}
    \mvec{P}_s  = m_s \int (\mvec{v} - \mvec{u}_s) (\mvec{v} - \mvec{u}_s) f_s d\mvec{v}.
\end{align}
Note that in the derivation of \eqr{\ref{eq:momentumEq}} from the first velocity moment of the Vlasov equation, we have utilized the zeroth moment of the Vlasov equation to eliminate the terms which involve the time evolution of the density.
We seek the drifts perpendicular to the magnetic field, so we take the cross product \eqr{\ref{eq:momentumEq}} with the magnetic field, $\mvec{B}$,
\begin{align}
    \frac{d\mvec{u}_s}{dt}\times\mvec{B} + \frac{1}{m_s n_s}\nabla\cdot\mvec{P}_s \times\mvec{B}
    =
    \frac{q_s}{m_s}\mvec{E}\times\mvec{B} 
    + \frac{q_s}{m_s} \underbrace{(\mvec{u}_s\times\mvec{B})\times\mvec{B}}_{(\mvec{B}\cdot \mvec{u}_s)\mvec{B} - |\mvec{B}|^2\mvec{u}_s}.
\end{align}
Rearranging, the perpendicular component of the bulk velocity is
\begin{align}
    \mvec{u}_{s\perp} =
    \frac{\mvec{E}\times\mvec{B}}{|\mvec{B}|^2}
    -\frac{\nabla\cdot\mvec{P}_s \times\mvec{B}}{q_s n_s |\mvec{B}|^2}
    -\frac{m_s}{q_s |\mvec{B}|^2} \frac{d\mvec{u}_s}{dt}\times\mvec{B}.
\end{align}

If the plasma is magnetized, or at least the electrons are as in Section~\ref{sec:electrons}, it is natural to split the pressure tensor as 
\begin{align}
    \mvec{P}_s = \mvec{P}^C_s + \mvec{\Pi}_s,
\end{align}
where
\begin{align}
    \mvec{P}^C_s = (\mvec{I}-\mvec{b}\mvec{b})p_{s,\perp} + \mvec{b}\mvec{b}p_{s,\parallel}
    = \mvec{I} p_{s,\perp} + \mvec{b}\mvec{b} (p_{s,\parallel}-p_{s,\perp}), 
\end{align}
is the Chew-Goldberger-Low (CGL) pressure tensor \citep{CGL:1956}, $\mvec{b} = \mvec{B}/|\mvec{B}|$ is the direction of the magnetic field, and $\mvec{\Pi}_s$ is the agyrotropic part of the pressure tensor. 
Note that $\mathrm{Tr}\thinspace (\mvec{P}_s^C) = 2 p_{s,\perp} + p_{s,\parallel} = 3 p_s$, where $p_s$ is the scalar pressure.
From this definition, we can also see that $\mathrm{Tr}\thinspace (\mvec{\Pi}_s) = 0$.
In the 1D-2V simulation of interest in this study, where $\mvec{B} = B_z(x) \hat{\mvec{z}}$, the trace of the pressure tensor is instead  $\mathrm{Tr}\thinspace (\mvec{P}_s^C) = 2 p_{s,\perp} = 2 p_s$ because we are not evolving the degree of freedom parallel to the magnetic field.
Thus, $p_{s,\perp} = p_s$ in this geometry, but for generality we will retain the subscript $\perp$ for the remainder of the derivation.

The divergence of the CGL pressure tensor is
\begin{align}
    \nabla\cdot \mvec{P}^C_s = \nabla p_{s,\perp}  + (p_{s,\parallel}-p_{s,\perp})
    \underbrace{\nabla\cdot(\mvec{b}\mvec{b})}_{ (\nabla\cdot \mvec{b}) \mvec{b} + \nabla_\parallel \mvec{b} }
    + \mvec{b}\nabla_\parallel (p_{s,\parallel}-p_{s,\perp}), \label{eq:divCGLPressure}
\end{align}
where $\nabla_\parallel \equiv \mvec{b}\cdot\nabla$. 
Hence, we can calculate the contribution of the CGL pressure tensor to the bulk drift
\begin{align}
    -\frac{\nabla\cdot\mvec{P}_s^C \times\mvec{B}}{q_s n_s |\mvec{B}|^2}
    =
    -\frac{\nabla p_{s,\perp} \times\mvec{B}}{q_s n_s |\mvec{B}|^2}  
    + (p_{s,\perp}-p_{s,\parallel}) \frac{\nabla_\parallel \mvec{b}\times\mvec{B}}{q_s n_s |\mvec{B}|^2},
\end{align}
where the terms in the direction of the magnetic field in \eqr{\ref{eq:divCGLPressure}} are eliminated upon taking the cross product with $\mvec{B}$.
Putting everything together, we obtain
\begin{align}
    \mvec{u}_{s\perp} =
    \frac{\mvec{E}\times\mvec{B}}{|\mvec{B}|^2}
    - \frac{\nabla p_{s,\perp} \times\mvec{B}}{q_s n_s |\mvec{B}|^2}  
    + (p_{s,\perp}-p_{s,\parallel}) \frac{\nabla_\parallel \mvec{b}\times\mvec{B}}{q_s n_s |\mvec{B}|^2}  
    - \frac{\nabla\cdot\mvec{\Pi}_s \times\mvec{B}}{q_s n_s |\mvec{B}|^2}
    - \frac{m_s}{q_s |\mvec{B}|^2} \frac{d\mvec{u}_s}{dt}\times\mvec{B}.
  \label{eq:uperp}
\end{align}

We now define the magnetization vector \citep{Hazeltine:1998},
\begin{align}
    \mvec{M}_s = - p_{s,\perp} \frac{\mvec{B}}{|\mvec{B}|^2},
\end{align}
which is a generalization of the definition in \eqr{\ref{eq:magnetization1x2v}}.
We note that 
\begin{align}
    \nabla\times\mvec{M}_s
    =
    \nabla\times\left( -p_{s,\perp} \frac{\mvec{B}}{|\mvec{B}|^2} \right)  
    =
    -\frac{\nabla p_{s,\perp}\times\mvec{B}}{|\mvec{B}|^2}
    -p_{s,\perp}\nabla\times\left( \frac{\mvec{B}}{|\mvec{B}|^2} \right),
\end{align}
so that we can rearrange \eqr{\ref{eq:uperp}}
\begin{align}
  \mvec{u}_{s\perp} =
  \frac{\mvec{E}\times\mvec{B}}{|\mvec{B}|^2}
  & +
  \frac{\nabla\times \mvec{M}_s} {q_s n_s}
  +
  \frac{p_{s,\perp}\nabla\times\left( \mvec{B}/|\mvec{B}|^2 \right)_\perp} {q_s n_s} \notag \\
  & + (p_{s,\perp}-p_{s,\parallel}) \frac{\nabla_\parallel \mvec{b}\times\mvec{B}}{q_s n_s |\mvec{B}|^2}  
  -\frac{\nabla\cdot\mvec{\Pi}_s \times\mvec{B}}{q_s n_s |\mvec{B}|^2}
  -\frac{m_s}{q_s |\mvec{B}|^2} \frac{d\mvec{u}_s}{dt}\times\mvec{B}.  
\end{align}
The first three terms, the $\mvec{E} \times \mvec{B}$ drift, the magnetization drift, and the $\nabla B$ drift\footnote{We can see this is identical to the $\nabla B$ drift definition employed in \eqr{\ref{eq:gradB1x2v}} for a magnetic field only in the $z$ direction with a bit of vector calculus, $\nabla \times (\hat{\mvec{z}}/B_z) = -\nabla B_z \times \hat{\mvec{z}}/B_z^2$.
} 
are the dominant three drifts in the 1D-2V perpendicular shock of interest in this study.
The other terms: the curvature drift, agyrotropy drift, and polarization drift are all either identically zero in this geometry or small.
For example, we demonstrated in Section~\ref{sec:electrons} that the polarization drift is small through the shock, and because the electron's adiabatic invariant is well conserved the agyrotropic component of the drift must be small. 

We note again that the combination of the bulk $\nabla B$ drift and the magnetization drift produce the familiar diamagnetic drift,
\begin{align}
    \mvec{u}_{\textrm{diamagnetic}} = - \frac{\nabla p_{s,\perp} \times\mvec{B}}{q_s n_s |\mvec{B}|^2},
\end{align}
and that an alternative interpretation of the results of Section~\ref{sec:electrons} is that the electron distribution's adiabatic invariant is conserved via the alignment of the diamagnetic drift with the motional electric field.
In other words, whereas for a single particle only the $\nabla B$ drift was important for that \emph{single particle} to heat adiabatically, the generalization to a distribution of particles leads to a bulk drift, the diamagnetic drift, which is a combination of the $\nabla B$ drift and the magnetization drift, being the principally important drift for the \emph{distribution} of particles to heat adiabatically. 


\section{Checking $\mu_e$ Conservation with a $m_i/m_e = 400$ Simulation}\label{app:massRatio400}

Here, we repeat some of the analysis of Section~\ref{sec:electrons} for a more realistic mass ratio simulation, $m_i/m_e = 400$, to determine a possible source for the slight disagreement between the energization due to the $\nabla B$ and the magnetization drifts and the energization, $\mvec{j}_e \cdot \mvec{E}$ computed from moments of the electron distribution function.
All other parameters are the same, e.g., box size, $L_x = 25 d_i$, plasma betas, $\beta_i = 1.3, \beta_e = 0.7$, and electron-electron collisionality, $\nu_{ee} = 0.01 \Omega_{ci}$.
Note that with the increased mass ratio, the ion-ion collisionality is commensurately reduced. 
In addition, because the ions are more massive and there is more scale separation between the ions and electrons, we have doubled the configuration space resolution to $N_x = 3072$ to keep $\Delta x \sim d_e/6$.
For computational convenience, we perform our analysis just after the shock is formed, $t = 4.3 \Omega_{ci}^{-1}$.

\begin{figure}
 \begin{center}
      \includegraphics[width=0.49\textwidth]{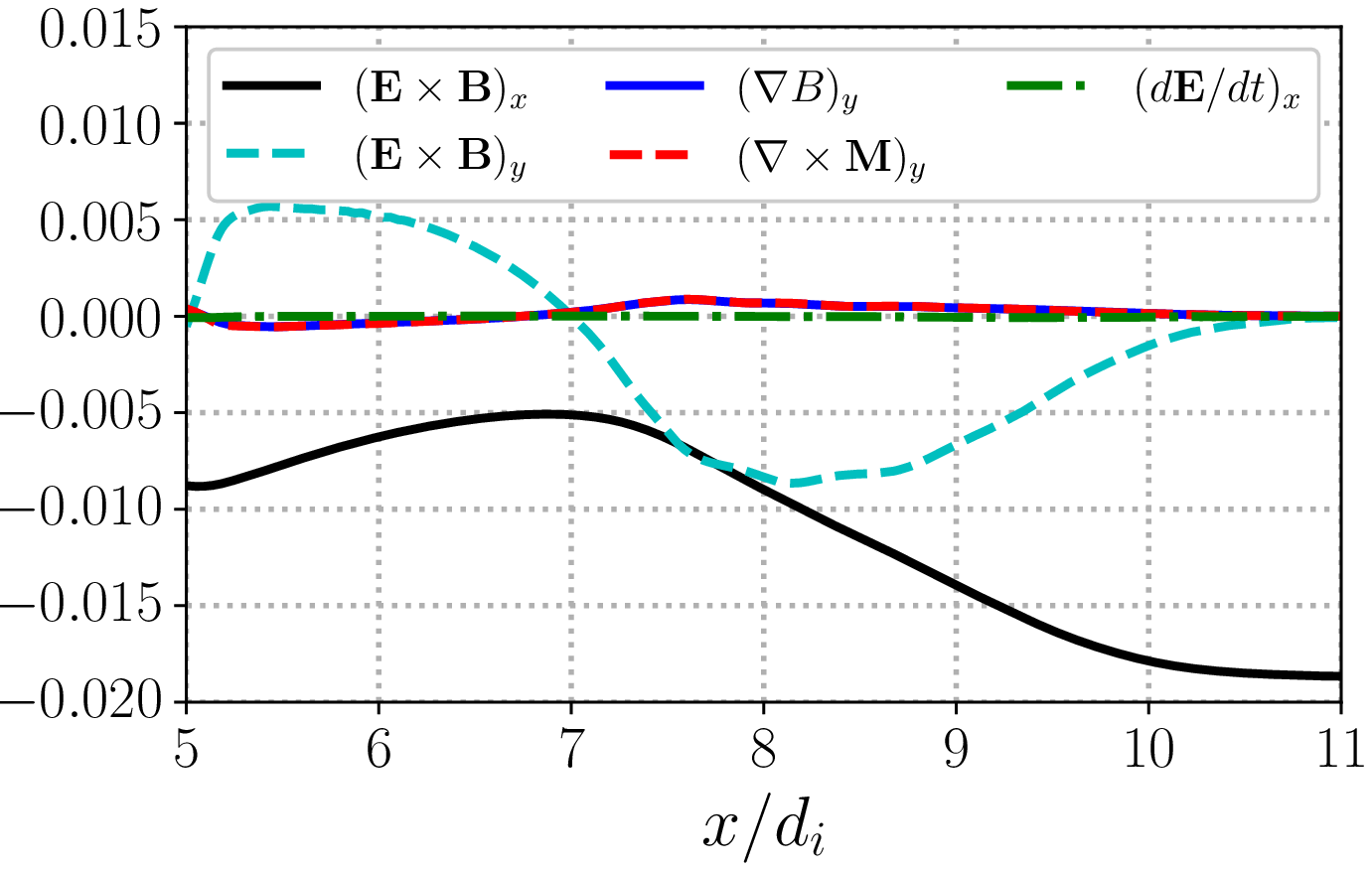}
      \includegraphics[width=0.49\textwidth]{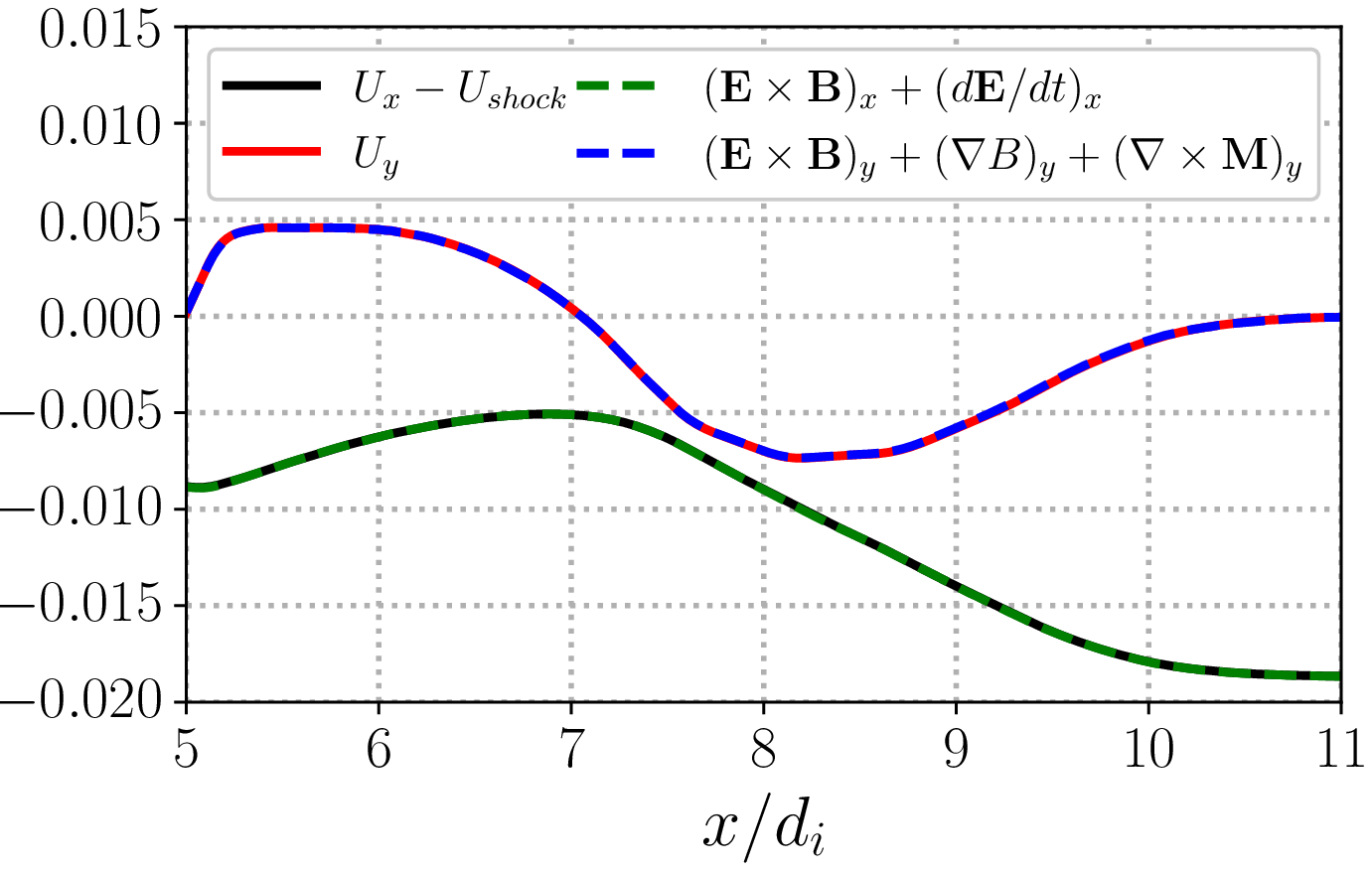}
   \end{center}
       \vskip -1.8in
\hspace*{0.35in} (a)\hspace*{2.15in} (b)
\vskip +1.7in
 \begin{center}
    \includegraphics[width=0.49\textwidth]{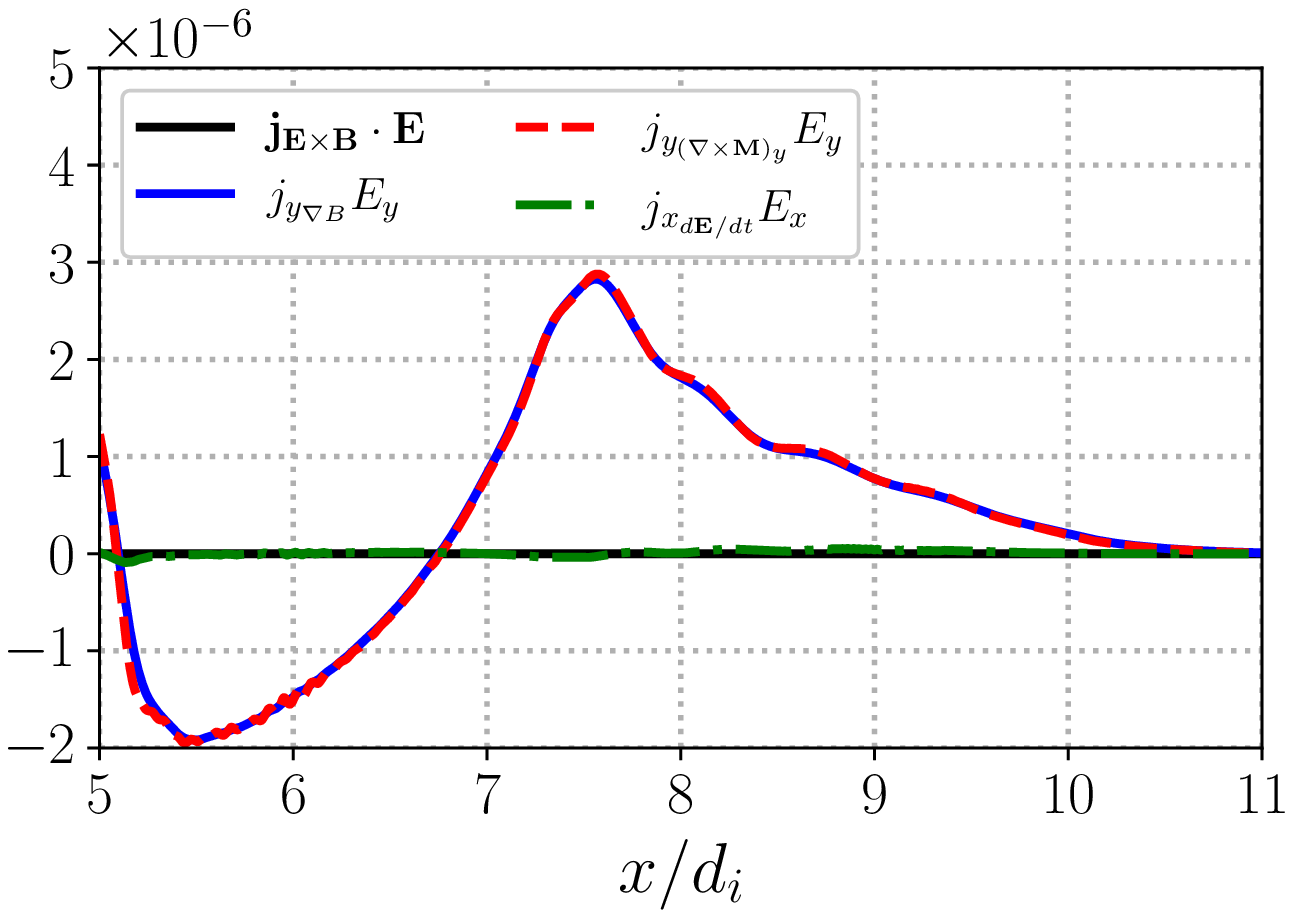}
    \includegraphics[width=0.49\textwidth]{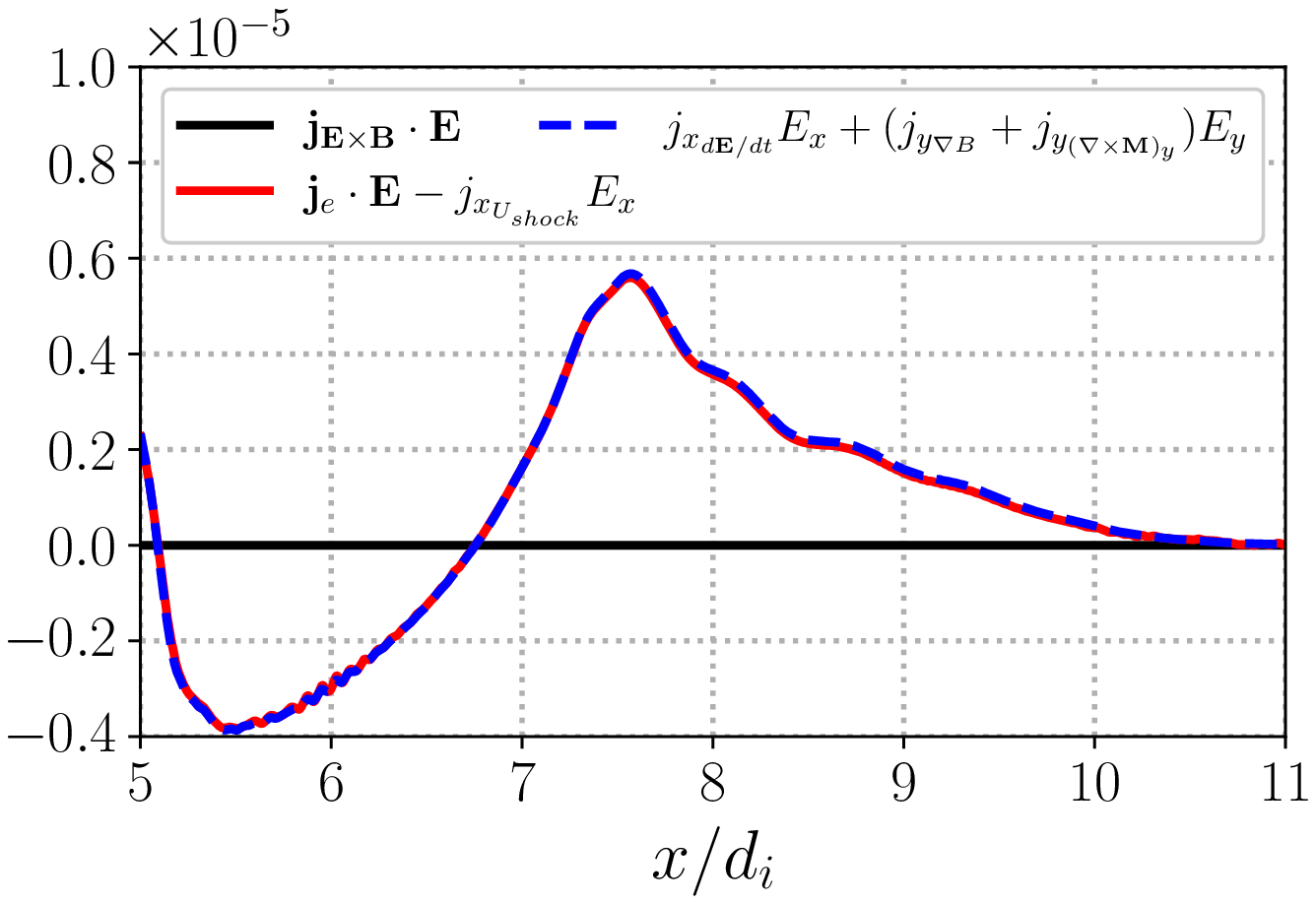}
    \end{center}
       \vskip -1.8in
\hspace*{0.35in} (c)\hspace*{2.15in} (d)
\vskip +1.7in
      \caption{A comparison of the strength of the major single-particle drifts through the shock (a), $\mvec{E} \times \mvec{B}$ in $x$ (black) and $y$ (green), the $\nabla B$ drift in $y$ (blue), magnetization drift in $y$ (red dashed), and the polarization drift in $x$ (magenta. dashed-dotted) for a $m_i/m_e = 400$. We check that these drifts sum to the total first moment computed from the electron distribution function (b) as well as determine how each of these drifts contributes to the overall energy exchange, $\mvec{j}_e \cdot \mvec{E}$ (c), and compare the $\mvec{j}_e \cdot \mvec{E}$ computed from these drifts to the total $\mvec{j}_e \cdot \mvec{E}$ computed from moments of the electron distribution function. As before in Figure~\ref{fig:drift-comp}, we sum the energy exchange due to $\mvec{E} \times \mvec{B}$ flows to demonstrate that this total energization is zero, as it should be. We note that the energization due to the combination of the $\nabla B$ and magnetization drifts more closely agrees with the energy exchange, $\mvec{j}_e \cdot \mvec{E}$ computed from moments of the electron distribution function in comparison to Figure~\ref{fig:drift-comp}.}
    \label{fig:drift-comp-mass400}
\end{figure}

\begin{figure}
    \centering
    \includegraphics[width=\textwidth]{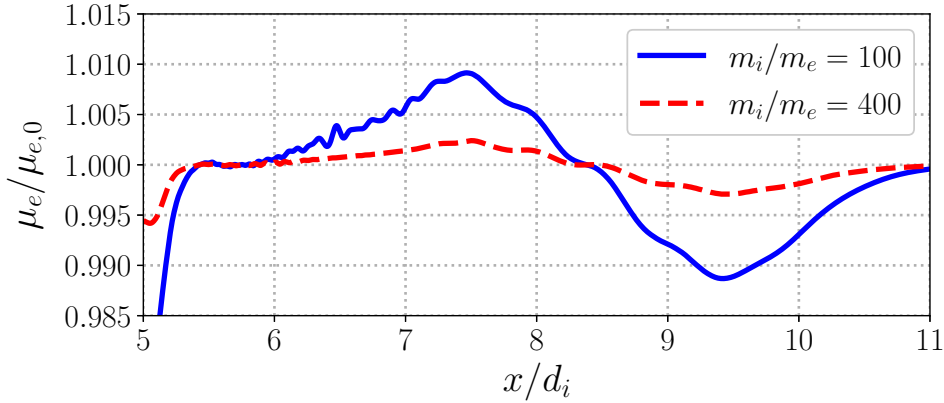}
      \caption{The electron adiabatic invariant, $\mu = T_\perp/B_z$, for the $m_i/m_e = 100$ simulation (blue) and $m_i/m_e = 400$ simulation (red dashed). The conservation of the adiabatic invariant is within $\sim 1$ percent in the $m_i/m_e = 100$ simulation, while the conservation is even better for the $m_i/m_e = 400$ simulation, suggesting that the $m_i/m_e = 400$ is even more strongly in the asymptotic limit of $\rho_e \ll L_{shock}$.}
    \label{fig:electronMu-mass400}
\end{figure}

We plot in Figure~\ref{fig:drift-comp-mass400} an identical figure to Figure~\ref{fig:drift-comp} for the $m_i/m_e = 400$ simulation to compare strengths of the same drifts of interest in Section~\ref{sec:electrons} in panel(a): $\mvec{E} \times \mvec{B}$ in $x$ and $y$, the $\nabla B$ drift in $y$, the magnetization drift, $\nabla \times \mvec{M}$, in $y$, and the polarization drift in $x$.
We also repeat the comparison of these drifts to the computed first moment from the electron distribution function (b), alongside a comparison of the amount of bulk energization arising from these drifts (c), and how it compares with the bulk energization, $\mvec{j}_e \cdot \mvec{E}$ computed from moments of the electron distribution function (d). 
We note that the agreement between the energization of the electrons arising solely from the alignment of the $\nabla B$ and the magnetization drifts with the motional electric field and the total $\mvec{j}_e \cdot \mvec{E}$ computed from moments of the electron distribution function is better than what was observed in Figure~\ref{fig:drift-comp} for the $m_i/m_e = 100$ simulation.
The more realistic mass ratio increases the scale separation between the shock-width, which remains $L_{shock} \sim d_i$, and the electron gyroradius, and thus we expect the electron adiabatic invariant to be more strongly conserved through the shock.
This stronger conservation is indeed the case, as we show in Figure~\ref{fig:electronMu-mass400} comparing $\mu_e$ computed from both the $m_i/m_e = 100$ and  $m_i/m_e = 400$ at the same time $t = 4.3 \Omega_{ci}^{-1}$.

\bibliography{abbrev.bib,transverse-simulation.bib}

\end{document}